\documentclass[aps,prb,reprint,floatfix,superscriptaddress]{revtex4-1}
\usepackage[english]{babel}
\usepackage[T1]{fontenc}
\usepackage{bbold}
\usepackage{epsfig}
\usepackage{amsmath}
\usepackage{hyphenat}
\usepackage{hyperref}
\usepackage{placeins}
\usepackage[justification=raggedright]{caption}
\usepackage{subcaption}	

\usepackage[usenames,dvipsnames]{color}

\usepackage{array}
\newcolumntype{C}[1]{>{\centering\arraybackslash}m{#1}}
\usepackage{overpic}

\newcommand{\comm}[1]{}

\begin{document}

\title{Interplay of disorder and interactions in subcritically tilted and anisotropic three-dimensional Weyl fermions}
\author{Tycho S. Sikkenk}
\author{Lars Fritz}
\affiliation{Institute for Theoretical Physics and Center for Extreme Matter and Emergent Phenomena, Utrecht University, Leuvenlaan 4, 3584 CE Utrecht, The Netherlands}

\begin{abstract}

We study the effects of disorder and Coulomb interactions on the physics of three-dimensional type-I Weyl fermions with tilted and anisotropic dispersions in a renormalization group approach. To lowest non-trivial loop order we show that the tendency of the Coulomb interactions to restore the symmetry of the dispersion in the semimetallic region of the phase diagram dominates the stabilization of the tilt and anisotropy favored by weak disorder. We argue that the topology of the renormalization flow of the disorder and Coulomb couplings is essentially determined by gauge invariance, so that these findings continue to hold qualitatively at any order in perturbation theory.  

\end{abstract}

\maketitle

\section{Introduction}

In electronic systems with band touching points the effecive low energy theory can often be expressed in terms of linearly dispersing Dirac fermions. Of these Dirac systems the two-dimensional material graphene, isolated only in 2004 ~\cite{Novoselov2004}, is the most prominent representative. In the three-dimensional Weyl subclass the massless Dirac fermions dissociate into pairs of Weyl fermions that can separate in momentum space ~\cite{Weyl1929}. Band structure calculations predict such excitations in several material families ~\cite{Weng2015weyl, Huang2016}, and they have been shown to exist in materials such as TaAs, NbAs, TaP and NbP ~\cite{Lv2015Experimental, Xu2015discovery, Xu2015discovery2, Yang2015weyl, Lv2015observation, Xu2016observation, Shekhar2015} in experiment. 

Weyl fermions exhibit interesting physical properties, such as the chiral anomaly and surface Fermi arcs, that have put them into the focus of intense research interest ~\cite{Parameswaran2014, Potter2014, Baum2015, Huang2015, Shekhar2015, Arnold2016, Zhang2016}. In the absence of perturbations, Weyl systems are semimetals with a density of states (DoS) that vanishes at the Weyl point. The corresponding nodes in the spectrum are sources and sinks of Berry curvature, and their resultant opposite topological charges imply that no gap can be opened in the spectrum without merging chiral partner modes. This makes the semimetallic (SM) phase remarkably robust to weak perturbations. 

Disorder constitues an irrelevant perturbation to the three-dimesional Weyl semimetal in the renormalization group (RG) sense. Consequently,  the system was subjected to perturbation theory based methods that showed that the SM phase prevails up to a critical disorder strength ~\cite{Fradkin1986, Altland2016, Sbierski2017}. Beyond this critical strength it enters a diffusive metallic (DM) phase characterized by a finite density of states at the Weyl point. This view is challenged by numerical results that call into question the existence of the SM phase on the basis of rare regions leading to an exponentially small but finite DoS at Fermi level ~\cite{Pixley2016}. More recent analytical works, however, insist on the stability of the SM phase and absence of these rare regions ~\cite{Buchhold2018}. Irrespective of the final resolution of this matter for intermediate energy scales the critical point between SM and DM phases should still control the physics, validating perturbative approaches. Even at high degrees of disorder Anderson localization cannot occur in a perfectly isolated Weyl cone due to the absence of available backscattering states. In realistic models comprising multiple pairs of chiral Weyl pairs backscattering processes connecting different cones are allowed, but they are supressed by the intercone distance in reciprocal space ~\cite{Altland2016}.

Furthermore, Coulomb interactions are a marginal perturbation in any number of dimensions. Due to the vanishing free density of states at the Fermi level in Weyl semimetals they are left unscreened  ~\cite{Kotov2012}. These long-ranged interactions decay quadratically in momentum space so that processes connecting well-separated Weyl nodes are surpressed. It is worth emphasizing that such interactions constitute but a particular version of quantum electrodynamics, whose central tennet is the preservation of the invariance under the $U(1)$ local gauge transformation. The coupling of the Weyl fermions with the photon mediating the Coulomb interactions should then be subject to Ward identities ~\cite{Ward1950, Peskin1995}.

In Weyl semimetal materials the Fermi velocity of the cone is much lower than the speed of light $c$. Consequently, Coulomb interactions are instantaneous on the time scale of the fermions and retardation effects are neglible ~\cite{Gonzalez1994}. Furthermore, unlike in particle physics, the effective Weyl fermions need not be Lorentz invariant. This allows for various distortions in the conical dispersion of condensed matter Weyl semimetals. Anisotropies are a common occurence in the Weyl spectrum and have several observable effects ~\cite{Trescher2015}. Tilts also appear frequently and have been predicted ~\cite{Trescher2015, Wurff2017, Das2019} and experimentally found ~\cite{Li2017} to produce clear signatures in various properties. In type-II Weyl cones the dispersion is tilted over, so that hole and particle pockets emerge in the Fermi surface besides the Weyl nodes by which the DoS develops a finite value ~\cite{Soluyanov2015typeii}. We here restrict to the subcritical tilts in type-I Weyl cones, which preserve the point-like nature of the Fermi surface. Within this model the effects of the tilt derive from increasing the number of states available at a given finite energy. 

In this paper, we present a renormalization group analysis of the interplay of disorder and interactions in three-dimensional Weyl fermions with a tilted and anisotropic dispersion. Including both disorder and electromagnetic perturbations stabilizes the SM phase in untilted isotropic cones ~\cite{Goswami2011}. Whereas the degree of tilting is increased by disorder ~\cite{Sikkenk2017}, Coulomb interaction tends to decrease it ~\cite{Detassis2017}. Coulomb interactions cause flow towards restoration of the isotropy of the Weyl dispersion, while disorder effects magnify the anisotropy between different momentum directions. The goal of this study is then to investigate the combined effect of these competing tendencies and the possible existence of new fixed points. 

The two main results of this study are: (I) despite competing tendencies to lowest loop order there is no new fixed point and like in the untilted case the SM phase is stabilized by the interplay of disorder and Coulomb interaction. (II) Using a field transformation to capture all RG generated terms we find a Ward identity by which this result holds to {\it all} orders in perturbation theory. 

\vspace{12 pt}

The main body of the paper is organized as follows. Sec.~\ref{sec:model} introduces the model and especially discusses the need for a field transformation to capture all terms generated under RG transformation. Sec.~\ref{sec:RGequations} presents the flow equations resulting from a lowest loop order expansion. Sec.~\ref{sec:RGinterpretation} follows with a discussion of the flow equations. We first treat various limiting cases analytically in Subsecs.~\ref{subsec:untilted_dis}-\ref{subsec:tilted_int} before numerically considering the full RG flow of the general theory in Subsec.~\ref{subsec:tilted_disint}. In Sec.~\ref{sec:WardIdentities} we establish a Ward identity of the model and discuss its implications for calculations at any order in perturbation theory. We finish with a conclusion in Sec.~\ref{sec:conc}.

\section{Model Setting}	\label{sec:model}

In this section we present our model and set it up for RG treatment. While we have relegated the technical details of the diagrammatic calculations to appendix ~\ref{appsec:diagrams}, they eventually result in the set flow equations presented in Sec.~\ref{sec:RGequations}. 

As argued in the introduction, the separation of Weyl cones in momentum space both represses processes that merge nodes of opposite topological charge, without which no gap can open in the dispersion, and subdues intercone disorder scattering and Coulomb interactions. We thus reasonably start from the generalized Bloch Hamiltonian of a single isolated Weyl fermion in $d+1$ dimensions, 
\begin{align}	\label{eq:ham}
\mathcal{H}' ({ \bf q}) =v' \left[ t' \, q_\parallel \sigma_0 + \chi \, \left( q_\parallel {\bf d} + \eta' \, {\bf q_\perp} \right) \cdot {\pmb \sigma} \right],
\end{align}
where the dot product implies summation over $d$ spatial dimensions. The Weyl matrices $\sigma_\mu$ are $2^{m-1}$ dimensional, where $m = \text{floor} \left[ (d+1)/2 \right]$, and satisfy the anticommuting Clifford algebra $\{ \sigma_\mu, \sigma_\nu \} = \delta_{\mu \nu}$ for $\mu, \nu = 0,1, \dots, d$. Here, $\sigma_0$ may be interpreted as the $2^{m-1}$ dimensional identity matrix. Note that in the case that $d=3$ the Weyl matrices reduce to the commonly known Pauli matrices. We have furthermore introduced a momentum parametrization where component $q_\parallel = {\bf d} \cdot {\bf q}$ is oriented along the unit vector ${\bf d}$ while ${\bf q_\perp} = {\bf q} - q_\parallel {\bf d}$ is a radial projection onto the $(d-1)$ dimensional perpendicular plane. The chirality of the Weyl node is determined by $\chi = \pm 1$. The Hamiltonian of Eq.~\eqref{eq:ham} results in a dispersion given by
\begin{align} 	\label{eq:tilt_disp}
E'_s ({\bf q}) = v' \left( t' q_\parallel  + s \sqrt{q_\parallel^2 + \eta'^2 q_\perp^2} \right) \,, 
\end{align}
where $s = \pm 1$ distinguishes electron and hole bands. The Fermi velocity is given by $v'$, while $\eta'$ controls the possible development of anisotropy between the parallel and perpendicular momentum directions. The parameter $t'$ tilts the Weyl cone in the direction of ${\bf d}$. Increasing $t' $ causes the band structure to tilt over until the fermion becomes dispersionless as $ t' \to 1$. This breakdown of the SM phase is also apparent in the divergence at overtilting of the DoS
 \begin{align} 	\label{eq:FreeDoS}
\rho_0 (\omega) = -\frac{1}{\pi} \int_{\bf q} \; \text{Im} \; \text{Tr} \; G'_0 (\omega + i 0^+, {\bf q})  \sim \frac{N \omega^2 }{v'^3 \eta'^2 (1-t'^2)^2},
\end{align} 
where we have used the propagator corresponding to the Weyl Hamiltonian of Eq.~\eqref{eq:ham}, 
\begin{align}
G_0'^{-1}	&= i \omega \, \sigma_0 - \mathcal{H}' \label{eq:P_FermionGreenFunc}\\
		&= (i \omega - v' t' \, q_\parallel) \sigma_0 - v' \chi (q_\parallel {\bf d} +  \eta' {\bf q_\perp} ) \cdot {\pmb \sigma}. \nonumber
\end{align}
In the remainder of this work, we concentrate on type-I Weyl cones with subcritical tilts $0 \leq t' < 1$. 

\vspace{12 pt}

We are interested in the behavior of Weyl fermions in a disordered background that is described by a quenched potential landscape $V ({\bf x})$ obeying a Gaussian white-noise distribution with
\begin{align} 	\label{eq:dis_distr}
\langle V ( {\bf x} ) \rangle = 0, \qquad \langle V ( {\bf x}) \, V ( {\bf x}' ) \rangle \sim \delta ( | {\bf x} - {\bf x}' |).
\end{align}
We average over the random potential using the replica trick, so preserving generic disorder properties ~\cite{Edwards1975, Altland2010}. This entails taking many copies of the theory, promoting the various disorder distributions to a collective field $V$ that is integrated over in the partition function with Gaussian weight
\begin{align}
S_V	= \frac{1}{2} \int_ { {\bf q} } V_{\bf q} \, V_{-{\bf q}},
\end{align}
and finally taking the number of replicas to zero in the limit. Suppressing the summed over replica indices, this results in a free fermion action 
\begin{align}	\label{eq:P_FermionAction}
S'_\psi = \int_{ {\bf q}, \omega} \psi'^\dagger_{{\bf q}, \omega} G_0'^{-1} \psi'_{{\bf q}, \omega} .
\end{align}
We couple the external disorder field to the density of the Weyl fermions according to
\begin{align}	 \label{eq:P_DisAction}
S'_\text{dis} = \int_{ {\bf q}, {\bf q}', \omega}  V_{{\bf q} - {\bf q}'} \psi^\dagger_{{\bf q}, \omega} \, \left( \Gamma \, \sigma_0 \right) \, \psi_{ {\bf q}',  \omega}.
\end{align}
Note that this approach is equivalent but technically differs from the more standard way of treating the disorder in which the field $V$ is integrated out explicitly, resulting in a four-fermion interaction term.

The Weyl fermions are furthermore coupled amongst each other by means of long-range Coulomb interactions. This is represented as
\begin{align}	\label{eq:P_CouAction}
S'_\text{Cou} = \int_{ {\bf q}, {\bf q}', \omega, \omega' } \varphi_{{\bf q} - {\bf q}', \omega - \omega'} \, \psi^\dagger_{{\bf q}, \omega} \, \left( i g \, \sigma_0 \right) \, \psi_{{\bf q}', \omega'}, 
\end{align}
where the fermions are interacting by means of a scalar gauge photon, whose free propagation is given by 
\begin{align}	\label{eq:photon_action} 
S_\varphi	= \frac{1}{2} \int_{ {\bf q}, \omega} \varphi_{{\bf q}, \omega} \, D_0^{-1} \varphi_{-{\bf q}, -\omega}. 
\end{align}
In Weyl semimetals the Fermi velocity is typically much smaller than the speed of light, $v' \ll c$, so that retardation effects in the Coulomb interaction can be safely neglected ~\cite{Gonzalez1994}. As a consequence, the bare photon propagator is taken to be
\begin{align}	\label{eq:photon_prop}
D_0^{-1} = q^{1-d+ {\bar \epsilon} } = \left( q_\parallel^2 + q_\perp^2 \right)^{ (1-d + {\bar \epsilon})/2}, 
\end{align}
where ${\bar \epsilon} \to 0$ is a dimensional regulator that is introduced for technical reasons. 

In real materials the dispersion can feature many different pairs of Weyl cones at comparable energies  ~\cite{Weng2015weyl, Lv2015observation}. In the above we have implicitly neglected the disorder scattering between different cones as it is surpressed by their momentum space distance ~\cite{Altland2016, Xiong2015}. The intercone Coulomb interaction, in three dimensions decaying as $\sim 1/q^2$, is similarly subdued by the cone separation and is thus also omitted ~\cite{Detassis2017}. As a consequence the model is composed of $N$ independent sectors evenly describing individual left- and right-handed Weyl fermions. 

\vspace{12 pt}

As was noted previously, perturbing the bare system $S'_\psi$ by disorder action $S'_\text{dis}$ generates a new term $\sim t' \, i \omega \, {\bf d} \cdot {\pmb \sigma}$ in the self-energy contribution once a finite tilt $t' > 0$ is included  ~\cite{Sikkenk2017}. This was also observed recently in the context of two dimensional Dirac fermions perturbed by various types of disorder ~\cite{Zhao2018, Yang2018}. In those works this issue was resolved by adding the term to the bare Green function by hand and treating it as a bona fide, stand-alone parameter of the theory. We here propose a different scheme to manage such terms, in which we absorb the anomalous contributions in a redefinition of the fermion field. This will have ramifications for both the bare parameters and the couplings of the theory we consider.

We transform the fermion field according to 
\begin{align}	\label{eq:field_transf}
\psi'_{{\bf q}, \omega} =  \hat{\lambda}^{1/2} \, \psi_{{\bf q}, \omega}, \qquad {\hat \lambda} =  \sigma_0 - \lambda \, \chi \, {\bf d} \cdot {\pmb \sigma}.	 
\end{align} 
This transformation matrix equals the identity at the beginning of our RG procedure and only perturbatively obtains a non-trivial structure. Under influence of this shift the free fermion action of Eq.~\eqref{eq:P_FermionAction} becomes
\begin{align}
S_{\psi} = \int_{ {\bf q}, \omega} \psi^\dagger_{{\bf q}, \omega} G_0^{-1} \psi_{{\bf q}, \omega}
\end{align}
with a modified (inverse) Green function
\begin{align}
&G_{0}^{-1} = \hat{\lambda}^{1/2} G_0'^{-1} \hat{\lambda}^{1/2}  	\label{eq:FermionGreenFunc} \\ 
& \quad =(i \omega \hat{\lambda} - v t q_\parallel) \sigma_0 - v \chi \left(q_\parallel {\bf d} +  \eta {\bf q_\perp} \right)  \cdot {\pmb \sigma}. \nonumber
\end{align}
This propagator has the same flavor as the original, with parameters that are related as 
\begin{align} 	\label{eq:ParamRedif}
v = v' \, (1 - t' \lambda), \quad \eta = \eta' \, \frac{ \sqrt{1-\lambda^2}}{1 - t' \lambda}, \quad t = \frac{t' - \lambda}{1 - t' \lambda},
\end{align}
or alternatively,
\begin{align}  	\label{eq:ParamReRedif}
v' = v \, \frac{1+ t \lambda }{1-\lambda^2}, \quad \eta' = \eta \, \frac{ \sqrt{1-\lambda^2}}{1+ t \lambda}, \quad t' = \frac{t+\lambda}{1+ t \lambda}.
\end{align}
Unlike in Eq.~\eqref{eq:P_FermionGreenFunc}, however, in the transformed propagator Eq.~\eqref{eq:FermionGreenFunc} the frequency $i \omega$ is supplemented with the matrix structure of the transformation that can absorb the contributions deriving from the disorder-induced self-energy. Since $\det{G_0} = \det{G'_0} / \det{ {\hat \lambda} }$, the poles of the Green function remain unchanged under transformation Eq.~\eqref{eq:field_transf} and the energy spectrum remains given by Eq.~\eqref{eq:tilt_disp}. Rather, the newly generated parameter $\lambda$ acts on the level of the quasiparticle weight attributed to the excitations of the system. 

The field transformation also impacts the coupling terms of the action, Eqs.~\eqref{eq:P_DisAction}-\eqref{eq:P_CouAction}. They become
\begin{align}
S_\text{dis} 	&= \int_{ {\bf q}, {\bf q}', \omega}  V_{{\bf q} - {\bf q}'} \psi^\dagger_{{\bf q}, \omega} \, \left( \Gamma \, {\hat \lambda} \right) \, \psi_{ {\bf q}',  \omega}, \label{eq:DisAction} \\
S_\text{Cou} 	&= \int_{ {\bf q}, {\bf q}', \omega, \omega' } \varphi_{{\bf q} - {\bf q}', \omega - \omega'} \, \psi^\dagger_{{\bf q}, \omega} \, \left( i g \, {\hat \lambda} \right) \, \psi_{{\bf q}', \omega'}. \label{eq:CouAction}
\end{align}
As the above equations show, one of the merits of the transformation procedure Eq.~\eqref{eq:field_transf} is that minimal coupling between frequency $i \omega$ on the one hand and gauge field $\varphi$ and external field $V$ on the other hand is respected by construction. The same parameter $\lambda$ is now present not only in the Green function Eq.~\eqref{eq:FermionGreenFunc} but also in both interacting parts of the action. We will find that it will obtain identical renormalizations in each of these sections.

\section{RG equations}	\label{sec:RGequations}

We study the action 
\begin{align} 	\label{eq:Action}
S_0 = S_{\psi} + S_V + S_\varphi, \qquad S_\text{int} = S_\text{dis} + S_\text{Cou}, 
\end{align}
in the framework of RG. Under anisotropic space-time rescaling 
\begin{align} 	\label{eq:SpacetimeRescaling}
\omega \to \mu^{+ z} \omega, \qquad {\bf q} \to \mu^{+1} {\bf q},
\end{align}
the parameters and fields change as $y_{i,0} \to y_i (\mu) =  y_{i,0} \, \mu^{+ [y_i]} \, Z_{y_i}^{-1}$, where $Z_{y_i} = 1 + \delta_{y_i}$. We determine the scaling dimensions of the fields from $S_0$ to be $[\psi_{{\bf q}, \omega}] = - (d + 2z)/2$ for the fermion field, $ [V_{\bf q}] = -d/2$ for the disorder field, $[\varphi_{{\bf q}, \omega}] = - (2d + z -1 - {\bar \epsilon} )/2 $ for the photon field and $[v]= z-1$ for the Fermi velocity. The other parameters are scale-invariant, {\it i.e.}, $[t]=[\lambda]=[\eta]=0$. 

The disorder coupling has dimension $[\Gamma] = z - d/2$ meaning for a free Weyl theory ($z=1$) it is marginal in $d=2$ and irrelevant in $d=3$. 
The Coulomb interaction mediated by the photons has scaling dimension
$[g]= (z-1 - {\bar \epsilon} )/2$, where ${\bar \epsilon} \to 0$ in the end. In that case Coulomb interactions are marginal irrespective of the number of spatial dimensions.
In the following we perform a double $\epsilon$-expansion around the marginal dimension, $d = 2$ and ${\bar \epsilon}=0$ by working in $d=2+\epsilon$. We keep $\epsilon$ and ${\bar \epsilon}$ finite throughout the calculation and take $\epsilon \to 1$, corresponding to three dimensions, and ${\bar \epsilon} \to 0$ in the end. Such dimensional regularization is known to respect gauge invariance~\cite{Peskin1995}.

We perform a one loop expansion in $\Gamma$ and Coulomb interaction strength $g$. Within our scheme the tilt $t'$, anisotropy $\eta'$ and the generated parameter $\lambda$ are treated non-perturbatively. This results in the set of diagrams presented in appendix ~\ref{appsec:diagrams}. The required counterterms and derived flow equations are set out in appendix ~\ref{appsec:Counterterms}. In appendix~\ref{appsec:ReexpBetaFncs} they are translated back to $\beta$ functions of the original model parameters appearing in the dispersion by using Eq.~\eqref{eq:ParamRedif}.

This results in a primary set of four coupled flow equations
\begin{align}
&\beta_{t'} = t' \left\{\frac{1}{\eta' \sqrt{1-t'^2} } \gamma'^2 -  \alpha' F_\parallel^{\eta'} \right\}, \label{eq:RG_tilt} \\
&\beta_{\eta'} = - \eta' \left\{ \frac{t'^2}{\eta' (1-t'^2)^{3/2}} \gamma'^2 + \alpha' \left( F_\parallel^{\eta'} - F_\perp^{\eta'} \right)  \right\},\\
& \beta_{\alpha'}	 = \alpha' \left\{ - {\bar \epsilon} + \frac{1}{\eta' \sqrt{1-t'^2} } \gamma'^2 - \alpha' F_\parallel^{\eta'} \right\}, 	\label{eq:RG_Cou}	\\
& \beta_{\gamma'} = \gamma' \left\{ -\frac{\epsilon}{2} + \frac{1}{\eta' \sqrt{1-t'^2} } \gamma'^2 - \alpha' F_\parallel^{\eta'} \right\}, 	\label{eq:RG_dis}
\end{align}
where  $\beta_{y_i} = \mathrm{d} y_i / \mathrm{d} \ln \mu $ for parameters $y_i$. We stress that the limits ${\bar \epsilon} \to 0$, $\epsilon \to 1$ should be taken in the end. Note also that we have introduced dimensionless couplings according to  
\begin{align}	\label{eq:CouplingRedif}
\gamma'^2	= \frac{\Omega_d \, \mu^\epsilon }{(2 \pi)^d \, v'^2} \, \Gamma^2, \qquad \alpha' = \frac{\Omega_d \, \mu^{ {\bar \epsilon} } }{4 (2 \pi)^d \, v'} \, g^2. 		
\end{align}
We have furthermore defined two functions that depend solely on the anisotropy $\eta'$, 
\begin{align}
F_\parallel^{\eta'}	&= \frac{4\eta' \left[ \text{EllipticE}(1-\eta'^{-2}) - \text{EllipticK}(1-\eta'^{-2}) \right]   }{\pi (1-\eta'^2)}, \label{eq:anisfuncpar} \\ 
F_\perp^{\eta'}		&=  -\frac{ 4 \left[ \text{EllipticE}(1-\eta'^2) - \text{EllipticK}(1-\eta'^2) \right]   }{\pi (1-\eta'^2)}. \label{eq:anisfuncperp}
\end{align}
When isotropy is restored, $\eta' =1$, these functions return $F_\parallel^{\eta'=1} = F_\perp^{\eta'=1} = 1$. 

\vspace{12 pt}

Apart from the coupled set Eqs.~\eqref{eq:RG_tilt}-\eqref{eq:RG_dis}, the behavior of the remaining parameters obey
\begin{align}
&\beta_{v'} = v' \left\{ z-1 - \frac{1}{\eta' \sqrt{1-t'^2} } \gamma'^2 + \alpha' F_\parallel^{\eta'} \right\}, \label{eq:RG_FermiVelocity}  \\
&\beta_\lambda	= t' \frac{1- \lambda^2}{\eta' (1-t'^2)^{3/2} } \gamma'^2.  \label{eq:RG_lambda}
\end{align}
Note from the last equation that the spontaneous generation of a finite value transformation parameter $\lambda$ is conditional on the simultaneous presence of non-zero tilt and disorder.

\section{Discussion of the RG equations} 	\label{sec:RGinterpretation}

The $\beta$ functions in Eqs.~\eqref{eq:RG_tilt}-\eqref{eq:RG_dis} form a closed set that describes the scale dependence of the general theory of a tilted and anisotropic Weyl semimetal under influence of disorder and interactions. In the following we indicate the initial parameters by a subscript zero in keeping with previously used terminology ~\cite{Sikkenk2017}. We stress that the field transformation Eq.~\eqref{eq:field_transf} is necessary only to account for perturbatively generated terms, so that we strictly have $\lambda_0 = 0$ initially. 

There are multiple parameter combinations for which the primary $\beta$ functions vanish simultaneously. These fixed points are indicated by subscript asterisk. The corresponding values for the secondary parameters $v'$ and $\lambda$ can then be obtained from their flow equations by substitution. The fixed points are characterized by a set of exponents that control the critical physics in their vicinity. The correlation length $\xi$ scales as $\xi \sim \delta^{- \nu}$, where $\delta$ corresponds to a linearization of the most relevant operator around the fixed point and $\nu$ is the correlation length exponent (CLE). For finite $\nu > 0$ the correlation length diverges when $\delta \to 0$ on approach to the fixed point, a critical indication the system is undergoing a phase transition. Technically, $\nu$ is the inverse of the most repulsive eigenvalue of the linearized RG transformation matrix $M_{ij} = \left. \partial \beta_{y_i}/ \partial y_j \right|_*$ at the fixed point. Since the parameter transformations Eq.~\eqref{eq:ParamReRedif} can become singular the CLE is best derived from the flow equations of the parameters presented in appendix \ref{appsec:Counterterms}. Another exponent is straightforwardly found from Eq.~\eqref{eq:RG_FermiVelocity}. Keeping the Fermi velocity $v'$ fixed requires a scale-dependent dynamical scaling exponent (DSE)
\begin{align}	 \label{eq:DSE}
z=1 + \left[ \frac{1}{\eta' \sqrt{1-t'^2} } \gamma'^2 - \alpha' F_\parallel^{\eta'} \right]_*  . 
\end{align}
at the critical point. This is highly significant as the DoS of the three-dimensional Weyl cone model scales with the energy away form the band touching point as 
\begin{align}
\rho (\omega) \sim | \omega |^{(3-z)/z}
\end{align}
in the SM phase, including close to the phase transition into the DM ~\cite{Kobayashi2014}. Both disorder and Coulomb interactions modify the DSE away from unity and could then lead to perturbative corrections to the square scaling of the free DoS in Eq.~\eqref{eq:FreeDoS}. 

\vspace{12 pt}

In order to interpret the flow produced by the RG equations it is instructive to first consider some of the limiting cases.


\subsection{Untilted, disordered case} \label{subsec:untilted_dis}

Firstly we investigate the untilted non-interacting model as in Refs.~\cite{Altland2016, Roy2014}, corresponding to initial values $\lambda_0 = t'_0 = \alpha'_0 = 0$. All the primary flow equations reduce to zero, except for the disorder $\beta$ function
\begin{align}
\beta_{\gamma'} = \gamma' \left\{ -\frac{\epsilon}{2} + \frac{1}{\eta_0'} \gamma'^2 \right\}.
\end{align}
Note that the presence of fermion anisotropy $\eta'$ is trivial only affecting the flow as a numerical factor and therefore omitted in our discussion. The disorder flow is typified by two distinct fixed points. First of all, there is the trivial attractive fixed point at $\gamma'_* =0$. Here we find that $\nu = 0$ and $z=1$, reflecting the irrelevance of the disorder perturbation. This fixed point is associated with the clean SM phase asymptotically described by the bare action Eq.~\eqref{eq:P_FermionAction}, in which the DoS scales quadratically with the energy. Secondly there is a non-trivial fixed point at intermediate disorder $\gamma_*' = \sqrt{\eta'_0 \epsilon/2}$. It is repulsive, separating the weakly disordered SM phase from the strongly disordered DM phase at critical value $\gamma'_{0,c} = \gamma'_*$. This SM-DM phase transition is characterized by a correlation length diverging with exponent $\nu$. For the dynamical critical exponent we find from Eq.~\eqref{eq:DSE} that $z = 3/2$. Close to the critical point the density of states is enhanced by strong disorder effects, scaling linearly away from the nodal point before becoming finite in the DM phase.

\subsection{Untilted, interacting case} \label{subsec:untilted_int}

\begin{figure}[h]
\centering
\includegraphics[width=0.4\textwidth]{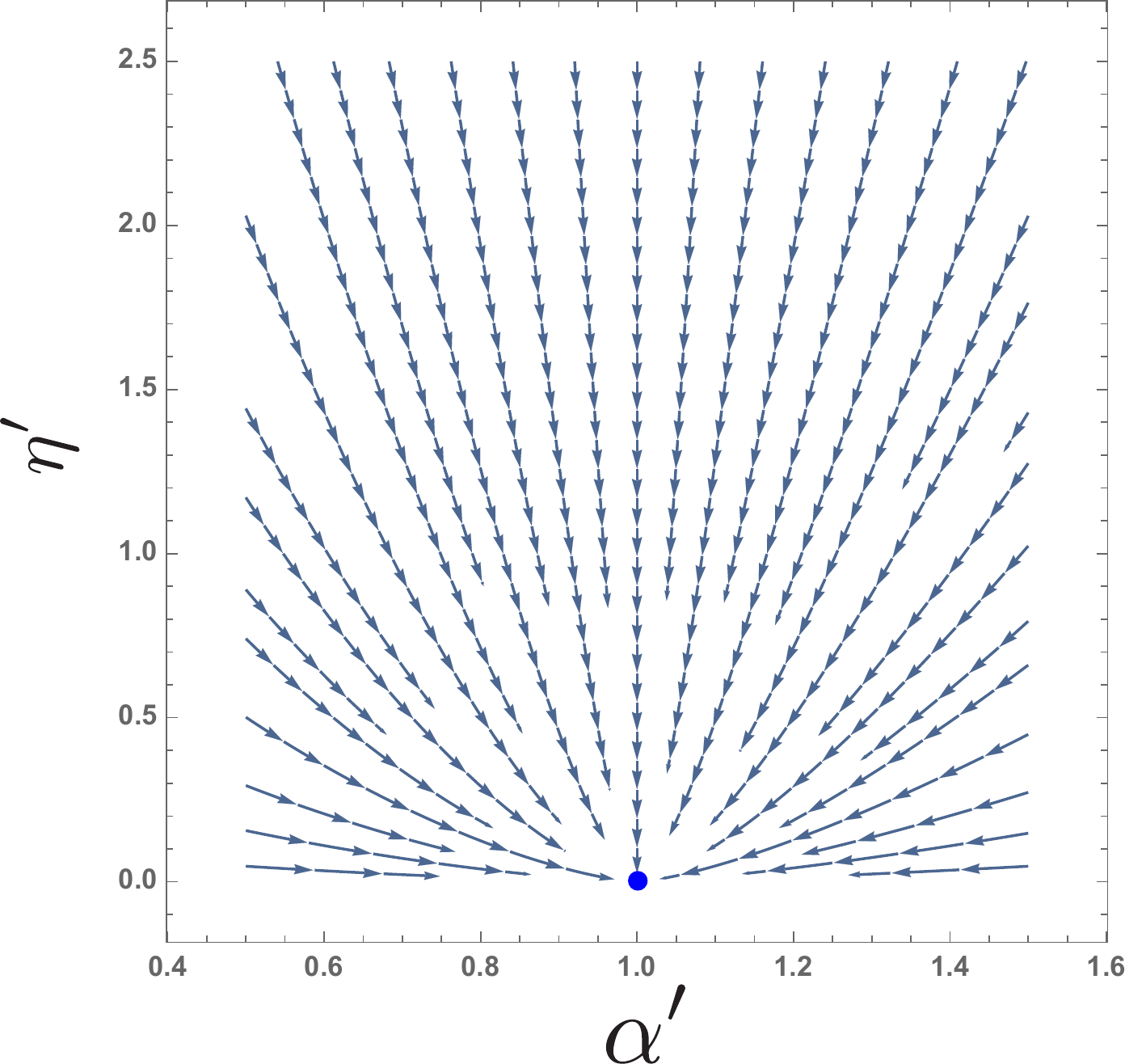}
  \caption{Streamplot of the flow deriving from equations \eqref{eq:UntiltedAnisCou_RG} of the untilted but anisotropic and interacting model. Trivial fixed point in blue.}
  \label{fig:Flow_UntiltedAnisCou}
\end{figure}

Alternatively we reflect on the basic model of an untitled Weyl cone with Coulomb interactions. Its purely isotropic limit was first studied in Ref.~\cite{Abrikosov1971}, and we here include the possibility of anisotropy in the Weyl fermion dispersion. We set out from the clean, untilted limit $t'_0 = \gamma'_0 = 0$. The primary $\beta$ functions \eqref{eq:RG_tilt}-\eqref{eq:RG_dis} reduce to
\begin{align}
\beta_{\eta'} = - \eta' \alpha' \left( F_\parallel^{\eta'} - F_\perp^{\eta'} \right), \quad \beta_{\alpha'}	 = -\alpha' \left\{ {\bar \epsilon} + \alpha' F_\parallel^{\eta'} \right\},\label{eq:UntiltedAnisCou_RG}
\end{align}
and the others vanishing. The two-dimensional flow described by these equations is presented in streamplot Fig.\ref{fig:Flow_UntiltedAnisCou}. It is invariably directed towards the trivial non-interacting fixed point $\alpha'_*=0$ where also the isotropy is restored, $\eta'_*= 1$. The corresponding critical exponents are $\nu = 0$ and $z=1$, reflecting the irrelevance of Coulomb interaction. The DoS receives logarithmic corrections that decrease it with compared to its free quadratic scaling behavior.

\subsection{Untilted, disordered, interacting case} \label{subsec:untilted_dis_int}

\begin{figure}[h]
\centering
\includegraphics[width=0.38\textwidth]{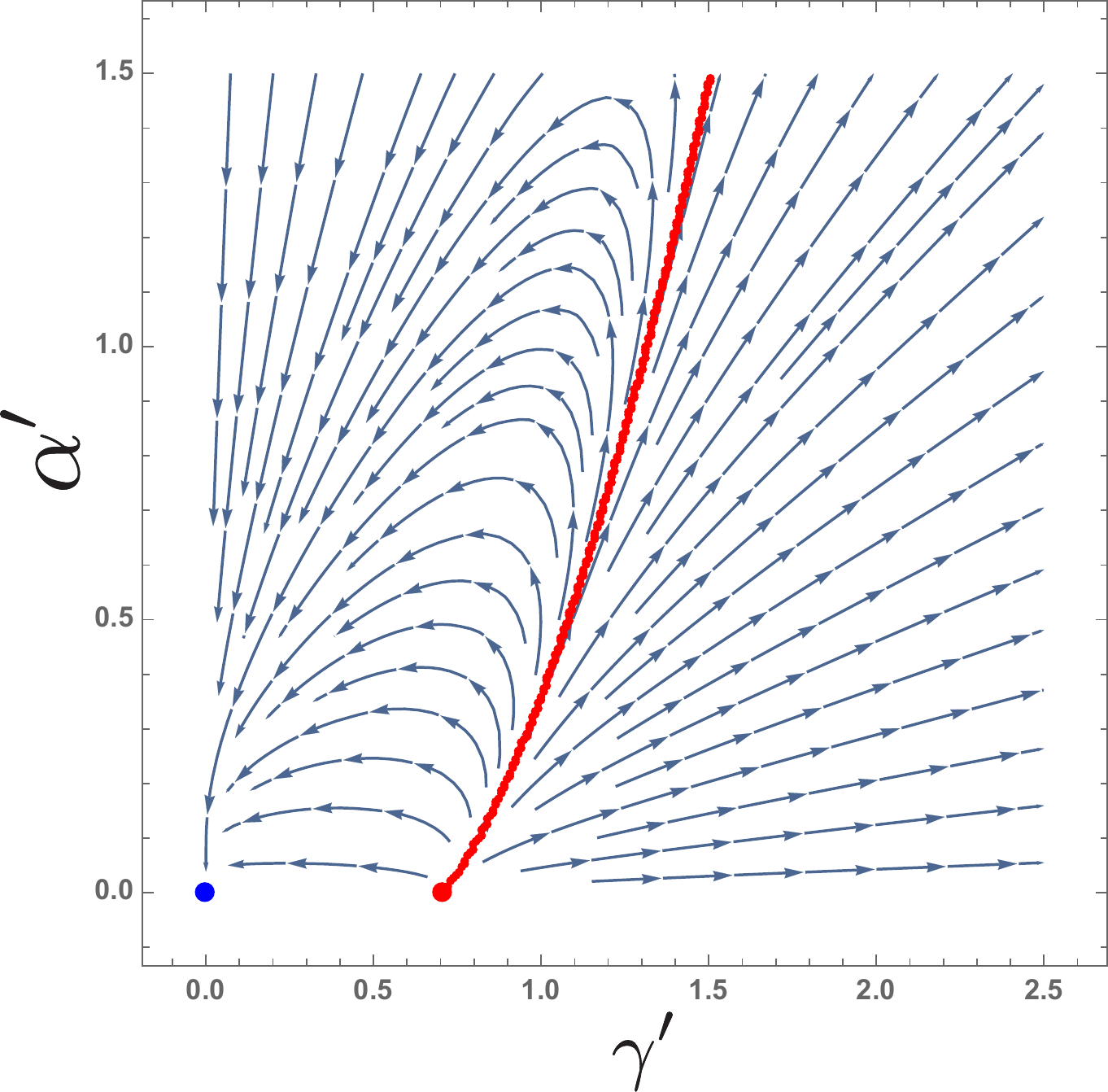}
  \caption{Streamplot of the flow corresponding to the isotropic limit $\eta'_0 =1$ of equations \eqref{eq:UntiltedAnisDisCou_RG} for the untilted model perturbed by disorder and interactions. Trivial fixed point in blue, numerical approximation of the phase boundary $\gamma'_{0,c}$ between SM and DM in red.}
  \label{fig:Flow_UntiltedIsoDisCou}
\end{figure}

Perturbing the free Weyl fermion model with both disorder and Coulomb interactions, the setup studied in Refs.~\cite{Goswami2011, Gonzalez2017} without anisotropy, means interplay effects might appear. The primary $\beta$ functions are 
\begin{align}
&\beta_{\eta'} = - \eta' \, \alpha' \left( F_\parallel^{\eta'} - F_\perp^{\eta'} \right),\\
& \beta_{\alpha'}	 = \alpha' \left\{ - {\bar \epsilon} + \frac{1}{\eta'  } \gamma'^2 - \alpha' F_\parallel^{\eta'} \right\}, \\
& \beta_{\gamma'} = \gamma' \left\{ -\frac{\epsilon}{2} + \frac{1}{\eta'} \gamma'^2 - \alpha' F_\parallel^{\eta'} \right\}.	\label{eq:UntiltedAnisDisCou_RG}
\end{align}
which produce the two-dimensional flow depicted in Fig.~\ref{fig:Flow_UntiltedIsoDisCou} in the isotropic case $\eta'_0=1$. Due to the different scaling dimensions of the perturbations these equations cannot simultaneously vanish at finite disorder and Coulomb interaction and there are no new fixed points. The anisotropy of the model influences the flow quantitatively but does not fundamentally change its topology as there is no competition in its $\beta$ function. Within the SM region all flow is directed towards the previously encountered trivial point $\alpha'_* =0$, $\gamma'_* = 0$ and $\eta'_*=1$ with exponents $\nu = 0$ and $z=1$ at which the model is asymptotically free, clean and isotropic. In the disorder-only limit there is also the non-trivial fixed point for $\alpha'_* = \alpha'_0 = 0$ and  $\gamma'_* = \sqrt{ \eta'_* \epsilon/ 2 }$ with $\eta'_* = \eta'_0$ that governs the phase transition into the DM state. It is perturbatively destabilized by the Coulomb interaction, extending into a phase boundary that expands the SM region towards higher disorder along which CLE is unchanged at $\nu = 1$~\cite{Goswami2011}.

\subsection{Tilted disordered case} 	\label{subsec:tilted_dis}

\begin{figure}[h]
\centering
\includegraphics[width=0.4\textwidth]{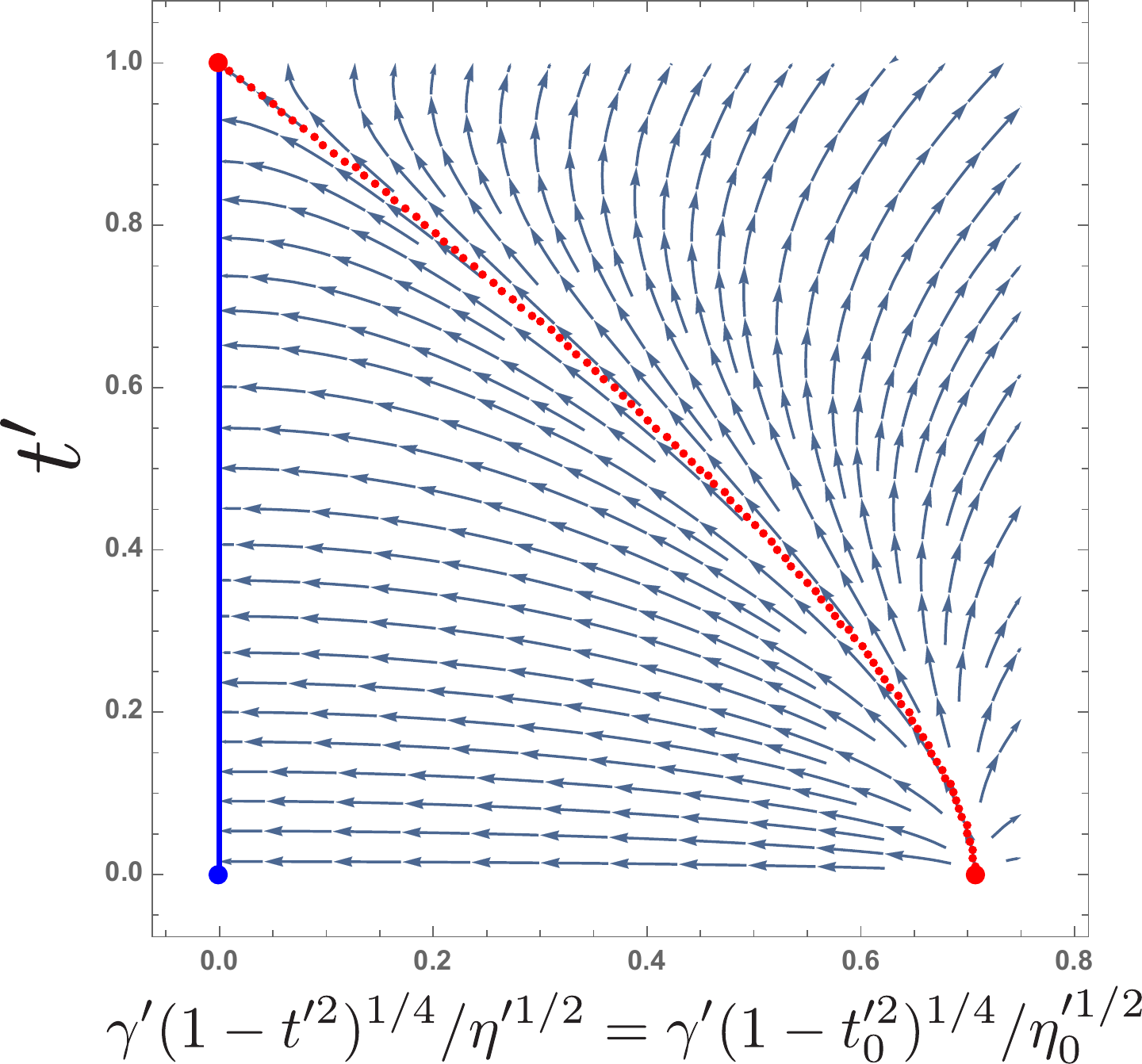}
  \caption{Streamplot of the flow deriving from equations \eqref{eq:TiltedDis_RG} of the tilted disordered model. Line of trivial fixed points at $0 \geq t'_* <1$ in blue, numerical approximation of the phase boundary $\gamma'_{0,c}$ between SM and DM in red.}
  \label{fig:Flow_TiltedDis}
\end{figure}

\begin{figure}[h]
\centering
\includegraphics[width=0.3\textwidth]{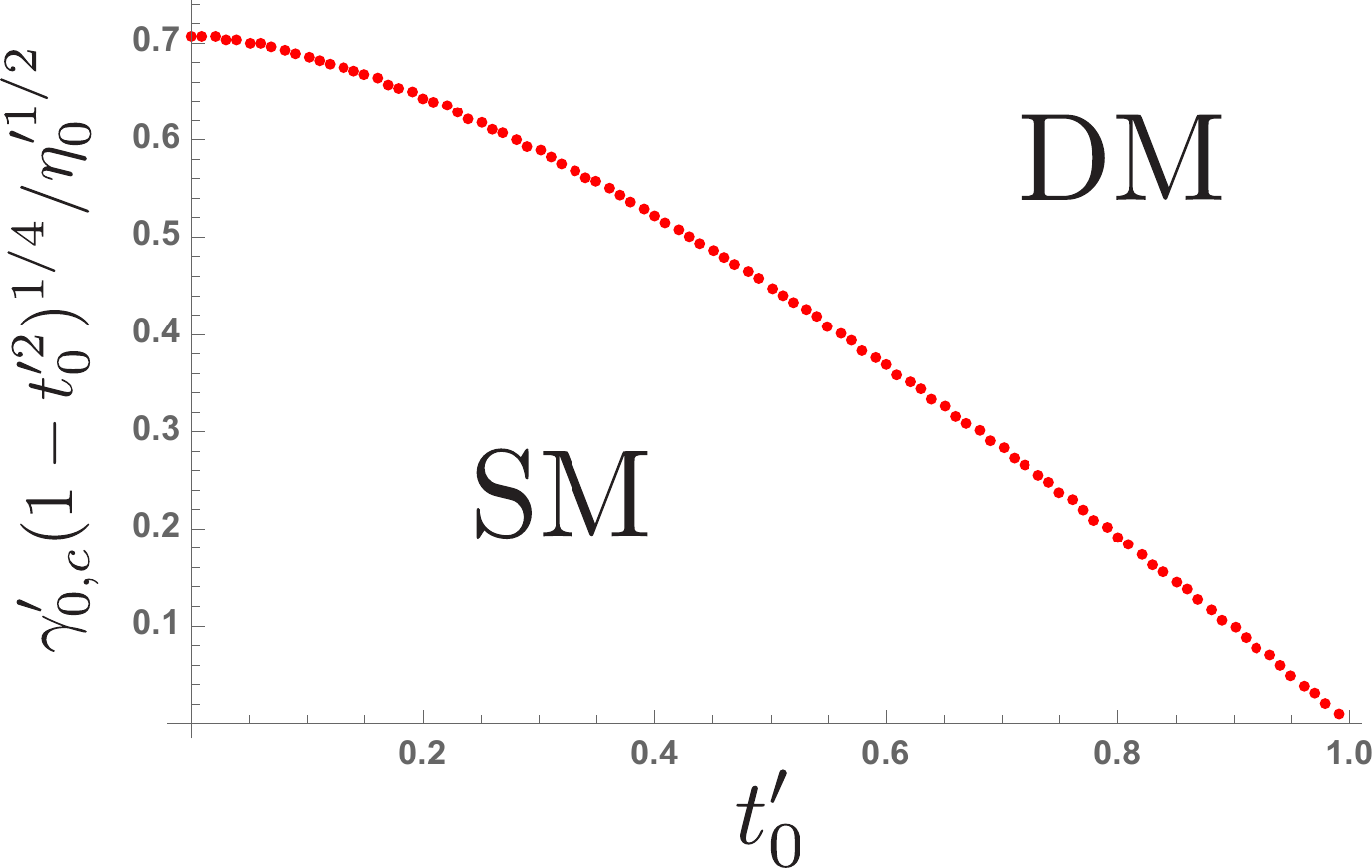}
  \caption{The critical initial disorder strength $\gamma'_{0,c}$ of the SM-DM phase transition for $\alpha'_0 = 0$ diminishes as initial tilt $t'_0$ increases.}
  \label{fig:CritDisTilt}
\end{figure}

\begin{figure}[h]
\centering
  \begin{subfigure}[b]{0.23\textwidth}
  \centering
    \includegraphics[width=\textwidth]{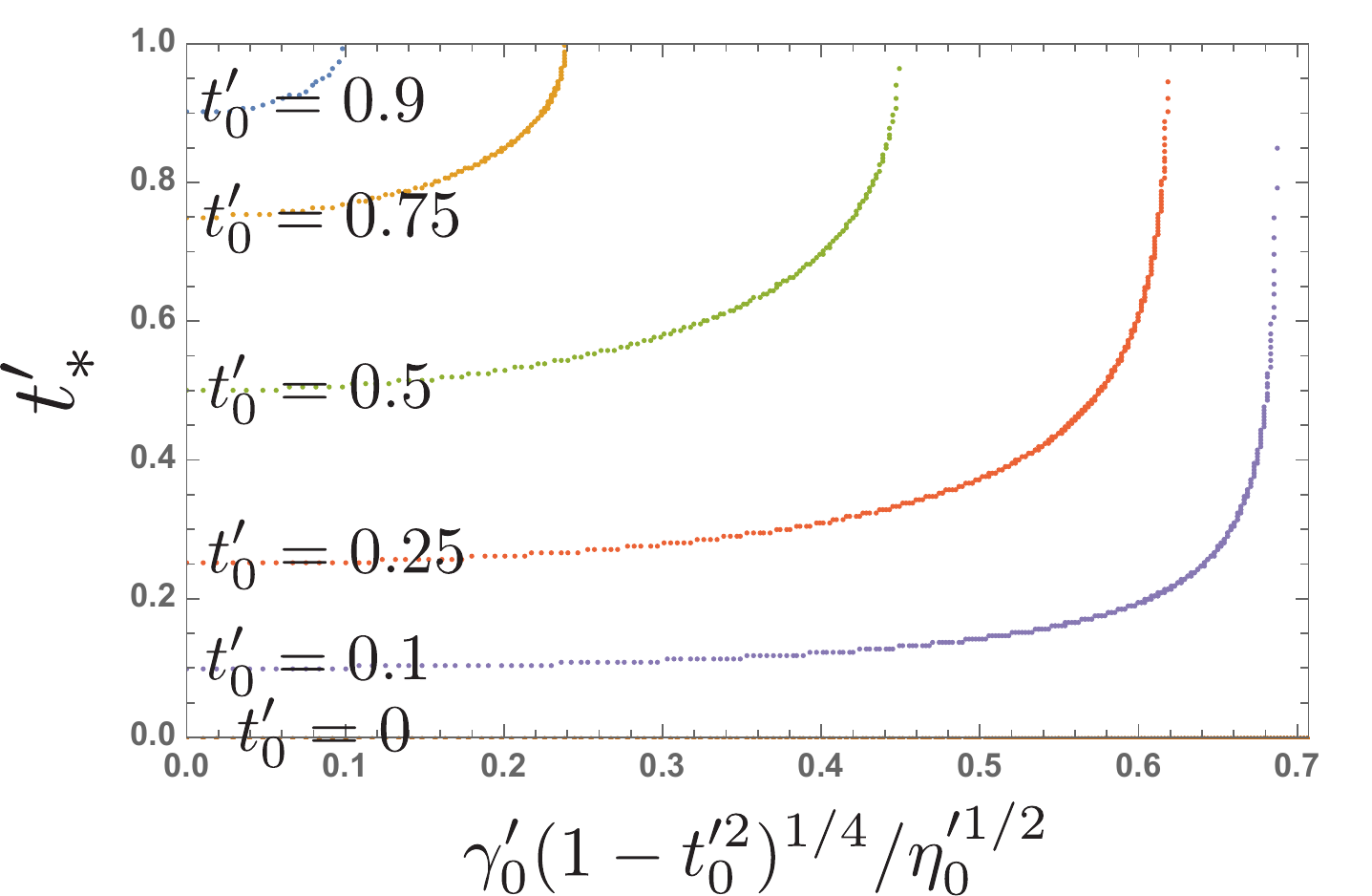}
    \caption{$t'_*$ as function of a parameter combination proportional to $\gamma'_0$ for arbitrary $\eta'_0$. }
    \label{fig:FinTilt_InDis}
  \end{subfigure}
  \;
  \begin{subfigure}[b]{0.23\textwidth}
  \centering
    \includegraphics[width=\textwidth]{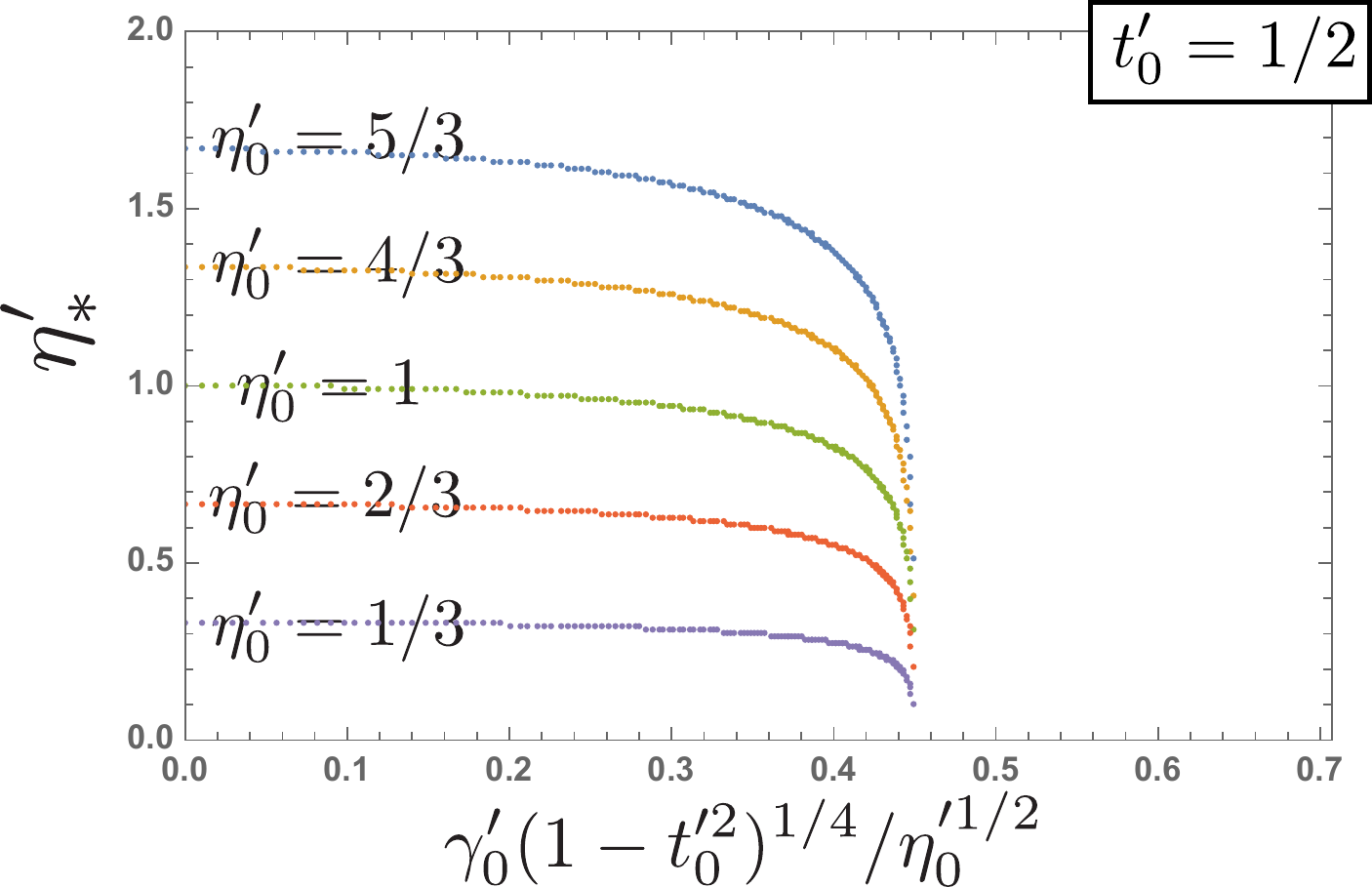}
    \caption{$\eta'_*$ as function of a parameter combination proportional to $\gamma'_0$ for $t'_0 = 1/2$. }
    \label{fig:FinAnis_InDis}
  \end{subfigure}
     \caption{Change of final tilt and anisotropy at the fixed line $\gamma'_* = 0$ as a function of the initial parameters of the tilted disordered model.}
\end{figure}

We next introduce a tilt into the model of a Weyl fermion perturbed by disorder, as studied before in Ref.~\cite{Sikkenk2017}. The $\beta$ functions of interest reduce to  
\begin{align}
&\beta_{t'} = t' \frac{1}{\eta' \sqrt{1-t'^2} } \gamma'^2, \\
&\beta_{\eta'} = - \eta' \frac{t'^2}{\eta' (1-t'^2)^{3/2} } \gamma'^2,\\
& \beta_{\gamma'} = \gamma' \left\{ -\frac{\epsilon}{2} + \frac{1}{\eta' \sqrt{1-t'^2} } \gamma'^2 \right\}. \label{eq:TiltedDis_RG}
\end{align}
By virtue of the identity $ \beta_{\eta'} / \eta' = - t' \beta_{t'} / (1-t'^2) $ the ratio $\eta'^2 / (1-t'^2) = \eta'^2_0 / (1-t'^2_0)$ is constant under renormalization group flow. Consequently the flow corresponding to this set of three differential equations can summarized in the two-dimensional streamplot Fig.~\ref{fig:Flow_TiltedDis}. The tilt shifts the boundary between semimetallic and diffusive metallic phases to lower critical disorder, see Fig.~\ref{fig:CritDisTilt}. Within the SM region, the flow is directed towards a line of clean fixed points at zero disorder $\gamma'_* = 0$ and finite tilts $t'_0 < t'_* < 1$ and anisotropies $0 < \eta'_* < \eta'_0$. This fixed line inherits its exponents $\nu = 0$ and $z=1$ from the trivial untilted and clean fixed point. The renormalized cone progessively tips over as the initial disorder approaches the critical value, see Fig.~\ref{fig:FinTilt_InDis}. Similarly, the final anisotropy at the disorder-free line of fixed points decreases after flowing from more disordered points, going to zero towards the phase boundary, see Fig.~\ref{fig:FinAnis_InDis}. The untilted nontrivial fixed point at $\gamma'_* =  \sqrt{ \eta'_0 \epsilon /2}$ with exponents $z=3/2$ and $\nu = 1$ is destabilized by the inclusion of a tilt term in favor of a new fixed point along the phase boundary at $\gamma'_* = (1-t'^2_*)^{1/4} \sqrt{\eta'_* \epsilon / 2} $ with $\eta'_* \rightarrow 0$ and $t'_* \rightarrow 1$. This new fixed point is however again characterized by critical exponents $z=3/2$ and $\nu= 1$. 

\subsection{Tilted interacting case} 	\label{subsec:tilted_int}

Alternatively there is the case of the tilted Weyl fermion perturbed by Coulomb interactions, whose isotropic case was studied in Ref.~\cite{Detassis2017}. The set of $\beta$ functions becomes
\begin{align}
&\beta_{t'} = -t'  \alpha' F_\parallel^{\eta'}, \\
&\beta_{\eta'} = - \eta' \alpha' \left( F_\parallel^{\eta'} - F_\perp^{\eta'} \right),\\
& \beta_{\alpha'}	 = -\alpha' \left\{ {\bar \epsilon} + \alpha' F_\parallel^{\eta'} \right\}.
\end{align}
The flow of the anisotropy and Coulomb interaction strength is independent of the tilt, and was previously depicted in the streamplot Fig.~\ref{fig:Flow_UntiltedAnisCou}. The interactions will inevitably renormalize the tilt downwards, asymptotically restoring an isotrpoic Weyl cone. All flow is towards towards the untilted isotropic non-interacting fixed point at $t'_*=0$, $\eta'_*=1$ and $\alpha'_*=0$ with exponents $\nu = 0$ and $z=1$.

\subsection{Full treatment: tilted, disordered and interacting case} 		\label{subsec:tilted_disint}

\begin{figure}[h]
\centering
  \begin{subfigure}[b]{0.23\textwidth}
  \centering
    \includegraphics[width=.9\textwidth]{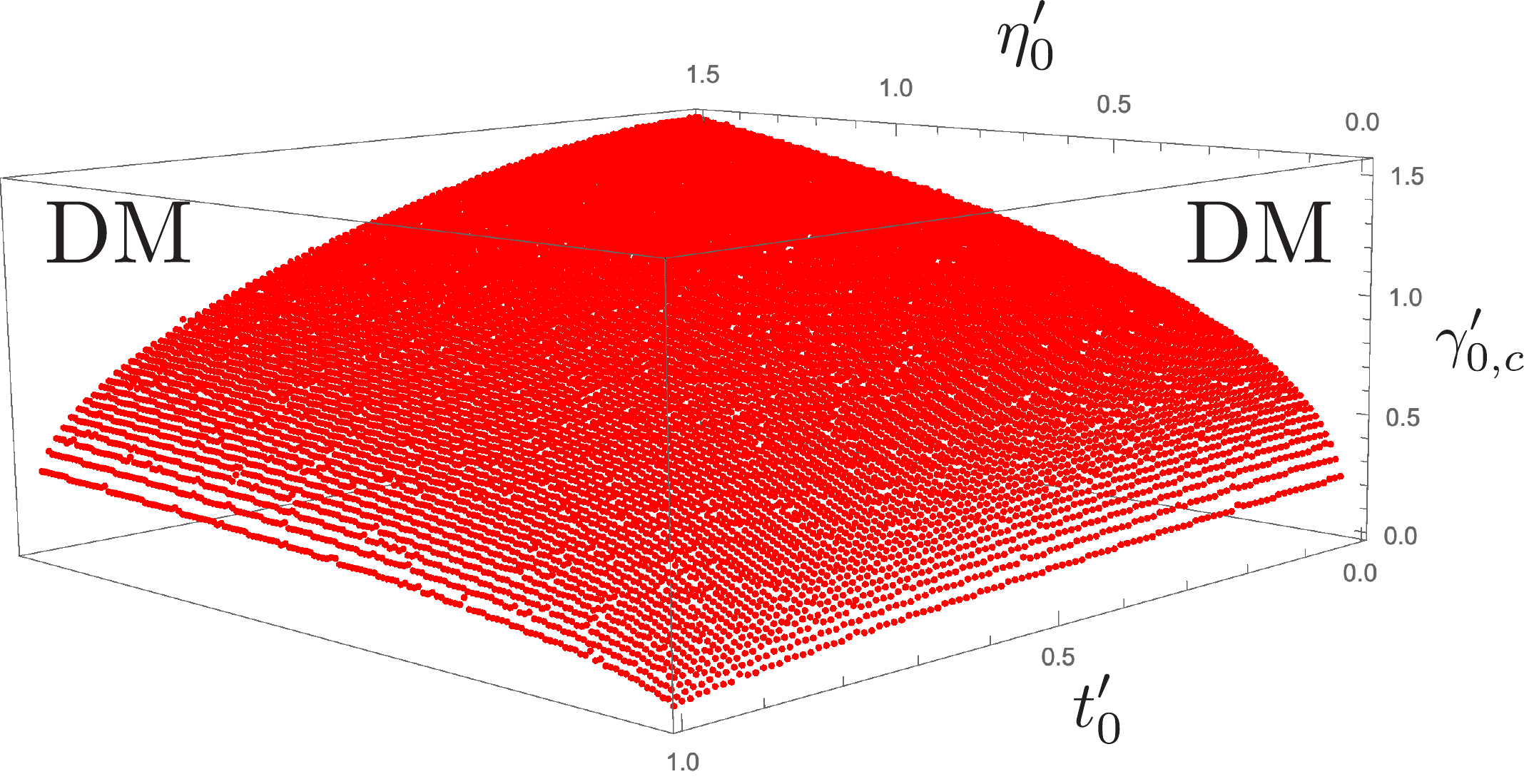}
    \caption{Front view. }
  \end{subfigure}
  \;
  \begin{subfigure}[b]{0.23\textwidth}
  \centering
    \includegraphics[width=.9\textwidth]{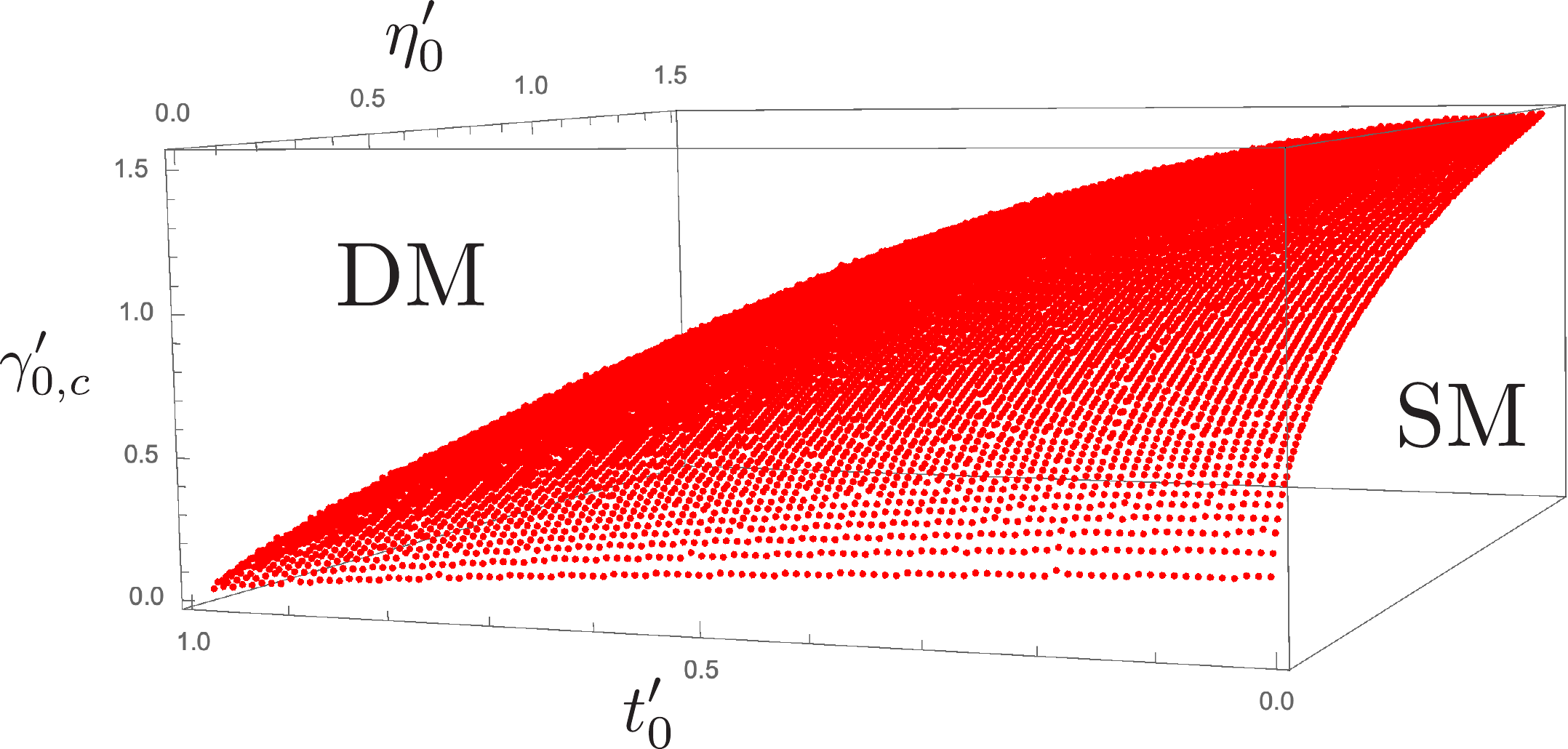}
    \caption{Side view. }  
  \end{subfigure}
     \caption{The critical initial disorder strength $\gamma'_{0,c}$ of the SM-DM phase transition as a function of initial tilt $t'_0$ and anisotropy $\eta'_0$ for fixed ratio $\alpha'/t' = \alpha'_0 / t'_0 = 1$.}
      \label{fig:CritDisIntTilt}
\end{figure}

We finally consider the full set of flow equations for the tilted anisotropic Weyl fermion under perturbation from disorder and Coulomb interactions, Eqs.~\eqref{eq:RG_tilt}-\eqref{eq:RG_dis}. These perturbations have competing effects, which play out in both the bare parameters and the couplings themselves. Since Coulomb interaction is marginal on tree level it holds that $\beta_{t'} / t' = \beta_{\alpha'} / \alpha' $ meaning the ratio $\alpha' / t' = \alpha'_0 / t'_0$ remains invariant under the RG transformation. A three-dimensional set of equations is then sufficient to capture the flow,
\begin{align}
&\beta_{t'} = t' \left\{\frac{1}{\eta' \sqrt{1-t'^2} } \gamma'^2 - t' \frac{\alpha'_0}{t'_0}  F_\parallel^{\eta'} \right\},  \\
&\beta_{\eta'} = - \eta' \left\{ \frac{t'^2}{\eta' (1-t'^2)^{3/2}} \gamma'^2 + t'  \frac{\alpha'_0}{t'_0} \left( F_\parallel^{\eta'} - F_\perp^{\eta'} \right)  \right\},\\
& \beta_{\gamma'} = \gamma' \left\{ -\frac{\epsilon}{2} + \frac{1}{\eta' \sqrt{1-t'^2} } \gamma'^2 - t' \frac{\alpha'_0}{t'_0}  F_\parallel^{\eta'} \right\}.
\end{align}
We numerically integrate these equations to study the SM-DM phase transition. Usefully we can use the spectator $\beta$ function for $\lambda$, Eq.~\eqref{eq:RG_lambda}, to determine the value of the phase transition line as it only vanishes at $\lambda_* =1$ for finite tilt and disorder. This procedure yields a two-dimensional phase boundary, depicted in Fig.~\ref{fig:CritDisIntTilt}, of the critical initial disorder $\gamma'_{0,c}$ versus initial tilt $t'_0$ and anisotropy $\eta'_0$. In its vicinity the physics of the model is controlled by the isotropic, untilted disordered but non-interacting fixed point with exponents $\nu = 1$ and $z \approx 3/2$. In the DM phase, for larger initial disorder content $\gamma'_0 > \gamma'_{0,c}$, the flow is directed towards strong disorder and ever larger Coulomb interactions. Within SM region $\gamma'_0 < \gamma'_{0,c}$ all flow is towards the trivial fixed point with $\nu=0$ and $z=1$, corresponding to the clean, free model. Due to the mismatch of the zeroth order scaling dimensions of the perturbations, ${\bar \epsilon} < \epsilon$, the Coulomb interactions scale out much more slowly at small couplings. As such, in the SM phase the Weyl cone is asymptotically upright and isotropic. The fixed line at $\gamma'_* = 0$ with finite $t'_* > 0$ and $\eta'_* \neq 1$ encountered in the non-interacting case $\alpha'_0 =0$ is destabilized when Coulomb interactions are included.

\section{Ward Identities and charge non-renormalization} 		\label{sec:WardIdentities}

From Eqs.~\eqref{eq:RG_Cou}-\eqref{eq:RG_dis} it is clear that the disorder and Coulomb interaction cannot have a fixed point at which these couplings are both finite. Although their first order corrections are identical, as the zeroth order scaling dimensions ${\bar \epsilon} \to 0, \epsilon \to 1$ differ simultaneous satisfaction of these equations is impossible.

We have found that this statement can be generalized, and that it so continues to hold to any order in perturbation theory. It is also independent of the presence of tilts and anisotropies in the dispersion. Disorder and Coulomb interaction couple in a very similar manner and have analogous perturbative expansions with diagrams of the same form. The polarization bubble, Fig.~\ref{fig:POLAR_CC}, is regular using our $d=2+\epsilon$ dimensional regularization scheme. Therefore all diagrams that include it are subleading and can be neglected. In other words, neither the photon nor the disorder propagator obtains any renormalization at any order in perturbation theory. At a given order $p$, corrections come from all (one-particle irreducible) permutations of $2 p$ vertices inserted on a single continuous fermion line. As such, all diagrams responsible for vertex renormalization can be exhaustively generated from self energy diagrams renormalizing the fermion propagator by introducing the suitable external vertex at all possible internal positions on the fermion line ~\cite{Ward1950}. Using the identity
\begin{align}
&\partial_{i \omega} G_0 ( i \omega - i \omega', {\bf k}-{\bf q} ) \\
& \quad = -  G_0 ( i \omega - i \omega', {\bf k}-{\bf q} ) \, {\hat \lambda} \, G_0 ( i \omega - i \omega', {\bf k}-{\bf q} ),	\nonumber
\end{align}
this might diagrammatically be depicted as 
\begin{align} 
\vcenter{\hbox{\includegraphics[width=1.6cm]{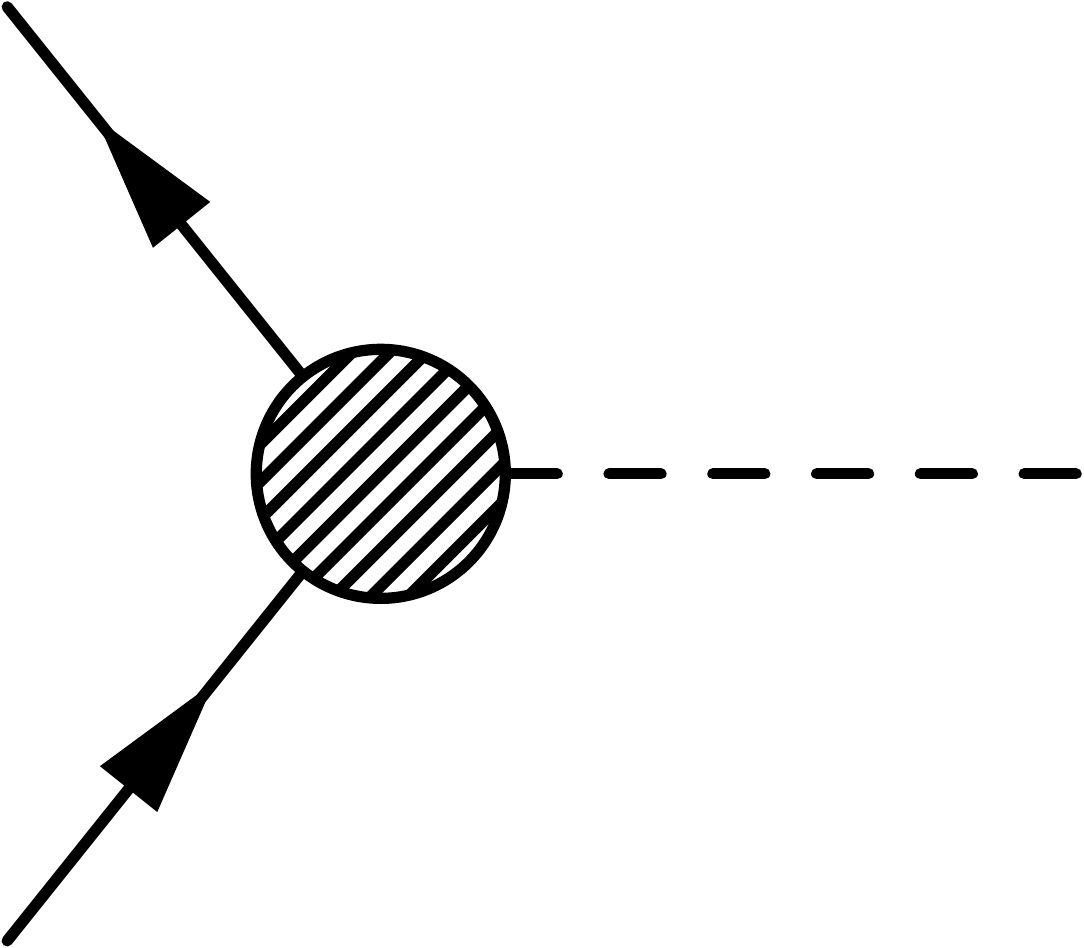}}}  &= - \Gamma \; \partial_{i \omega} \left( \vcenter{\hbox{\includegraphics[width=1.6cm]{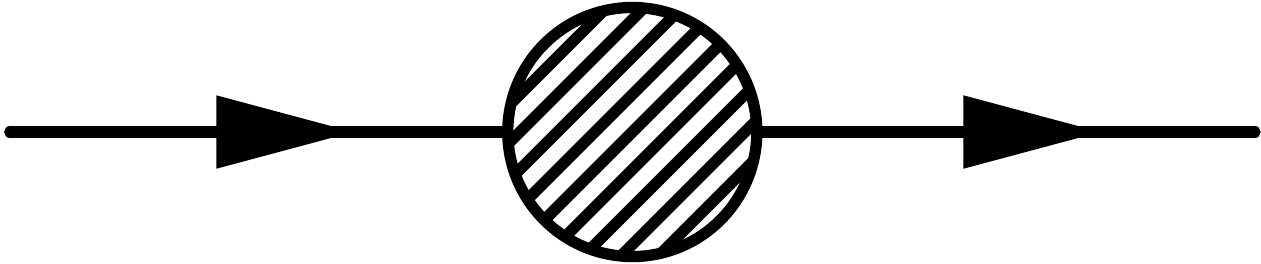}}} \right), 	\label{eq:DiagramRelations} \\
\vcenter{\hbox{\includegraphics[width=1.6cm]{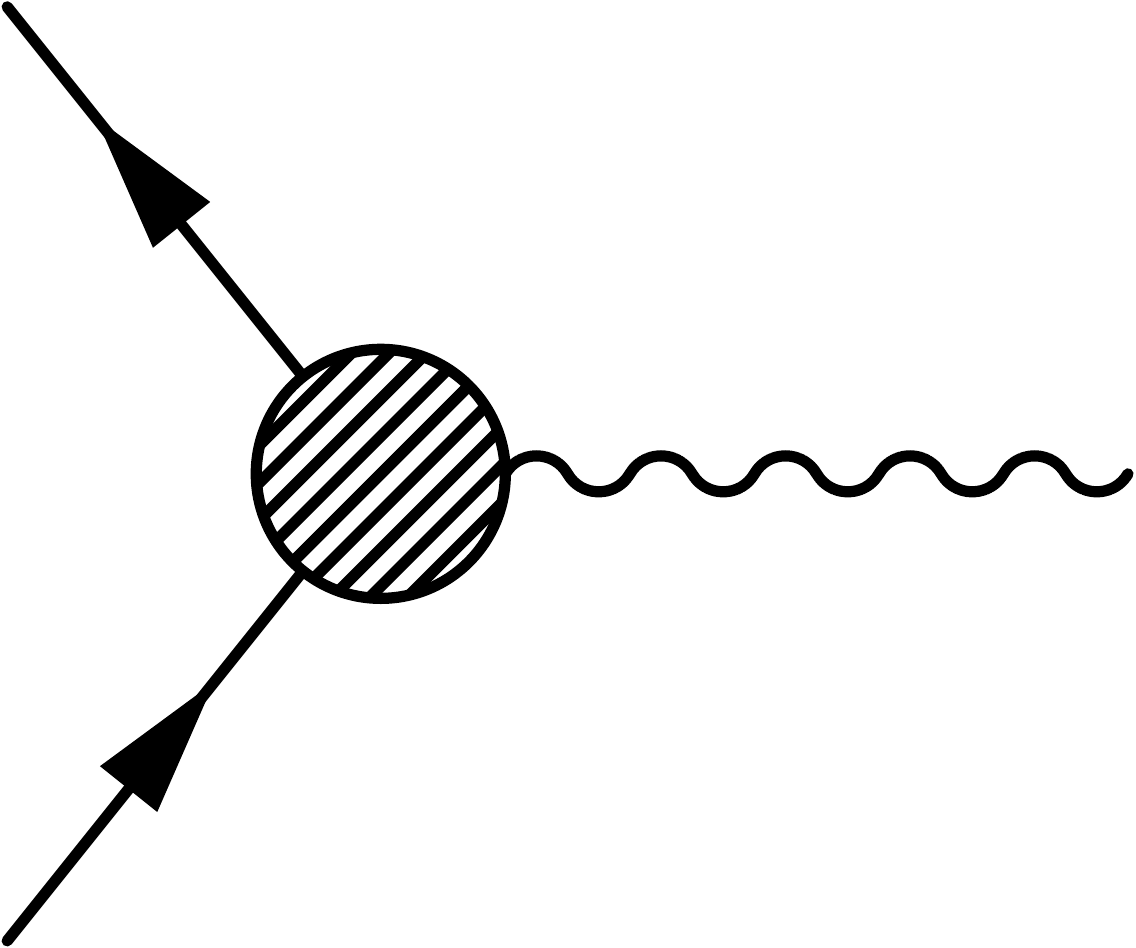}}}  &= - (ig) \; \partial_{i \omega} \left( \vcenter{\hbox{\includegraphics[width=1.6cm]{SELF_Aggregate.pdf}}} \right), \nonumber 
\end{align} 
where the shaded areas represent the sum of all (one-particle irreducible) subdiagrams that can be used to connect the external legs and the derivative is understood to be taken with respect to the external frequency and applying the product rule. Due to the relations Eq.~\eqref{eq:DiagramRelations} vertex corrections will be exactly cancelled by the couterterms $\delta_\psi$ for the fermion field and $\delta_\lambda$ for the field transformation parameter by the counter terms $\sim i \omega$ from the self-energy corrections. We then have identically vanishing counterterms $\delta_\Gamma = \delta_g = 0$. Such charge non-renormalization has been observed before in the 2d context of graphene by Ref.~\cite{Vozmediano2010}. A consequence $\delta_{\gamma'} = \delta_{\alpha'} = - \delta_{v'}$, at all orders in perturbation. The $\beta$ functions of the dimensionless couplings can thus only differ in the tree level scaling dimensions ${\bar \epsilon} \to 0$, $\epsilon \to 1$, and a intermediate fixed point at finite disorder and Coulomb interaction cannot exist up to any order in perturbation theory.

\vspace{12 pt}

In more general terms the non-renormalization of $\Gamma$ and $g$ can be identified as Ward identity deriving from the gauge invariance of the model. At its core it is but a particular manifestation of quantum electrodynamics, of which such symmetries are a defining characteristic. The Lagrangian contains a term of the form 
\begin{align}	 
&\psi_{ {\bf q}, \omega}^\dagger \left[ \, i \omega \, \delta_{{\bf q}- {\bf q}'} \delta_{\omega - \omega'} + i g \, \varphi_{ {\bf q}-{\bf q}', \omega - \omega' }  \right. 	\label{eq:MinCoupledTerm} \\
& \qquad \qquad \quad + \left. \Gamma \, V_{ {\bf q} - {\bf q}'} \delta(\omega - \omega') \, \right] \, {\hat \lambda} \, \psi_{ {\bf q'}, \omega'},	 \nonumber 
\end{align} 
which can be obtained from the free Green function in Eq.~\ref{eq:FermionGreenFunc} by a minimal coupling procedure. Here the disorder field $V$ acts as an external field, which should ordinarily couple to the fermion fields with equal charge $g$ as the (scalar) gauge field $\varphi$ mediating the electromagnetic interactions. This term would then be protected during renormalization flow by Ward identities, {\it i.e.}, gauge invariance, to all orders in perturbation theory, guaranteeing that $\delta_\Gamma = \delta_g = 0$. However as the zeroth order scaling of $V_{\bf q} \delta(\omega)$ does not match that of counterpart $\varphi_{\bf q, \omega}$, relative compensation between the dimensions of the couplings $\Gamma$ and $g$ is necessary. It is thus that the $\beta$ functions of $\gamma'$ and $\alpha'$ can only ever differ by the tree level scaling dimensions up to all orders in perturbation theory. Note furthermore that this argument applies to all type-I Weyl fermions including the untilted case~\cite{Goswami2011}, irrespective of possible tilts or anisotropies in their dispersion, as the ${\hat \lambda}$ field transformation preserves minimal coupling by construction.

\section{Conclusion} 		\label{sec:conc}

Within this paper we studied type-I Weyl fermions including anisotropies and tilt and their physics if exposed to disorder and Coulomb interactions from an RG perspective. On a technical level, we find that a new term is generated under renormalization group flow, which we incorporate by means of the field transformation ${\hat \lambda}$ of Eq.~\eqref{eq:field_transf}. This transformation respects the minimal coupling between frequency $i \omega$ on the one hand and gauge field $\varphi$ and external field $V$ on the other hand. It also has ramifications on the other parameters of the model, as set out in Eq.~\eqref{eq:ParamRedif}. 

Without tilts or anisotropies, disorder and Coulomb perturbations result in $\beta$ functions that are the same except for the tree level scaling dimensions. Besides the attractive trivial fixed point, with exponents $\nu =0$ and $z=1$ there is only the repulsive non-interacting fixed point at finite disorder. It  governs the SM-DM phase transition with critical exponents $\nu =1$ and $z = 3/2$. There cannot be an intermediate fixed point at both finite disorder and finite Coulomb interaction ~\cite{Goswami2011}. 

Including a tilt and anisotropies does not lead to new terms in the coupling $\beta$ functions, but only modifies them. It does not change the qualitative behavior and no new intermediate fixed point develops. Numerical integration shows that there is still a critical disorder value at which the system transitions from a weakly interacting and weakly disordered phase into a DM phase at strong interactions. It is a function of the initial values of the parameters of the model, decreasing as a function of the tilt and anisotropy but increasing with the Coulomb interation.

Within the SM phase the flow is directed towards smaller couplings. Since Coulomb interaction is marginal it eventually dominates disorder perturbation which is irrelevant. Therefore, in the SM region, the flow is directed towards the attractive trivial fixed point at which the cone is upright and isotropic~\cite{Detassis2017}. The attractive line of fixed points at which the tilt and anisotropy reach finite values to which the flow is directed in the tilted disordered model is particular to the complete absence of Coulomb interactions. When this is included it is immediately destabilized in favor of the trivial fixed point. 

Importantly, we have found that these findings hold to all orders in perturbation theory, as the relation between frequency $i \omega$, gauge field $\phi$ and external disorder field $V$ corresponds to a minimal coupling which is protected by a Ward identity. Including a tilt does not change this as the field transformation Eq.~\eqref{eq:field_transf} uniformly affects the terms in Eq.\ref{eq:MinCoupledTerm}. As a result, there cannot be any renormalization of the coupling parameters. The $\beta$ functions of the disorder and Coulomb interactions only differ because of their different scaling dimensions. This implies that an intermediate fixed point at which both couplings are finite cannot exist. There is only the disorder driven phase transition into the DM phase from the SM phase, where Coulomb effects will dominate due to the due to the fact that they are less irrelevant.

\vspace{12pt}
{\it{Acknowledgments:}}
T.S.S. thanks S. Kooi and E. van der Wurff for insightful discussions. L. F. acknowledges former collaborations on this subject with F. Detassis and S. Grubinskas, who was also involved at an early stage of the project. T.S.S. acknowledges funding from the Swaantje Mondt Fund for a research visit during the writing of this paper. This work is part of the D-ITP consortium, a program of the Netherlands Organisation for Scientific Research (NWO) that is funded by the Dutch Ministry of Education, Culture and Science (OCW).

\newpage
\appendix 

\section{Perturbative analysis} \label{appsec:diagrams}

Under anisotropic space-time rescaling $\omega \to \mu^{+ z} \omega$, ${\bf q} \to \mu^{+1} {\bf q}$ the parameters and fields change as $y_{i,0} \to y_i (\mu) =  y_{i,0} \, \mu^{+ [y_i]} \, Z_{y_i}^{-1}$.  Going beyond tree level in perturbation theory in the interactions of Eq.~\eqref{eq:Action} will result in diagrammatic divergences that are to be cancelled by the inclusion of counterterms as $Z_{y_i} = 1 + \delta_{y_i}$. The divergences arising from the disorder perturbation are captured by regularization of the number of spatial dimensions $d=2 + \epsilon$ ~\cite{Roy2014}. Because the Coulomb interaction offers a marginal perturbation to the tree-level Weyl fermion indepedent of $d$, an additional expansion in ${\bar \epsilon}$ appearing the power of its propagator is required to absorb the resulting divergences.
  
\vspace{12 pt}

\begin{figure}[h]
\centering
  \begin{subfigure}[b]{0.15\textwidth}
  \centering
    \includegraphics[width=0.95\textwidth]{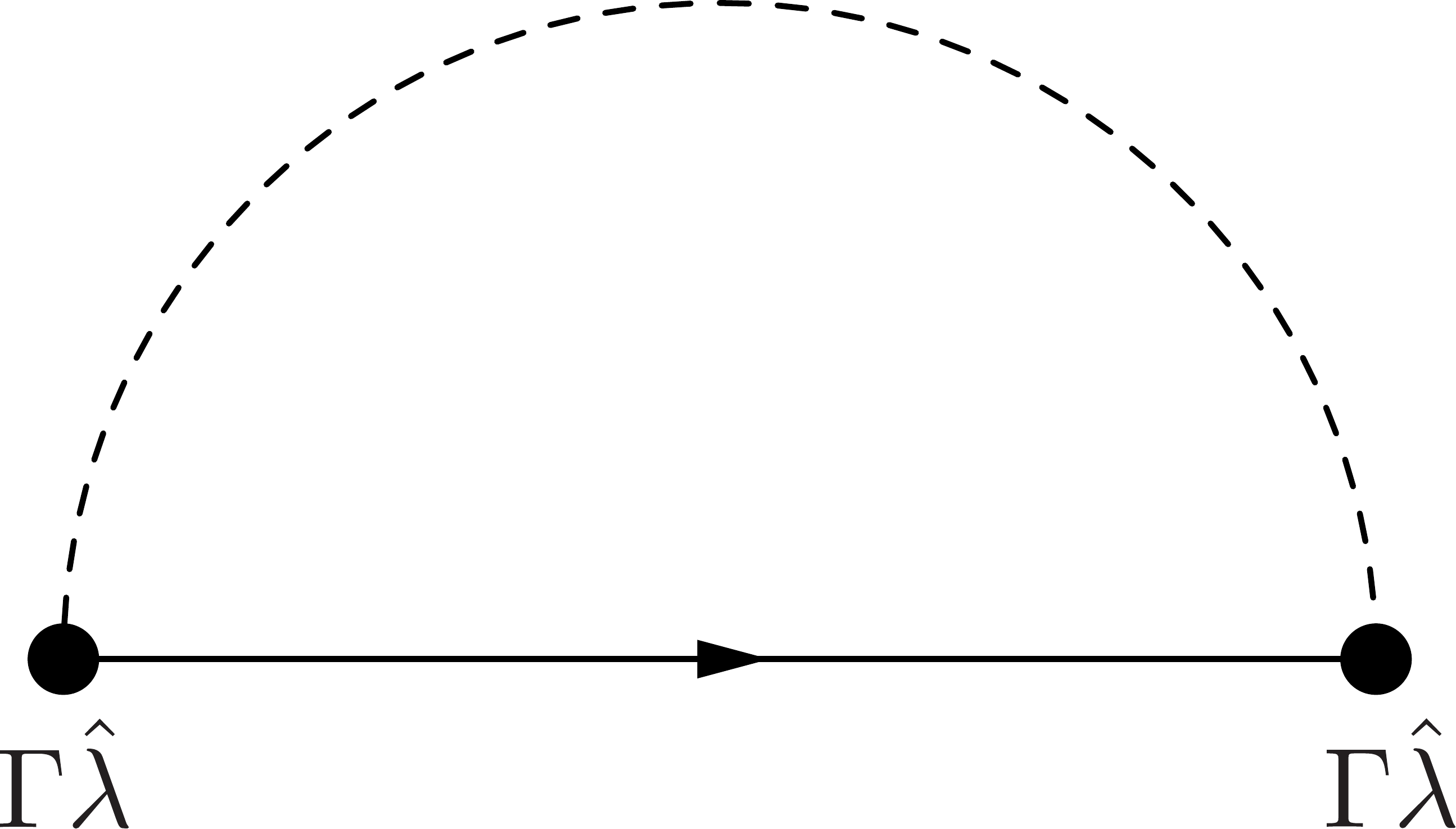}
    \caption{$\Sigma_\text{dis}(i \omega,{\bf k})$}
    \label{fig:SELF_dd}
  \end{subfigure}
  \;
  \begin{subfigure}[b]{0.15\textwidth}
  \centering
    \includegraphics[width=0.95\textwidth]{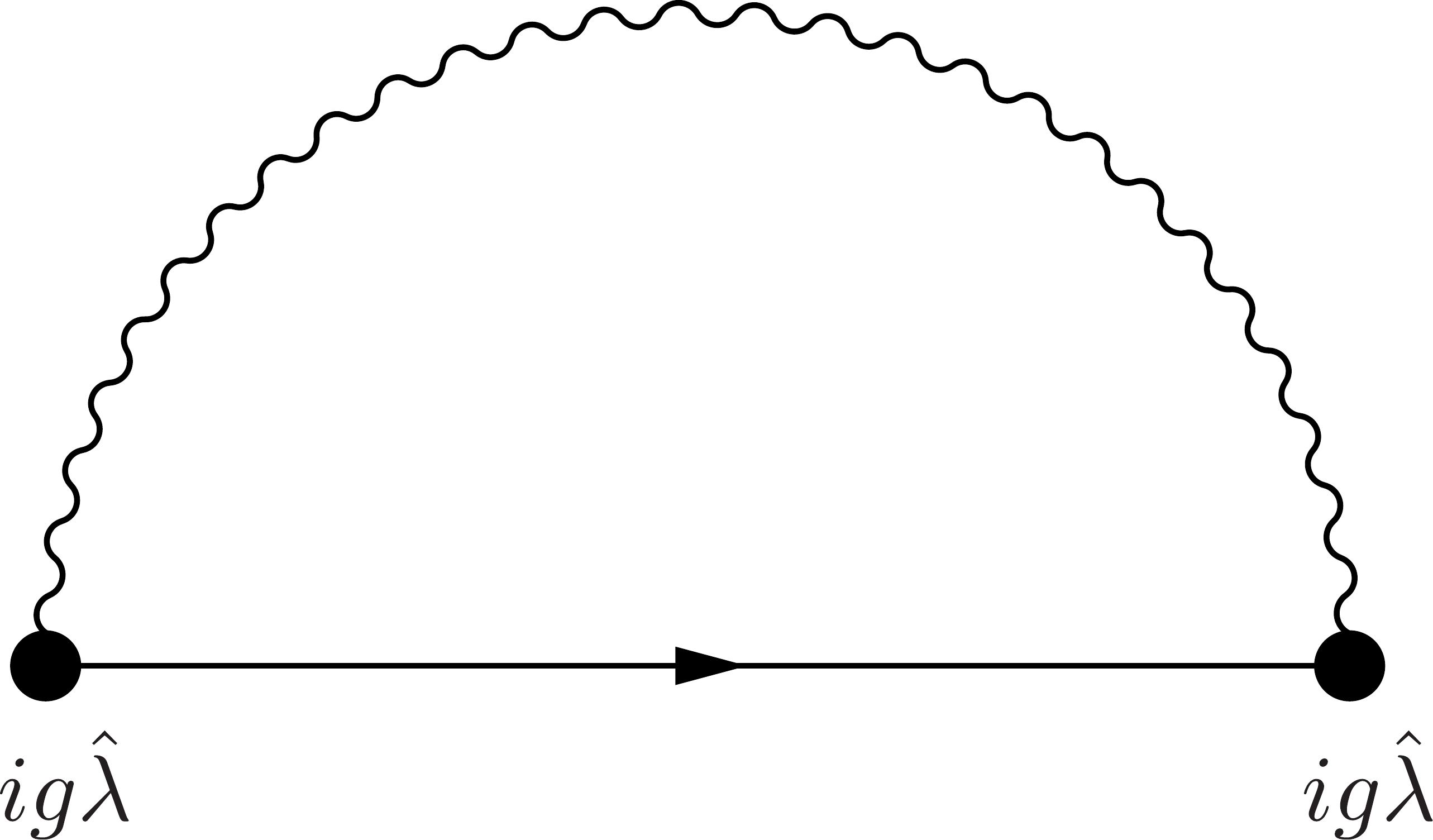}
    \caption{$\Sigma_\text{Cou}(i \omega,{\bf k})$}
    \label{fig:SELF_CC}
  \end{subfigure}
  \;
  \begin{subfigure}[b]{0.15\textwidth}
  \centering
    \includegraphics[width=0.95\textwidth]{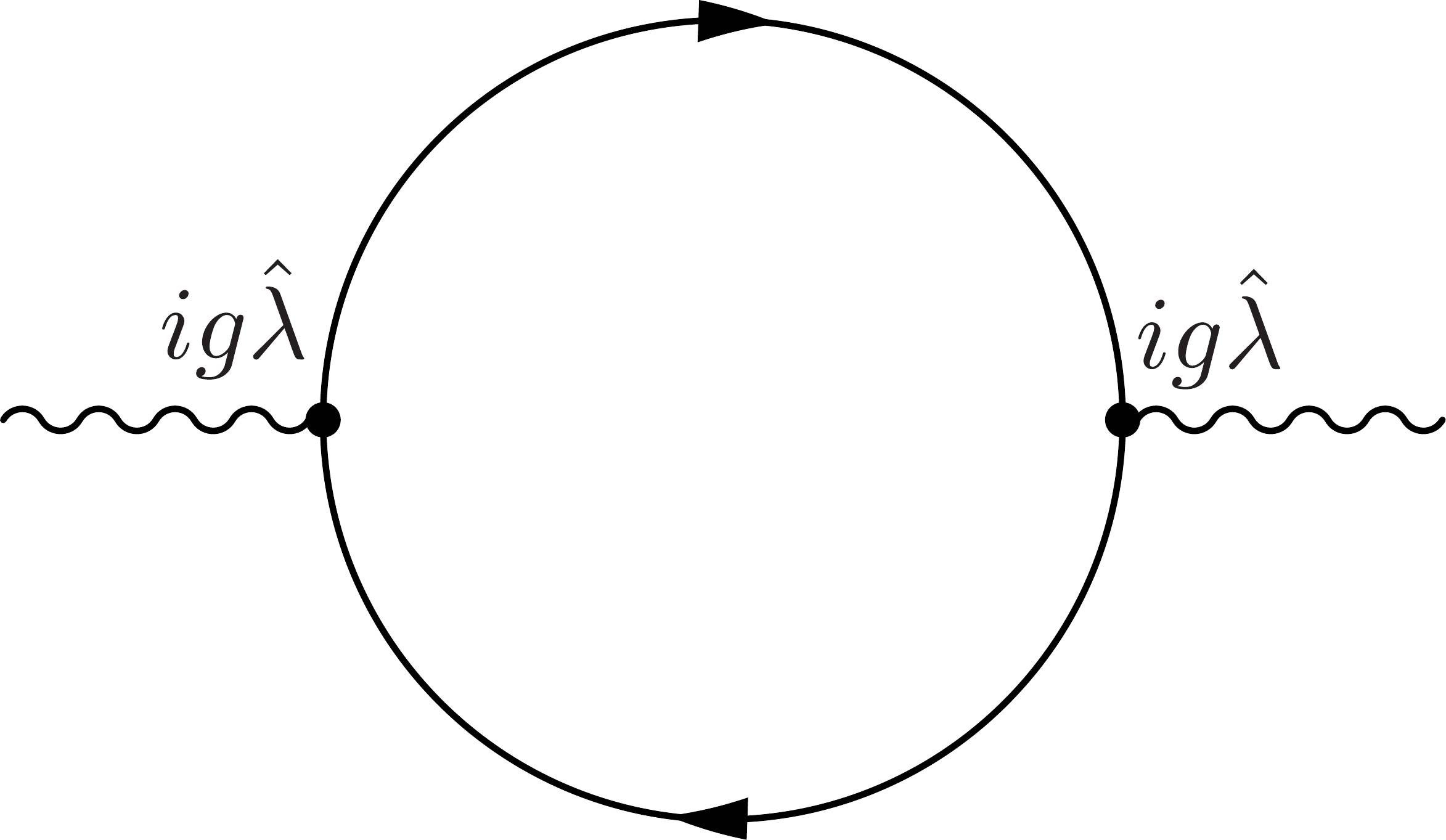}
    \caption{$\Pi ( i \omega, {\bf k})$}
    \label{fig:POLAR_CC}
  \end{subfigure}
   \caption{Self-energy corrections to the fermion Green function $G_0$ and polarization contributing to renormalization of photon propagator $D_0$.}
   \label{fig:SELFS}
\end{figure}

\begin{figure}[h] 
\centering
  \begin{subfigure}[b]{0.12\textwidth}
  \centering
    \includegraphics[width=0.95\textwidth]{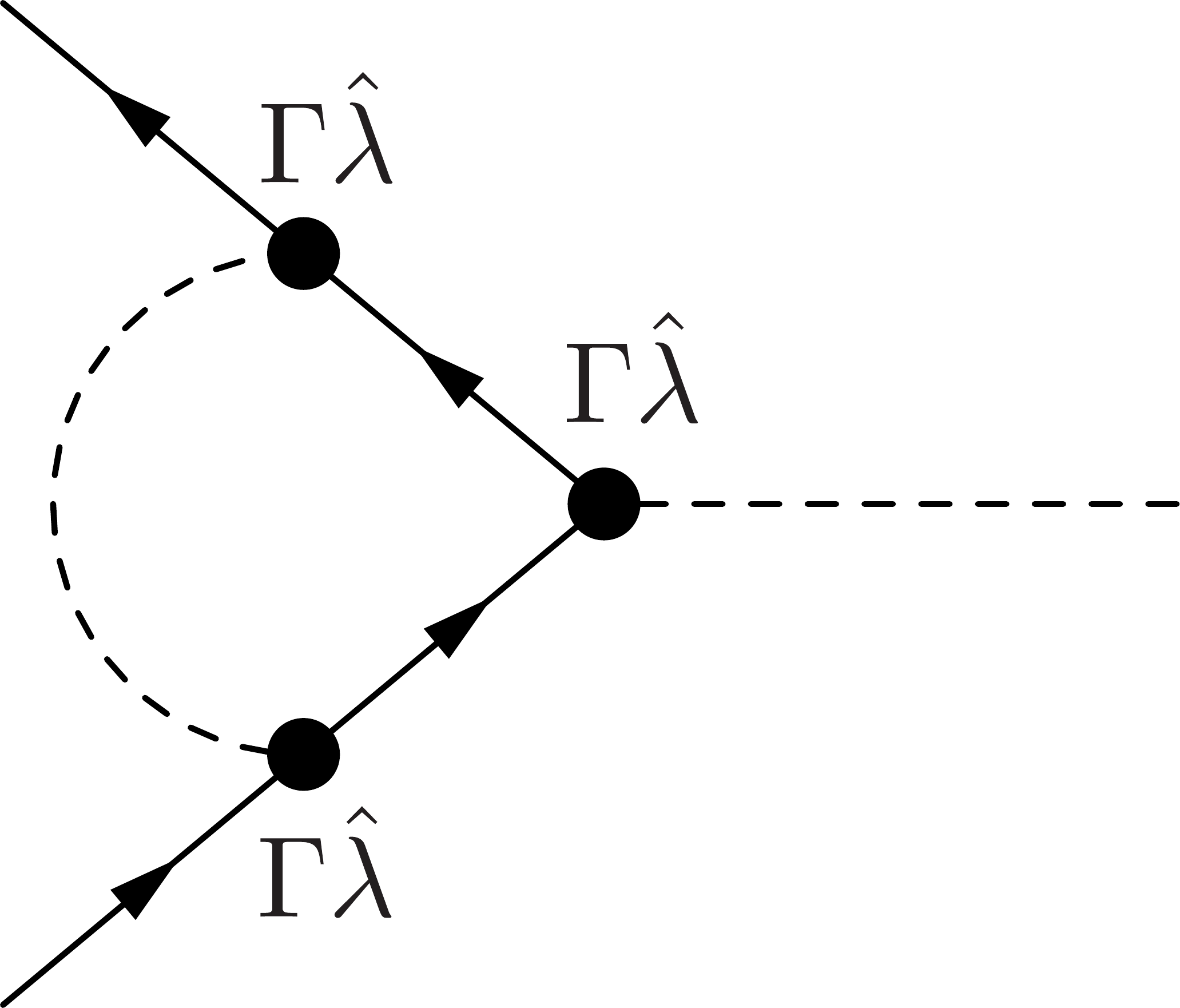}
    \caption{}
    \label{fig:VERTEX_ddd}
  \end{subfigure}
  \;
  \begin{subfigure}[b]{0.12\textwidth}
  \centering
    \includegraphics[width=0.95\textwidth]{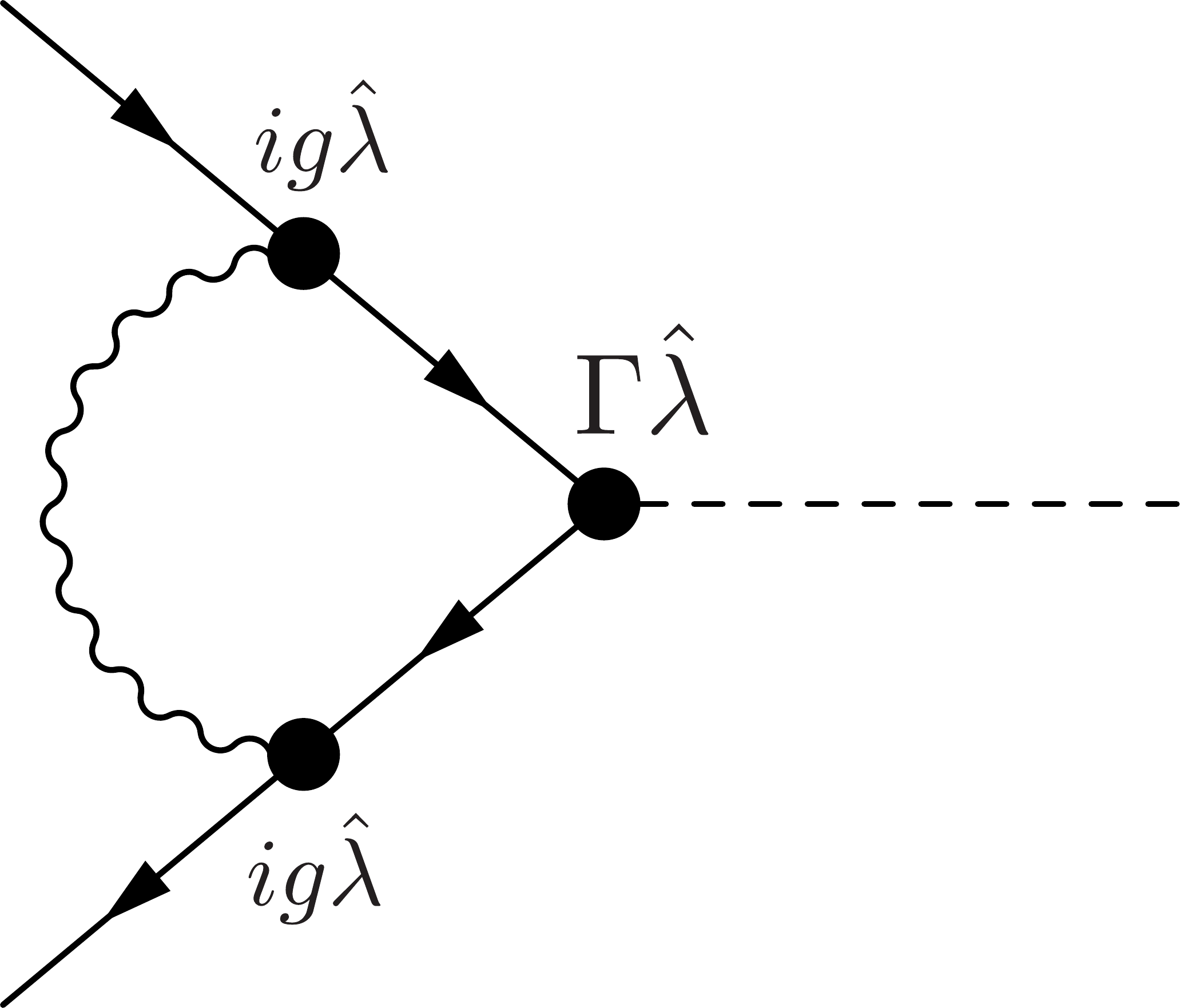}
    \caption{}
    \label{fig:VERTEX_CdC}
  \end{subfigure}
   \;
  \begin{subfigure}[b]{0.175\textwidth}
  \centering
    \includegraphics[width=0.95\textwidth]{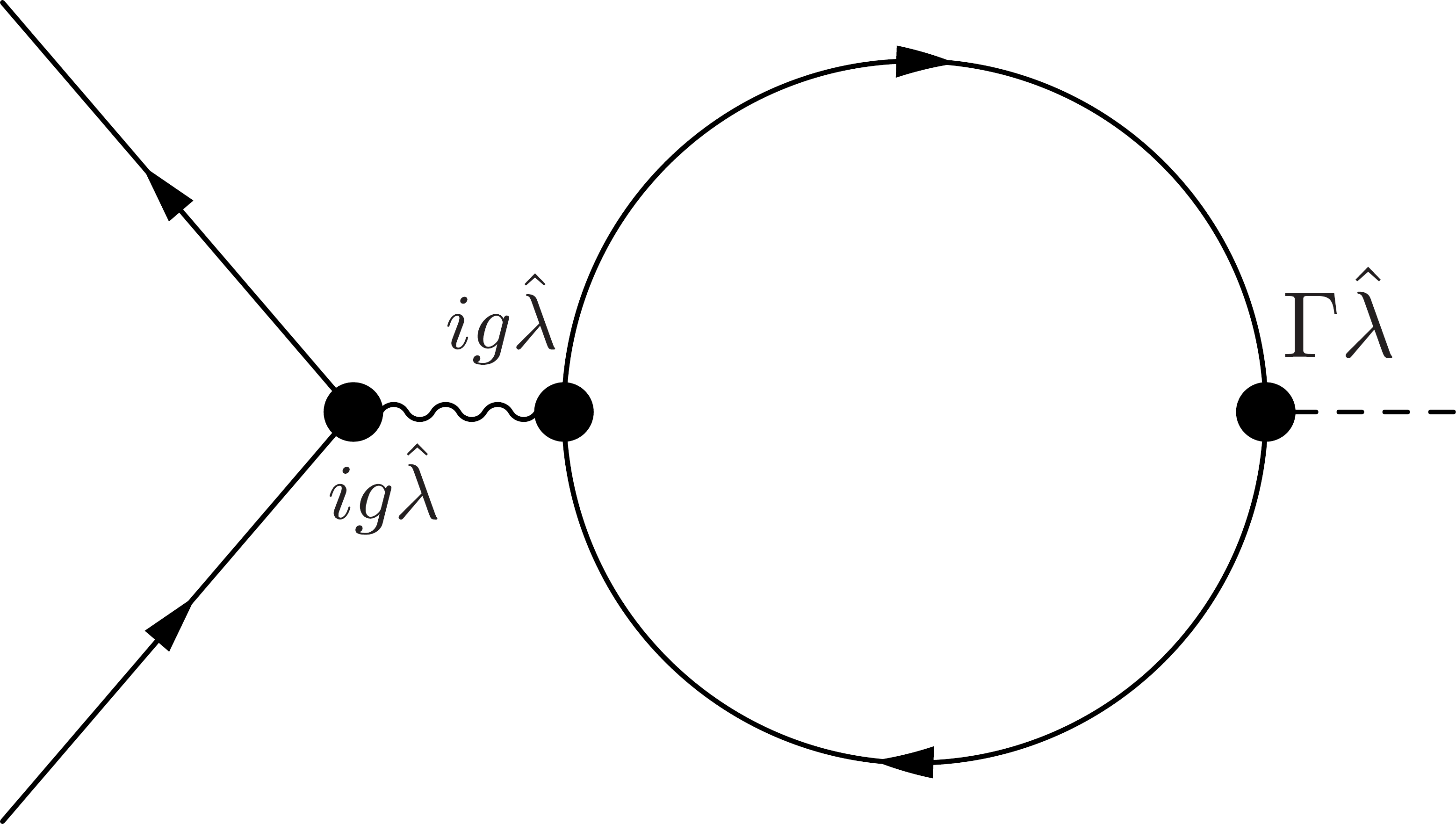}
    \caption{}
    \label{fig:VERTEX_dfermionloop}
  \end{subfigure}
   \caption{Vertex corrections to disorder coupling $\Gamma_\theta$.}
  \label{fig:VERTICES_d}
\end{figure}

\begin{figure}[h]
  \begin{subfigure}[b]{0.2\textwidth}
  \centering
    \includegraphics[width=0.6\textwidth]{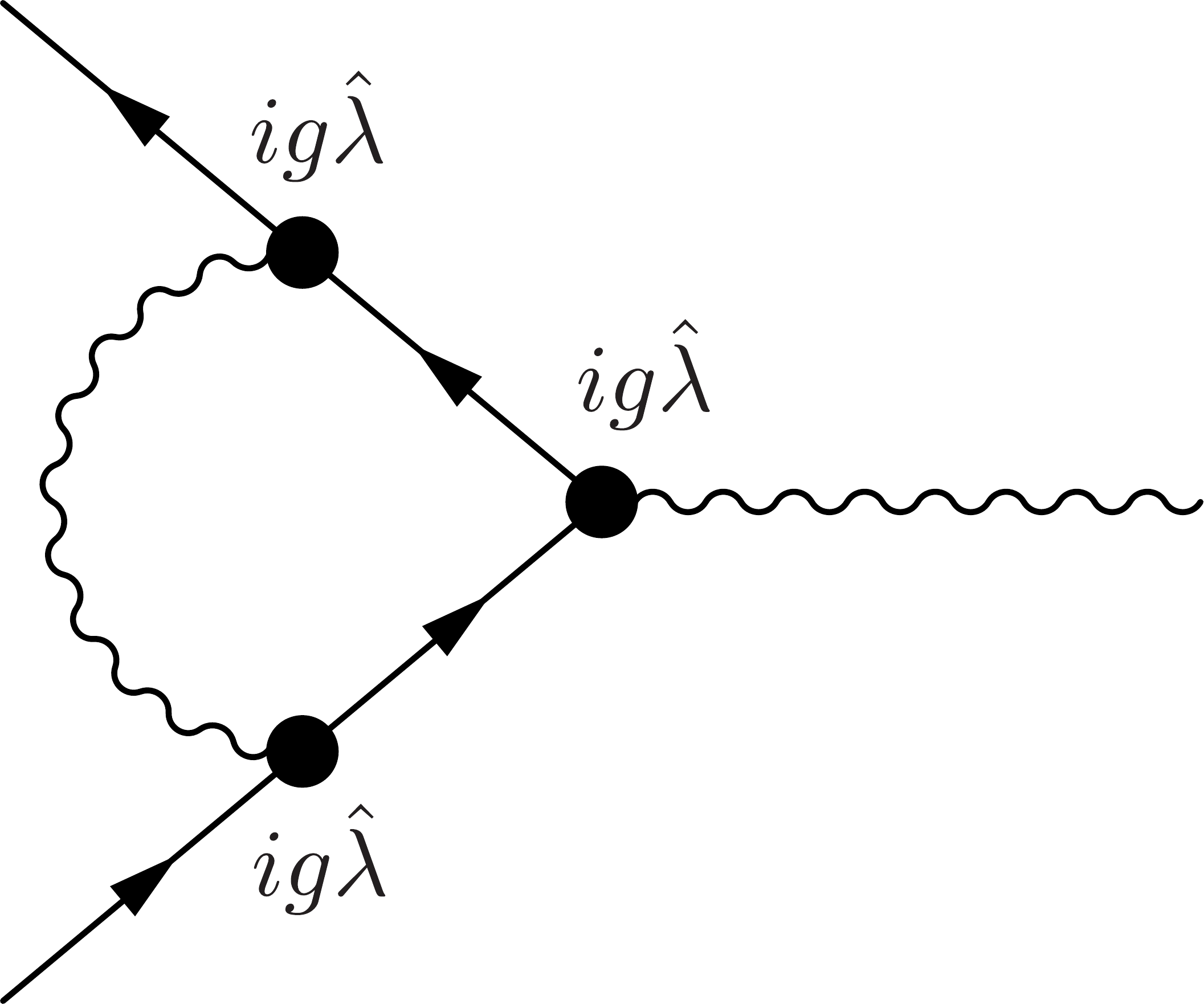}
    \caption{}
    \label{fig:VERTEX_CCC}
  \end{subfigure}
  \;
  \begin{subfigure}[b]{0.2\textwidth}
  \centering
    \includegraphics[width=0.6\textwidth]{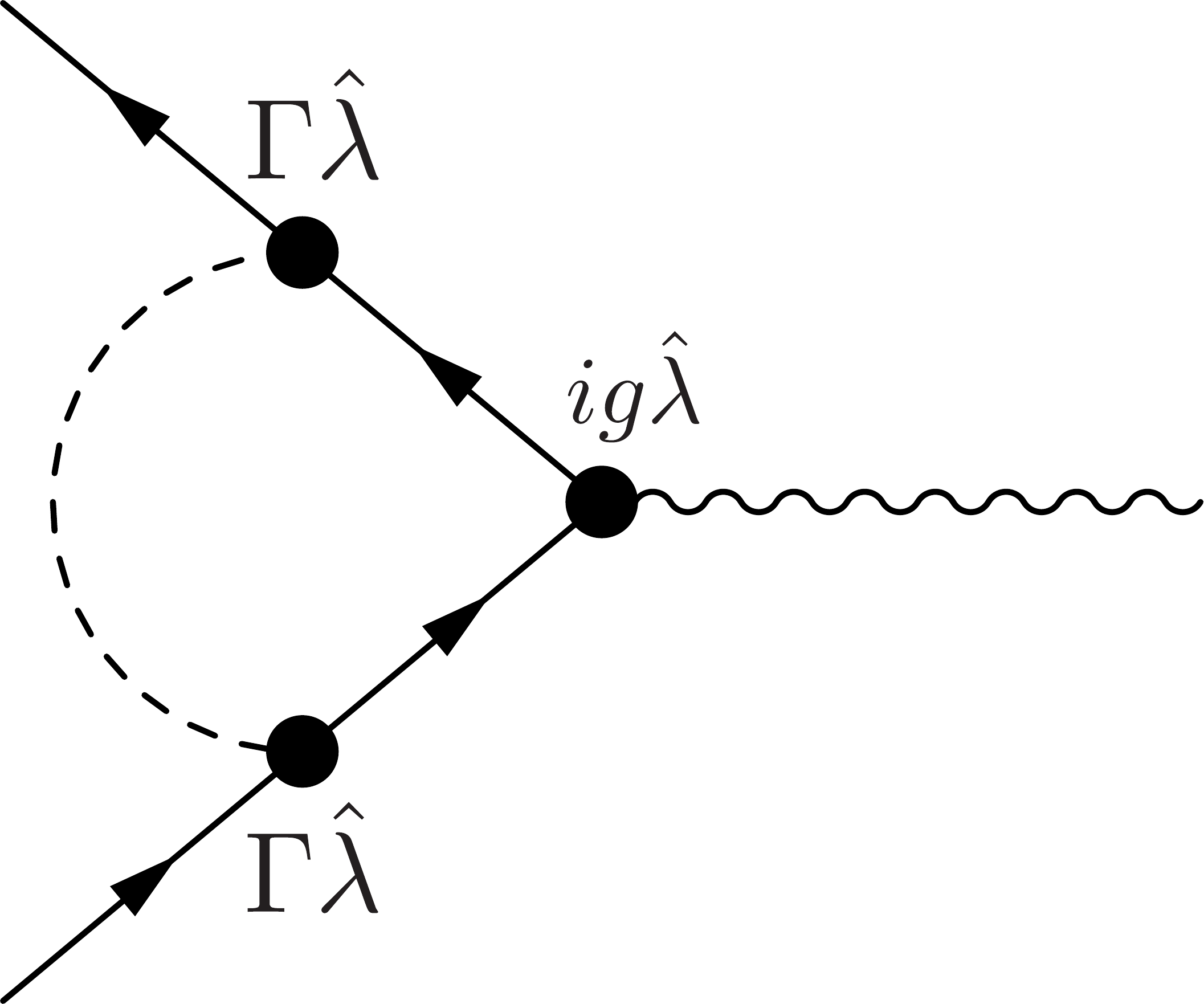}
    \caption{}
    \label{fig:VERTEX_dCd}
  \end{subfigure}
  \caption{Vertex corrections to the Coulomb coupling $g$.}
  \label{fig:VERTICES_C}
\end{figure}

Contributing diagrams are listed in Figs.(\ref{fig:SELFS})-(\ref{fig:VERTICES_C}). We have adhered to the convention to represent the fermion propagator $G_0$ as arrowed line, the photon propagator $D_0$ as wavy line and propagation of the disorder field by a dashed line. Vertices indicated by $\Gamma {\hat \lambda}$ derive from the disorder part Eq.~\eqref{eq:DisAction} of the interacting action. Vertices indicated by $i g {\hat \lambda}$ come from the Coulomb part Eq.~\eqref{eq:CouAction}. Note here that all those graphs that have some dependence on the number of field replicas will vanish in the replica limit. Practically, this implies diagrams with a fermion loop connected purely by disorder legs can be safely neglected. 

The derivation of these diagrammatic divergences is presented below. In calculating their divergences, we have often employed the generalized Feynman trick
\begin{align}
A^{-n} B^{-r} = \frac{ \Gamma[n+r] }{\Gamma[n] \Gamma[r] } \int_0^\infty {\mathrm d} u \, \frac{u^{n-1} }{ \left( u A+B \right)^{n+r}}
\end{align}
to handle the different powers of the denominators of the Weyl fermion and Coulomb boson propagators. Notationally, it has proved useful to define dimensionless couplings 
\begin{align}
\gamma^2	= \frac{\Omega_d \, \mu^\epsilon }{(2 \pi)^d \, v^2} \, \Gamma^2, \qquad \alpha = \frac{\Omega_d \, \mu^{ {\bar \epsilon} } }{4 (2 \pi)^d \, v} \, g^2, 	\label{appeq:CouplingRedif}
\end{align} 
to shorten commonly occuring expressions. 

\vspace{12 pt}

We first investigate the diagrams in Fig.(\ref{fig:SELFS}), which will cause the renormalization of the parameters of the bare actions. The self-energy deriving from the disorder interaction yields 
\begin{align}
&\text{Fig.(\ref{fig:SELF_dd})} = \Sigma_\text{dis}(i\omega,{\bf k})\\
							&\, = \Gamma^2 \int_{\bf q} {\hat \lambda} \, G_{0} (i\omega,{\bf {\bf q}}) \, {\hat \lambda} \nonumber \\
							&\, = - \frac{1}{2} \, \Gamma\left[ 1-\frac{d}{2} \right] i \omega \, \gamma^2 \, \frac{1+t \lambda}{\eta (1-t^2)^{3/2} } \, {\hat \lambda} (\sigma_0 - t \chi {\bf d} \cdot {\pmb \sigma} ) {\hat \lambda} \nonumber \\
							&\,= \left( \frac{1}{\epsilon} \right) i \omega \, \gamma^2 \frac{1+ t \lambda}{\eta (1-t^2)^{3/2} } \left[ \left( (1+ \lambda^2) + 2 t \lambda \right) \sigma_0 \right. \nonumber \\
							&\qquad - \left. \left( t (1+\lambda^2)+2 \lambda \right) \chi {\bf d} \cdot {\pmb \sigma} \right] + O (\epsilon^0). \nonumber
\end{align}

On the other hand, the Coulomb self-energy reads
\begin{align}
&\text{Fig.(\ref{fig:SELF_CC})} = \Sigma_\text{Cou}(i\omega,{\bf k})\\
							&\, =  (ig)^2 \int_{\omega', {\bf q} } {\hat \lambda} \, G_{0} (i\omega - i \omega' ,{\bf k} - {\bf q} ) \, {\hat \lambda} \, D_{0} (i\omega', {\bf q} ) \nonumber \\
							&\,= - 2 \, \Gamma \left[ - \frac{ {\bar \epsilon} }{2} \right] \, v \, \alpha \, {\hat \lambda} \left[ - F_\parallel^{\eta'} \frac{ k_\parallel }{1-\lambda^2} (\lambda \sigma_0 + \chi {\bf d} \cdot {\pmb \sigma} ) \right. \nonumber \\
							&\qquad - \left. \frac{ \eta } { 1 + t \lambda} F_\perp^{\eta'} \chi {\bf k}_\perp \cdot {\pmb \sigma} \right] {\hat \lambda} \nonumber \\
							&\,= \left( \frac{1}{ {\bar \epsilon} } \right) \, v \, \alpha \left[ F_\parallel^{\eta'} k_\parallel (\lambda \sigma_0 - \chi {\bf d} \cdot {\pmb \sigma}) \right. \nonumber \\
							&\qquad - \left. \eta \frac{1-\lambda}{1+ t \lambda} F_\perp^{\eta'} \chi {\bf k}_\perp \cdot {\pmb \sigma} \right] + O ({\bar \epsilon}^0), \nonumber
\end{align}
where $F_\parallel^{\eta'}$ and $F_\perp^{\eta'}$ are functions of the anisotropy parameter $\eta'$,
\begin{align}
&F_\parallel^{\eta'} = \frac{2}{\pi} \int_0^\infty {\mathrm d}u \; u^{ ({\bar \epsilon}-1)/2} (1+u)^{-3/2} (1+ \eta'^2 u)^{ (1-d)/2} \nonumber \\
&\quad \left( {k_\parallel^2 + \eta'^2 k_\perp^2}  \right)^{ - {\bar \epsilon}/2} \left( \frac{k_\parallel^2}{1+u} + \frac{\eta'^2 k_\perp^2}{1+\eta'^2 u}  \right)^{ {\bar \epsilon}/2},	 \label{Appeq:anisfuncparInt} \\
&F_\perp^{\eta'} = \frac{2}{\pi} \int_0^\infty {\mathrm d}u \; u^{ ({\bar \epsilon}-1)/2} (1+u)^{-1/2} (1+ \eta'^2 u)^{ -(1+d)/2} \nonumber \\
&\quad \left( {k_\parallel^2 + \eta'^2 k_\perp^2}  \right)^{ - {\bar \epsilon}/2} \left( \frac{k_\parallel^2}{1+u} + \frac{\eta'^2 k_\perp^2}{1+\eta'^2 u}  \right)^{ {\bar \epsilon}/2}.	 \label{Appeq:anisfuncperpInt}
\end{align}
Note that these functions are not divergent under ${\bar \epsilon} \to 0$. At zeroth order in ${\bar \epsilon}$, the integrals in Eqs.~\eqref{Appeq:anisfuncparInt}-\eqref{Appeq:anisfuncperpInt} can be done explicitly and functions reduce to the definitions given in Eqs.~\eqref{eq:anisfuncpar}-\eqref{eq:anisfuncperp} of the main body. For the polarisation we find the expression
\begin{align}
&\text{Fig.(\ref{fig:POLAR_CC})} = \Pi( i \omega, {\bf k}) \\
 			&\, = - (ig)^2 \int_{\omega', {\bf q} } \text{Tr}\left[ {\hat \lambda} \, G_{0} (i\omega', {\bf q}) \, {\hat \lambda} \,G_{0} (i\omega' + i \omega,  {\bf q} + {\bf k})\right] \nonumber \\
 			&\,= -  2^{m-d} N   \alpha \left( \frac{d-1}{d} \right) \Gamma \left[ \frac{d-1}{2} \right] \Gamma \left[ \frac{3-d}{2} \right] \nonumber \\
 			&\qquad \quad \frac{1-\lambda^2}{1+ t \lambda} (v \eta)^{1-d} \mu^{d-3} k'^2  \nonumber \\
 			&\,= O (\epsilon^0). \nonumber
\end{align}

Then there are the diagrams in Fig.(\ref{fig:VERTICES_d}), which will renormalize the disorder interaction strength $\Gamma$. The disorder-only leading order correction yields
\begin{align}
&\text{Fig.(\ref{fig:VERTEX_ddd})} = \Gamma^3 \int_{\bf q} \left[  {\hat \lambda} \, G_{0}(i \omega, {\bf q}) \right]^2 \, {\hat \lambda}  \nonumber \\
			&= \frac{1}{2} \Gamma \left[1 - \frac{d}{2} \right] (d-1) \Gamma \gamma^2 \frac{1+t \lambda}{\eta (1-t^2)^{3/2} } {\hat \lambda} ( \sigma_0 - t \chi {\bf d} \cdot {\pmb \sigma} ) {\hat \lambda} \nonumber \\
			&= - \left( \frac{1}{\epsilon} \right) \Gamma \, \gamma^2 \frac{1+ t \lambda}{\eta (1-t^2)^{3/2} } \left[ \left( (1+ \lambda^2) + 2 t \lambda \right) \sigma_0 \right. \nonumber \\
							&\qquad - \left. \left( t (1+\lambda^2)+2 \lambda \right) \chi {\bf d} \cdot {\pmb \sigma} \right] + O (\epsilon^0), 
\end{align}
whereas the mixed disorder-Coulomb diagram results in
\begin{align}
& \text{Fig.(\ref{fig:VERTEX_CdC})} \nonumber \\
&= \Gamma (ig)^2 \int_{\omega', {\bf q} } D_{0}(i \omega', {\bf q}) \left[ {\hat \lambda} \, G_{0} (i\omega - i \omega', {\bf k} - {\bf q}) \right]^2\, {\hat \lambda} \,   \nonumber \\
&= 0.
\end{align}
Another perturbative contribution to the disorder vertex comes from the fermion loop diagram
\begin{align}
& \text{Fig.(\ref{fig:VERTEX_dfermionloop})} =  - (ig)^2 \Gamma {\hat \lambda}  \int_{\omega', {\bf q} }  D_{0}(i \omega, {\bf k}) \nonumber \\
		 	&\qquad \quad \text{Tr} \left[{\hat \lambda} \, G_{0} (i\omega', {\bf q}) \, {\hat \lambda} \,G_{0} (i\omega' + i \omega,  {\bf q}+{\bf k})\right] \nonumber \\
		 	&= -  2^{m-d} N \alpha \left( \frac{d-1}{d} \right) \Gamma {\hat \lambda} \left[ \frac{d-1}{2} \right] \Gamma \left[ \frac{3-d}{2} \right] \nonumber \\
 			&\qquad \quad \frac{1-\lambda^2}{1+ t \lambda} (v \eta)^{1-d} \mu^{d-3}  \nonumber \\
 			&= O (\epsilon^0).
\end{align}

Lastly there are the diagrams in Fig.(\ref{fig:VERTICES_C}), which source the renormalization of the Coulomb interaction strength $g$. The purely Coulombic diagram vanishes identically,
\begin{align}
& \text{Fig.(\ref{fig:VERTEX_CCC})} \nonumber \\
&= (ig)^3 \int_{\omega', {\bf q} } D_{0}(i \omega', {\bf q}) \left[ {\hat \lambda} \, G_{0} (i\omega - i \omega', {\bf k} - {\bf q}) \right]^2  {\hat \lambda}  \nonumber \\
&= 0, 
\end{align}
in a restatement of gauge invariance. The mixed diagram however results in a finite contribution of the form
\begin{align}
&\text{Fig.(\ref{fig:VERTEX_dCd})} = (ig) \Gamma^2 \int_{\omega', {\bf q} } \delta ( \omega - \omega')   \left[ {\hat \lambda} \, G_{0}(i \omega', {\bf q}) \right]^2 \, {\hat \lambda} \nonumber \\
			&= \frac{1}{2} \Gamma \left[1 - \frac{d}{2} \right] (d-1) (ig) \gamma^2 \frac{1+t \lambda}{\eta (1-t^2)^{3/2} } {\hat \lambda} ( \sigma_0 - t \chi {\bf d} \cdot {\pmb \sigma} ) {\hat \lambda} \nonumber \\
			&= - \left( \frac{1}{\epsilon} \right) (ig) \, \gamma^2 \frac{1+ t \lambda}{\eta (1-t^2)^{3/2} } \left[ \left( (1+ \lambda^2) + 2 t \lambda \right) \sigma_0 \right. \nonumber \\
							&\qquad - \left. \left( t (1+\lambda^2)+2 \lambda \right) \chi {\bf d} \cdot {\pmb \sigma} \right] + O (\epsilon^0).
\end{align}
Note that we might furthermore consider a pututative diagram in which an internal disorder line interpolates between an external Coulomb line and vertex point by means of an intermediate fermion loop. This however will have a momentum dependent result that is irrelevant in the RG sense and is therefore neglected. 

\section{Counterterms and $\beta$ functions} \label{appsec:Counterterms}

We now include counterterms to cancel the divergences in the diagrams of the perturbative expansion in the couplings. The renormalized self energy becomes
\begin{align}
& \Sigma_R (i \omega, {\bf q} ) = \Sigma_{\text dis} (i \omega, {\bf q} ) +  \Sigma_{\text Cou} (i \omega, {\bf q} ) \nonumber \\
&\quad - \left\{ \left( \delta_\psi i \omega - (\delta_\psi + \delta_v + \delta_v ) v t q_\parallel \right) \sigma_0 - \chi  \left( ( \delta_\psi + \delta_v ) v q_\parallel {\bf d} \right. \right. \nonumber \\
&\qquad + \left. \left. (\delta_\psi + \delta_v + \delta_\eta) v \eta {\bf q}_\perp + (\delta_\psi + \delta_\lambda) i \omega \lambda {\bf d} \right) \cdot {\pmb \sigma} \right\}  \nonumber \\
&= i \omega \sigma_0 \left\{ \left( \frac{1}{\epsilon} \right) \gamma^2 \frac{1 + t \lambda}{\eta (1-t^2)^{3/2} }  \left( (1+\lambda^2) + 2 t \lambda \right) - \delta_\psi \right\} \nonumber \\
&\quad - v t q_\parallel \sigma_0 \left\{ - \left( \frac{1}{ {\bar \epsilon} } \right) \alpha \frac{\lambda}{t} F_\parallel^{\eta'} - \left( \delta_\psi + \delta_v + \delta_t \right) \right\} \nonumber \\
&\quad - v \chi q_\parallel {\bf d} \cdot {\pmb \sigma} \left\{ \left( \frac{1}{ {\bar \epsilon} } \right) \alpha F_\parallel^{\eta'} - \left( \delta_\psi + \delta_v \right) \right\} \nonumber \\
&\quad - v \eta \chi {\bf q}_\perp \cdot {\pmb \sigma} \left\{ \left( \frac{1}{ {\bar \epsilon} } \right) \alpha \frac{1-\lambda^2}{1+t \lambda} F_\perp^{\eta'} - \left( \delta_\psi + \delta_v + \delta_\eta \right) \right\} \nonumber \\
&\quad - i \omega \lambda \chi {\bf d} \cdot {\pmb \sigma} \left\{ \left( \frac{1}{\epsilon} \right) \gamma^2 \frac{1+t \lambda}{\eta (1-t^2)^{3/2} } \left( t (1+\lambda^2) + 2 \lambda \right) \right. \nonumber \\
&\qquad \quad - \left. \left( \delta_\psi + \delta_\lambda \right) \right\} \nonumber \\
&=0.
\end{align}
Because the polarization diagram is regular under our renormalization scheme the photon field counterterm vanishes along the lines of
\begin{align}
& \Pi_R (i \omega, {\bf q} ) = \Pi (i \omega, {\bf q} ) - 2 \delta_\varphi q^2 = 0. 
\end{align}
The renormalization deriving form the vertex correction diagrams can be counteracted as 
\begin{align}
& \vcenter{\hbox{\includegraphics[width=1.6cm]{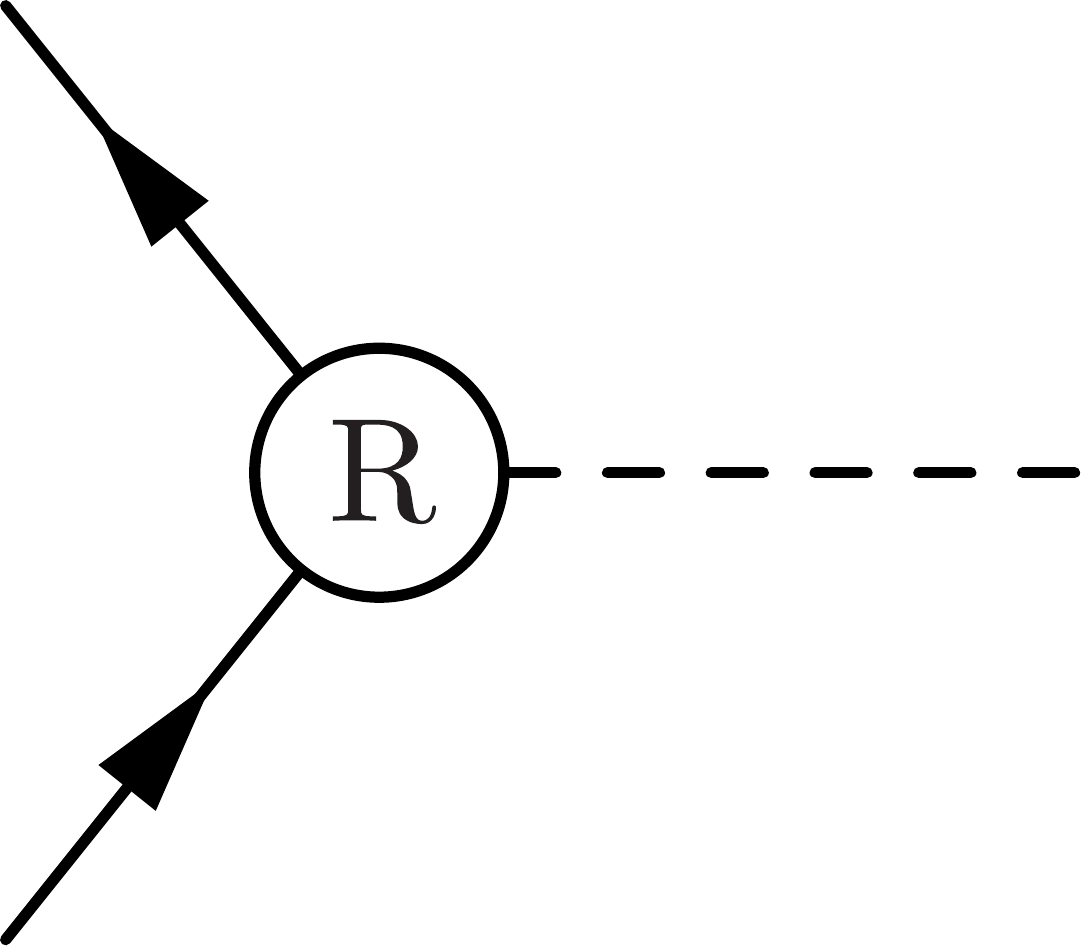}}} \quad = \text{Figs.(\ref{fig:VERTICES_d})} \nonumber  \\
&\quad + \left\{ \left( \delta_\psi + \delta_\Gamma \right) \Gamma \sigma_0 - \left( \delta_\psi + \delta_\Gamma + \delta_\lambda \right) \Gamma \lambda \chi {\bf d} \cdot {\pmb \sigma} \right\}  \nonumber \\
&=\Gamma \sigma_0 \left\{ - \left( \frac{1}{\epsilon} \right) \gamma^2 \frac{1+t \lambda}{\eta (1-t^2)^{3/2} } \left( (1+ \lambda^2) + 2 t \lambda) \right) \right. \nonumber \\ 
&\qquad + \left. \left( \delta_\psi + \delta_\Gamma \right) \right\} \nonumber \\
&\quad -\Gamma \lambda \chi {\bf d} \cdot {\pmb \sigma} \left\{ - \left( \frac{1}{\epsilon} \right) \gamma^2 \frac{1+t \lambda}{\eta (1-t^2)^{3/2} } \left( t (1+\lambda^2) + 2 \lambda \right) \right. \nonumber \\
&\qquad + \left. \left( \delta_\psi + \delta_\Gamma + \delta_\lambda \right) \right\} \nonumber \\
&=0.
\end{align}
and  
\begin{align}
& \vcenter{\hbox{\includegraphics[width=1.6cm]{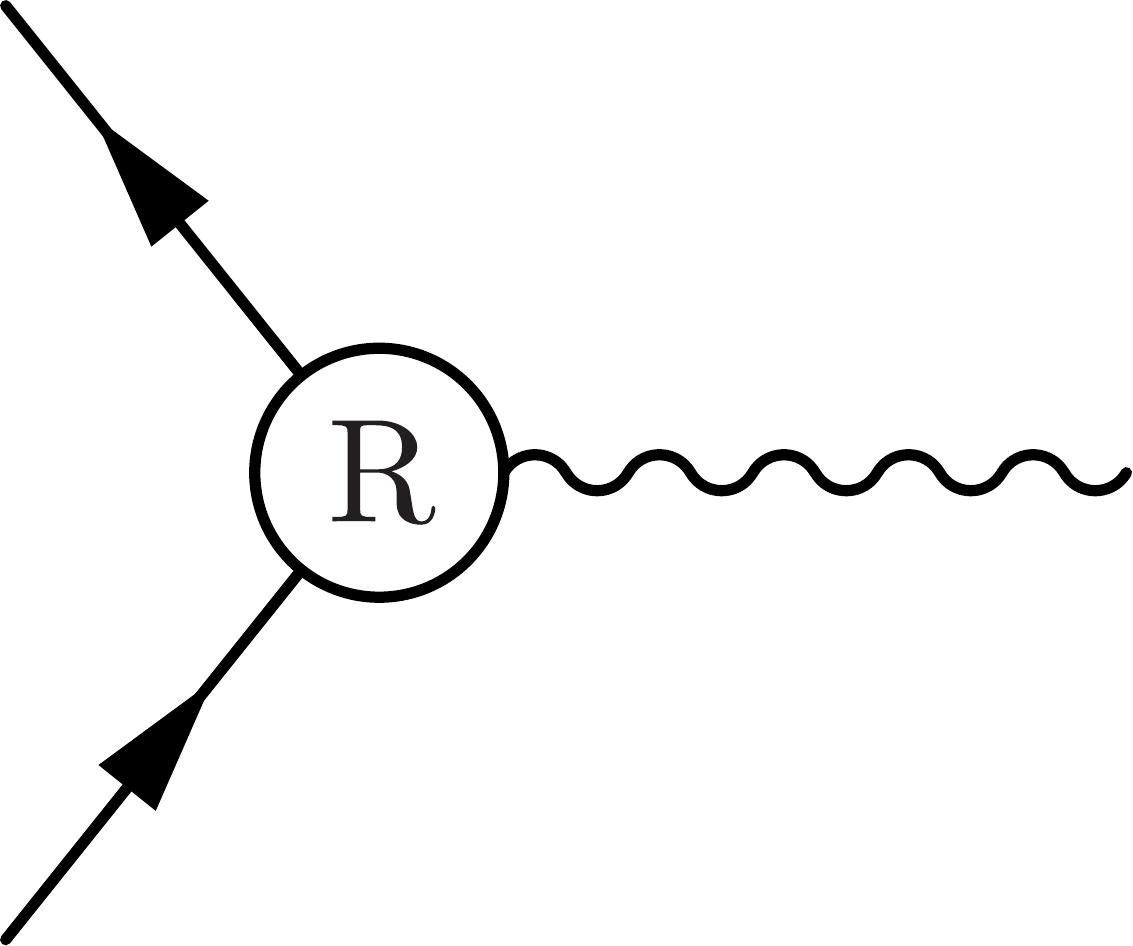}}} \quad = \text{Figs.(\ref{fig:VERTICES_C})}  \nonumber \\
&\quad + \left\{ \left( \delta_\psi + \delta_\varphi + \delta_g \right) ig \sigma_0 - \left( \delta_\psi + \delta_\varphi + \delta_g + \delta_\lambda \right) ig \lambda \chi {\bf d} \cdot {\pmb \sigma} \right\}  \nonumber \\
&= ig \sigma_0 \left\{ - \left( \frac{1}{\epsilon} \right) \gamma^2 \frac{1+t \lambda}{\eta (1-t^2)^{3/2} } \left( (1+ \lambda^2) + 2 t \lambda) \right) \right. \nonumber \\ 
&\qquad + \left. \left( \delta_\psi + \delta_\varphi + \delta_g \right) \right\} \nonumber \\
&\quad- ig \lambda \chi {\bf d} \cdot {\pmb \sigma} \left\{ - \left( \frac{1}{\epsilon} \right) \gamma^2 \frac{1+t \lambda}{\eta (1-t^2)^{3/2} } \left( t (1+\lambda^2) + 2 \lambda \right) \right. \nonumber \\
&\qquad + \left. \left( \delta_\psi + \delta_\varphi + \delta_g + \delta_\lambda \right) \right\} \nonumber \\
&=0.
\end{align}
Consequently, for the fields of the theory we find counterterms
\begin{align}
&\delta_\psi = \left( \frac{1}{\epsilon} \right) \gamma^2 \frac{1+t \lambda}{\eta (1-t^2)^{3/2} } \left( (1+ \lambda^2) + 2 t \lambda) \right), \\
&\delta_\varphi = 0,
\end{align}
with which we also derive the terms neede to nullify the diagrammatic contributions to the tree level parameters,
\begin{align}
&\delta_v 	= \left( \frac{1}{ {\bar \epsilon} } \right) \alpha F_\parallel^{\eta'} - \delta_\psi \\
&\,=  \left( \frac{1}{ {\bar \epsilon} } \right) \alpha F_\parallel^{\eta'} -  \left( \frac{1}{\epsilon} \right) \gamma^2 \frac{1+t \lambda}{\eta (1-t^2)^{3/2} } \left( (1+ \lambda^2) + 2 t \lambda) \right) \nonumber \\
&\delta_t = - \left( \frac{1}{ {\bar \epsilon} } \right) \alpha \frac{\lambda}{t} F_\parallel^{\eta'} - \delta_\psi - \delta_v, \\
&\,=  - \left( \frac{1}{ {\bar \epsilon} } \right) \alpha \left( \frac{\lambda}{t} + 1 \right) F_\parallel^{\eta'}, \nonumber \\
&\delta_\eta = \left( \frac{1}{ {\bar \epsilon} } \right) \alpha \frac{1-\lambda^2}{1+ t \lambda} F_\perp^{\eta'} - \delta_\psi - \delta_v \\
&\,= -\left( \frac{1}{ {\bar \epsilon} } \right) \alpha \left( F_\parallel^{\eta'} - \frac{1-\lambda^2}{1+ t \lambda} F_\perp^{\eta'} \right). \nonumber
\end{align}
Because of gauge invariance the field counterterms will be exactly sufficient to cancel the divergences coming from the vertex diagrams so that the coupling strenghts are not renormalized and their counterterms vanish.  
\begin{align}
&\delta_g = \left( \frac{1}{\epsilon} \right) \gamma^2 \frac{1+t \lambda}{\eta (1-t^2)^{3/2} } \left( (1+ \lambda^2) + 2 t \lambda) \right) - \delta_\psi - \delta_\phi \nonumber \\
&\, = 0,\\
&\delta_\Gamma = \left( \frac{1}{\epsilon} \right) \gamma^2 \frac{1+t \lambda}{\eta (1-t^2)^{3/2} } \left( (1+ \lambda^2) + 2 t \lambda) \right) - \delta_\psi \nonumber \\
&\, = 0.
\end{align}
Note that because $\delta_\phi = \delta_g = \delta_\Gamma =0$, we also find consistently from the renormalization of the self energy and both the interaction vertices that
\begin{align}
&\delta_\lambda = \left( \frac{1}{\epsilon} \right) \gamma^2 \frac{1+t \lambda}{\eta (1-t^2)^{3/2} } \left( t (1+ \lambda^2) + 2 \lambda) \right) - \delta_\psi \nonumber \\
&\,= \left( \frac{1}{\epsilon} \right) \gamma^2 \frac{1+t \lambda}{\eta (1-t^2)^{3/2} }(t + \lambda) (1-\lambda^2). 
\end{align}

Beyond first order in perturbation, the flow equations become 
\begin{align}
\beta_{y_i} 	&= \mu \frac{ {\mathrm d} y_i } { {\mathrm d} \mu } = y_i \left( \mu^{- [ y_i] } \mu \frac{  {\mathrm d} \mu^{ [y_i] } } {  {\mathrm d} \mu } + \mu \frac{ Z_{y_i}^{-1} } { {\mathrm d} \mu } + y_{i,0}^{-1} \mu \frac{ \mathrm{d} y_{i,0} } { {\mathrm d} \mu } \right) \nonumber \\
			&= y_i \left( [y_i] - \sum_{j \neq i}  \frac{ {\mathrm d} \ln Z_{y_i} } { {\mathrm d} y_j } \beta_{y_j} \right) \left(1+ y_i  \frac{ {\mathrm d} Z_{y_i} } { {\mathrm d} y_i } \right)^{-1} \nonumber \\
			&= y_i  [y_i]  - y_i \sum_j y_j [y_j] \frac{ {\mathrm d} \delta_{y_i} } { {\mathrm d} y_j } + 
O ( \text{perturb.}^2)
\end{align}

Using that the only non-vanishing scaling dimensions are 
\begin{align*}
[v] = z-1, \; [\gamma] = [\Gamma] - [v] = - \frac{\epsilon}{2}, \; [\alpha] = 2 [g]-[v] = - {\bar \epsilon},
\end{align*}
and that all the counterterms are independent of $v$, i.e. ${\mathrm d}  \delta_{y_i} / {\mathrm d}  v = 0$ $\forall y_i$, this simplifies to
\begin{align*}
\beta_{y_i} = y_i [y_i] + y_i \left\{ \frac{ \epsilon}{2} \, \gamma \frac{ {\mathrm d} y_i } { {\mathrm d} \gamma } + {\bar \epsilon} \,\alpha \frac{ {\mathrm d} y_i } { {\mathrm d} \alpha } \right \} + O ( g^2, \Gamma^2, g \Gamma).
\end{align*}
Applying this formula to the parameters of the model yields
\begin{align}
\beta_v 		&= v \left\{ z - 1 - \gamma^2 \frac{(1+ t \lambda) \left( (1+ \lambda^2) + 2 t \lambda \right)}{\eta (1-t^2)^{3/2} }  + \alpha F_\parallel^{\eta'} \right\}, \\
\beta_t 		&= - \alpha \left(t + \lambda \right) F_\parallel^{\eta'}, \\
\beta_\lambda	&=  \gamma^2 \frac{(1+ t \lambda)(t + \lambda)}{\eta (1-t^2)^{3/2} } (1-\lambda^2), \\
\beta_\eta		&= - \eta \alpha \left( F_\parallel^{\eta'} - F_\perp^{\eta'} \right), \\
\beta_\alpha	&= \alpha \left\{- {\bar \epsilon} + \gamma^2 \frac{(1+ t \lambda) \left( (1+ \lambda^2) + 2 t \lambda \right)}{\eta (1-t^2)^{3/2} } - \alpha F_\parallel^{\eta'} \right\}, \\
\beta_\gamma	&= \gamma \left\{ - \frac{\epsilon}{2} + \gamma^2 \frac{(1+ t \lambda) \left( (1+ \lambda^2) + 2 t \lambda \right)}{\eta (1-t^2)^{3/2} } - \alpha F_\parallel^{\eta'} \right\}.
\end{align}

\section{Re-expressing the flow equations} \label{appsec:ReexpBetaFncs}

Would proceed to analyze the set of equations presented in the previous section of the appendix, finding the flow's fixed points and then examining afterwards what these look like in terms of the original parameters of the theory by using Eq.~\eqref{eq:ParamReRedif},
\begin{align*}
v' = v \, \frac{1+ t \lambda }{1-\lambda^2}, \quad \eta' = \eta \, \frac{ \sqrt{1-\lambda^2}}{1+ t \lambda}, \quad t' = \frac{t+\lambda}{1+ t \lambda}. 
\end{align*}
We take an alternative strategy in which we use this equation to directly translate back the $\beta$ functions into the language of the original model parameters as we have found this to significantly simplify their structure. The flow equations become
\begin{align}
\beta_v 		&= v \left\{ z - 1 - \gamma^2 \frac{(1+ t' \lambda) (1- t' \lambda)^2}{\eta' (1-t'^2)^{3/2} }  + \alpha F_\parallel^{\eta'} \right\}, \\
\beta_t 		&= - t' \frac{1-\lambda^2}{1- t' \lambda} \, \alpha F_\parallel^{\eta'}, \\
\beta_\lambda	&=  t' \frac{(1-\lambda^2)(1- t' \lambda)^2}{\eta' (1-t'^2)^{3/2} }  \gamma^2, \\
\beta_\eta		&= - \eta \alpha \left( F_\parallel^{\eta'} - (1 - t' \lambda) F_\perp^{\eta'} \right), \\
\beta_\alpha	&= \alpha \left\{- {\bar \epsilon} + \gamma^2 \frac{(1+ t' \lambda) (1-t' \lambda)^2 }{\eta' (1-t'^2)^{3/2} } - \alpha F_\parallel^{\eta'} \right\}, \\
\beta_\gamma	&= \gamma \left\{ - \frac{\epsilon}{2} + \gamma^2 \frac{(1+ t' \lambda) (1-t' \lambda)^2 }{\eta' (1-t'^2)^{3/2} } - \alpha F_\parallel^{\eta'} \right\}.
\end{align}
The $\beta$ functions for the original Fermi velocity $v'$, fermion anisotropy $\eta'$ and tilt $t'$ are straightforward combinations of the above through redefinitions Eq.~\eqref{eq:ParamReRedif}. This yields
\begin{align}
& \beta_{v'} = v' \left\{ \frac{\beta_v}{v}+ \frac{\lambda}{1+t \lambda} \beta_t + \frac{t + 2 \lambda + t \lambda^2}{(1+ t \lambda) (1-\lambda^2)} \beta_\lambda \right\} \\
&\quad = v' \left\{ \frac{\beta_v}{v}+ \frac{\lambda (1- t' \lambda)}{1-\lambda^2} \beta_t + \frac{t' + \lambda}{1- \lambda^2} \beta_\lambda \right\} \nonumber \\
&\quad = v' \left\{ z - 1 - \frac{ (1-t' \lambda)^2}{ \eta' (1-t'^2)^{1/2} } \gamma^2 + (1-t' \lambda) \alpha F_\parallel^{\eta'}  \right\}, 	\nonumber \\
&\beta_{t'} = t' \left\{ \frac{1-\lambda^2}{(t+\lambda)(1+t \lambda)} \beta_t + \frac{1-t^2}{(t+\lambda)(1+t \lambda)} \beta_\lambda \right\} \\
&\quad = t' \left\{ \frac{(1- t' \lambda)^2}{t' (1-\lambda^2)} \beta_t + \frac{1-t'^2}{t' (1-\lambda^2)}  \beta_\lambda \right\}  \nonumber \\
&\quad = -t' \left\{ (1-t' \lambda) \alpha F_\parallel^{\eta'} - \frac{ (1-t' \lambda)^2}{ \eta' (1-t'^2)^{1/2} } \gamma^2 \right\},  \nonumber \\
&\beta_{\eta'} = \eta' \left\{ \frac{\beta_\eta}{\eta} - \frac{\lambda}{1+ t \lambda} \beta_t - \frac{t + \lambda}{(1+t \lambda)(1-\lambda^2)} \beta_\lambda \right\} \\
&\quad = \eta' \left\{ \frac{\beta_\eta}{\eta} - \frac{\lambda (1 - t' \lambda)}{1-\lambda^2} \beta_t - \frac{t'}{1-\lambda^2} \beta_\lambda \right\} \nonumber \\
&\quad = - \eta' \left\{ (1-t' \lambda) \left( F_\parallel^{\eta'} - F_\perp^{\eta'} \right) + t' \frac{ (1-t' \lambda)^2}{ \eta' (1-t'^2)^{3/2} } \gamma^2 \right\}. \nonumber
\end{align}

We can simplify further with by redefining the couplings to those set out in Eq.~\eqref{eq:CouplingRedif}  of the main body,
\begin{align*}
\gamma'^2	= \frac{\Omega_d \, \mu^\epsilon }{(2 \pi)^d \, v'^2} \, \Gamma^2, \qquad \alpha' = \frac{\Omega_d \, \mu^{ {\bar \epsilon} } }{4 (2 \pi)^d \, v'} \, g^2. 	
\end{align*}
In terms of these, we find
\begin{align}
&\beta_{v'} = v' \left\{ z-1 - \frac{1}{\eta' \sqrt{1-t'^2} } \gamma'^2 + \alpha' F_\parallel^{\eta'} \right\},  \\
&\beta_{t'} = -t' \left\{ \alpha' F_\parallel^{\eta'} - \frac{1}{\eta' \sqrt{1-t'^2} } \gamma'^2 \right\},  \\
&\beta_{\eta'} = - \eta' \left\{ \alpha' \left( F_\parallel^{\eta'} - F_\perp^{\eta'} \right) + \frac{t'^2}{\eta' (1-t'^2)^{3/2}} \gamma'^2 \right\},\\
&\beta_\lambda	= t' \frac{1- \lambda^2}{\eta' (1-t'^2)^{3/2} } \gamma'^2. 
\end{align}
The flow of the redefined couplings themselves is determined by the equations
\begin{align}
& \beta_{\alpha'}	 = \alpha' \left( \frac{\beta_\alpha}{\alpha} + \frac{\beta_v}{v} - \frac{\beta_{v'}}{v'} \right) \\
&\quad = \alpha' \left\{ - {\bar \epsilon} + \frac{1}{\eta' \sqrt{1-t'^2} } \gamma'^2 - \alpha' F_\parallel^{\eta'} \right\}, \nonumber  \\
& \beta_{\gamma'} = \gamma' \left( \frac{\beta_\gamma}{\gamma} + \frac{\beta_v}{v} - \frac{\beta_{v'} }{v'} \right)\\
&\quad = \gamma' \left\{ -\frac{\epsilon}{2} + \frac{1}{\eta' \sqrt{1-t'^2} } \gamma'^2 - \alpha' F_\parallel^{\eta'} \right\}. \nonumber
\end{align}
Taken together these form the set of four coupled equations Eqs.~\eqref{eq:RG_tilt}-\eqref{eq:RG_dis} and two further decoupled equations Eqs.~\eqref{eq:RG_FermiVelocity}-\eqref{eq:RG_lambda}.

\newpage

\begin{thebibliography}{44}%
\makeatletter
\providecommand \@ifxundefined [1]{%
 \@ifx{#1\undefined}
}%
\providecommand \@ifnum [1]{%
 \ifnum #1\expandafter \@firstoftwo
 \else \expandafter \@secondoftwo
 \fi
}%
\providecommand \@ifx [1]{%
 \ifx #1\expandafter \@firstoftwo
 \else \expandafter \@secondoftwo
 \fi
}%
\providecommand \natexlab [1]{#1}%
\providecommand \enquote  [1]{``#1''}%
\providecommand \bibnamefont  [1]{#1}%
\providecommand \bibfnamefont [1]{#1}%
\providecommand \citenamefont [1]{#1}%
\providecommand \href@noop [0]{\@secondoftwo}%
\providecommand \href [0]{\begingroup \@sanitize@url \@href}%
\providecommand \@href[1]{\@@startlink{#1}\@@href}%
\providecommand \@@href[1]{\endgroup#1\@@endlink}%
\providecommand \@sanitize@url [0]{\catcode `\\12\catcode `\$12\catcode
  `\&12\catcode `\#12\catcode `\^12\catcode `\_12\catcode `\%12\relax}%
\providecommand \@@startlink[1]{}%
\providecommand \@@endlink[0]{}%
\providecommand \url  [0]{\begingroup\@sanitize@url \@url }%
\providecommand \@url [1]{\endgroup\@href {#1}{\urlprefix }}%
\providecommand \urlprefix  [0]{URL }%
\providecommand \Eprint [0]{\href }%
\providecommand \doibase [0]{http://dx.doi.org/}%
\providecommand \selectlanguage [0]{\@gobble}%
\providecommand \bibinfo  [0]{\@secondoftwo}%
\providecommand \bibfield  [0]{\@secondoftwo}%
\providecommand \translation [1]{[#1]}%
\providecommand \BibitemOpen [0]{}%
\providecommand \bibitemStop [0]{}%
\providecommand \bibitemNoStop [0]{.\EOS\space}%
\providecommand \EOS [0]{\spacefactor3000\relax}%
\providecommand \BibitemShut  [1]{\csname bibitem#1\endcsname}%
\let\auto@bib@innerbib\@empty
\bibitem [{\citenamefont {Novoselov}\ \emph {et~al.}(2004)\citenamefont
  {Novoselov}, \citenamefont {Geim}, \citenamefont {Morozov}, \citenamefont
  {Jiang}, \citenamefont {Zhang}, \citenamefont {Dubonos}, \citenamefont
  {Grigorieva},\ and\ \citenamefont {Firsov}}]{Novoselov2004}%
  \BibitemOpen
  \bibfield  {author} {\bibinfo {author} {\bibfnamefont {K.~S.}\ \bibnamefont
  {Novoselov}}, \bibinfo {author} {\bibfnamefont {A.~K.}\ \bibnamefont {Geim}},
  \bibinfo {author} {\bibfnamefont {S.~V.}\ \bibnamefont {Morozov}}, \bibinfo
  {author} {\bibfnamefont {D.}~\bibnamefont {Jiang}}, \bibinfo {author}
  {\bibfnamefont {Y.}~\bibnamefont {Zhang}}, \bibinfo {author} {\bibfnamefont
  {S.~V.}\ \bibnamefont {Dubonos}}, \bibinfo {author} {\bibfnamefont {I.~V.}\
  \bibnamefont {Grigorieva}}, \ and\ \bibinfo {author} {\bibfnamefont {A.~A.}\
  \bibnamefont {Firsov}},\ }\href {\doibase 10.1126/science.1102896} {\bibfield
   {journal} {\bibinfo  {journal} {Science}\ }\textbf {\bibinfo {volume}
  {306}},\ \bibinfo {pages} {666} (\bibinfo {year} {2004})},\ \Eprint
  {http://arxiv.org/abs/https://science.sciencemag.org/content/306/5696/666.full.pdf}
  {https://science.sciencemag.org/content/306/5696/666.full.pdf} \BibitemShut
  {NoStop}%
\bibitem [{\citenamefont {Weyl}(1929)}]{Weyl1929}%
  \BibitemOpen
  \bibfield  {author} {\bibinfo {author} {\bibfnamefont {H.}~\bibnamefont
  {Weyl}},\ }\href {\doibase 10.1007/BF01339504} {\bibfield  {journal}
  {\bibinfo  {journal} {Zeitschrift für Physik}\ }\textbf {\bibinfo {volume}
  {56}},\ \bibinfo {pages} {330} (\bibinfo {year} {1929})}\BibitemShut
  {NoStop}%
\bibitem [{\citenamefont {Weng}\ \emph {et~al.}(2015)\citenamefont {Weng},
  \citenamefont {Fang}, \citenamefont {Fang}, \citenamefont {Bernevig},\ and\
  \citenamefont {Dai}}]{Weng2015weyl}%
  \BibitemOpen
  \bibfield  {author} {\bibinfo {author} {\bibfnamefont {H.}~\bibnamefont
  {Weng}}, \bibinfo {author} {\bibfnamefont {C.}~\bibnamefont {Fang}}, \bibinfo
  {author} {\bibfnamefont {Z.}~\bibnamefont {Fang}}, \bibinfo {author}
  {\bibfnamefont {B.~A.}\ \bibnamefont {Bernevig}}, \ and\ \bibinfo {author}
  {\bibfnamefont {X.}~\bibnamefont {Dai}},\ }\href {\doibase
  10.1103/PhysRevX.5.011029} {\bibfield  {journal} {\bibinfo  {journal} {Phys.
  Rev. X}\ }\textbf {\bibinfo {volume} {5}},\ \bibinfo {pages} {011029}
  (\bibinfo {year} {2015})}\BibitemShut {NoStop}%
\bibitem [{\citenamefont {Huang}\ \emph {et~al.}(2016)\citenamefont {Huang},
  \citenamefont {McCormick}, \citenamefont {Ochi}, \citenamefont {Zhao},
  \citenamefont {Suzuki}, \citenamefont {Arita}, \citenamefont {Wu},
  \citenamefont {Mou}, \citenamefont {Cao}, \citenamefont {Yan} \emph
  {et~al.}}]{Huang2016}%
  \BibitemOpen
  \bibfield  {author} {\bibinfo {author} {\bibfnamefont {L.}~\bibnamefont
  {Huang}}, \bibinfo {author} {\bibfnamefont {T.~M.}\ \bibnamefont
  {McCormick}}, \bibinfo {author} {\bibfnamefont {M.}~\bibnamefont {Ochi}},
  \bibinfo {author} {\bibfnamefont {Z.}~\bibnamefont {Zhao}}, \bibinfo {author}
  {\bibfnamefont {M.-T.}\ \bibnamefont {Suzuki}}, \bibinfo {author}
  {\bibfnamefont {R.}~\bibnamefont {Arita}}, \bibinfo {author} {\bibfnamefont
  {Y.}~\bibnamefont {Wu}}, \bibinfo {author} {\bibfnamefont {D.}~\bibnamefont
  {Mou}}, \bibinfo {author} {\bibfnamefont {H.}~\bibnamefont {Cao}}, \bibinfo
  {author} {\bibfnamefont {J.}~\bibnamefont {Yan}},  \emph {et~al.},\ }\href
  {\doibase 10.1038/nmat4685} {\bibfield  {journal} {\bibinfo  {journal}
  {Nature Mater.}\ }\textbf {\bibinfo {volume} {15}},\ \bibinfo {pages}
  {1155—} (\bibinfo {year} {2016})}\BibitemShut {NoStop}%
\bibitem [{\citenamefont {Lv}\ \emph {et~al.}(2015{\natexlab{a}})\citenamefont
  {Lv}, \citenamefont {Weng}, \citenamefont {Fu}, \citenamefont {Wang},
  \citenamefont {Miao}, \citenamefont {Ma}, \citenamefont {Richard},
  \citenamefont {Huang}, \citenamefont {Zhao}, \citenamefont {Chen} \emph
  {et~al.}}]{Lv2015Experimental}%
  \BibitemOpen
  \bibfield  {author} {\bibinfo {author} {\bibfnamefont {B.~Q.}\ \bibnamefont
  {Lv}}, \bibinfo {author} {\bibfnamefont {H.~M.}\ \bibnamefont {Weng}},
  \bibinfo {author} {\bibfnamefont {B.~B.}\ \bibnamefont {Fu}}, \bibinfo
  {author} {\bibfnamefont {X.~P.}\ \bibnamefont {Wang}}, \bibinfo {author}
  {\bibfnamefont {H.}~\bibnamefont {Miao}}, \bibinfo {author} {\bibfnamefont
  {J.}~\bibnamefont {Ma}}, \bibinfo {author} {\bibfnamefont {P.}~\bibnamefont
  {Richard}}, \bibinfo {author} {\bibfnamefont {X.~C.}\ \bibnamefont {Huang}},
  \bibinfo {author} {\bibfnamefont {L.~X.}\ \bibnamefont {Zhao}}, \bibinfo
  {author} {\bibfnamefont {G.~F.}\ \bibnamefont {Chen}},  \emph {et~al.},\
  }\href {\doibase 10.1103/PhysRevX.5.031013} {\bibfield  {journal} {\bibinfo
  {journal} {Phys. Rev. X}\ }\textbf {\bibinfo {volume} {5}},\ \bibinfo {pages}
  {031013} (\bibinfo {year} {2015}{\natexlab{a}})}\BibitemShut {NoStop}%
\bibitem [{\citenamefont {Xu}\ \emph {et~al.}(2015{\natexlab{a}})\citenamefont
  {Xu}, \citenamefont {Belopolski}, \citenamefont {Alidoust}, \citenamefont
  {Neupane}, \citenamefont {Bian}, \citenamefont {Zhang}, \citenamefont
  {Sankar}, \citenamefont {Chang}, \citenamefont {Yuan}, \citenamefont {Lee}
  \emph {et~al.}}]{Xu2015discovery}%
  \BibitemOpen
  \bibfield  {author} {\bibinfo {author} {\bibfnamefont {S.-Y.}\ \bibnamefont
  {Xu}}, \bibinfo {author} {\bibfnamefont {I.}~\bibnamefont {Belopolski}},
  \bibinfo {author} {\bibfnamefont {N.}~\bibnamefont {Alidoust}}, \bibinfo
  {author} {\bibfnamefont {M.}~\bibnamefont {Neupane}}, \bibinfo {author}
  {\bibfnamefont {G.}~\bibnamefont {Bian}}, \bibinfo {author} {\bibfnamefont
  {C.}~\bibnamefont {Zhang}}, \bibinfo {author} {\bibfnamefont
  {R.}~\bibnamefont {Sankar}}, \bibinfo {author} {\bibfnamefont
  {G.}~\bibnamefont {Chang}}, \bibinfo {author} {\bibfnamefont
  {Z.}~\bibnamefont {Yuan}}, \bibinfo {author} {\bibfnamefont {C.-C.}\
  \bibnamefont {Lee}},  \emph {et~al.},\ }\href {\doibase
  10.1126/science.aaa9297} {\bibfield  {journal} {\bibinfo  {journal}
  {Science}\ }\textbf {\bibinfo {volume} {349}},\ \bibinfo {pages} {613}
  (\bibinfo {year} {2015}{\natexlab{a}})}\BibitemShut {NoStop}%
\bibitem [{\citenamefont {Xu}\ \emph {et~al.}(2015{\natexlab{b}})\citenamefont
  {Xu}, \citenamefont {Alidoust}, \citenamefont {Belopolski}, \citenamefont
  {Yuan}, \citenamefont {Bian}, \citenamefont {Chang}, \citenamefont {Zheng},
  \citenamefont {Strocov}, \citenamefont {Sanchez}, \citenamefont {Chang} \emph
  {et~al.}}]{Xu2015discovery2}%
  \BibitemOpen
  \bibfield  {author} {\bibinfo {author} {\bibfnamefont {S.-Y.}\ \bibnamefont
  {Xu}}, \bibinfo {author} {\bibfnamefont {N.}~\bibnamefont {Alidoust}},
  \bibinfo {author} {\bibfnamefont {I.}~\bibnamefont {Belopolski}}, \bibinfo
  {author} {\bibfnamefont {Z.}~\bibnamefont {Yuan}}, \bibinfo {author}
  {\bibfnamefont {G.}~\bibnamefont {Bian}}, \bibinfo {author} {\bibfnamefont
  {T.-R.}\ \bibnamefont {Chang}}, \bibinfo {author} {\bibfnamefont
  {H.}~\bibnamefont {Zheng}}, \bibinfo {author} {\bibfnamefont {V.~N.}\
  \bibnamefont {Strocov}}, \bibinfo {author} {\bibfnamefont {D.~S.}\
  \bibnamefont {Sanchez}}, \bibinfo {author} {\bibfnamefont {G.}~\bibnamefont
  {Chang}},  \emph {et~al.},\ }\href {\doibase 10.1038/nphys3437} {\bibfield
  {journal} {\bibinfo  {journal} {Nature Physics}\ }\textbf {\bibinfo {volume}
  {11}},\ \bibinfo {pages} {748} (\bibinfo {year}
  {2015}{\natexlab{b}})}\BibitemShut {NoStop}%
\bibitem [{\citenamefont {Yang}\ \emph {et~al.}(2015)\citenamefont {Yang},
  \citenamefont {Liu}, \citenamefont {Sun}, \citenamefont {Peng}, \citenamefont
  {Yang}, \citenamefont {Zhang}, \citenamefont {Zhou}, \citenamefont {Zhang},
  \citenamefont {Guo}, \citenamefont {Rahn} \emph {et~al.}}]{Yang2015weyl}%
  \BibitemOpen
  \bibfield  {author} {\bibinfo {author} {\bibfnamefont {L.}~\bibnamefont
  {Yang}}, \bibinfo {author} {\bibfnamefont {Z.}~\bibnamefont {Liu}}, \bibinfo
  {author} {\bibfnamefont {Y.}~\bibnamefont {Sun}}, \bibinfo {author}
  {\bibfnamefont {H.}~\bibnamefont {Peng}}, \bibinfo {author} {\bibfnamefont
  {H.}~\bibnamefont {Yang}}, \bibinfo {author} {\bibfnamefont {T.}~\bibnamefont
  {Zhang}}, \bibinfo {author} {\bibfnamefont {B.}~\bibnamefont {Zhou}},
  \bibinfo {author} {\bibfnamefont {Y.}~\bibnamefont {Zhang}}, \bibinfo
  {author} {\bibfnamefont {Y.}~\bibnamefont {Guo}}, \bibinfo {author}
  {\bibfnamefont {M.}~\bibnamefont {Rahn}},  \emph {et~al.},\ }\href {\doibase
  10.1038/nphys3425} {\bibfield  {journal} {\bibinfo  {journal} {Nature
  Physics}\ }\textbf {\bibinfo {volume} {11}},\ \bibinfo {pages} {728}
  (\bibinfo {year} {2015})}\BibitemShut {NoStop}%
\bibitem [{\citenamefont {Lv}\ \emph {et~al.}(2015{\natexlab{b}})\citenamefont
  {Lv}, \citenamefont {Xu}, \citenamefont {Weng}, \citenamefont {Ma},
  \citenamefont {Richard}, \citenamefont {Huang}, \citenamefont {Zhao},
  \citenamefont {Chen}, \citenamefont {Matt}, \citenamefont {Bisti} \emph
  {et~al.}}]{Lv2015observation}%
  \BibitemOpen
  \bibfield  {author} {\bibinfo {author} {\bibfnamefont {B.}~\bibnamefont
  {Lv}}, \bibinfo {author} {\bibfnamefont {N.}~\bibnamefont {Xu}}, \bibinfo
  {author} {\bibfnamefont {H.}~\bibnamefont {Weng}}, \bibinfo {author}
  {\bibfnamefont {J.}~\bibnamefont {Ma}}, \bibinfo {author} {\bibfnamefont
  {P.}~\bibnamefont {Richard}}, \bibinfo {author} {\bibfnamefont
  {X.}~\bibnamefont {Huang}}, \bibinfo {author} {\bibfnamefont
  {L.}~\bibnamefont {Zhao}}, \bibinfo {author} {\bibfnamefont {G.}~\bibnamefont
  {Chen}}, \bibinfo {author} {\bibfnamefont {C.}~\bibnamefont {Matt}}, \bibinfo
  {author} {\bibfnamefont {F.}~\bibnamefont {Bisti}},  \emph {et~al.},\ }\href
  {\doibase 10.1038/nphys3426} {\bibfield  {journal} {\bibinfo  {journal}
  {Nature Physics}\ }\textbf {\bibinfo {volume} {11}},\ \bibinfo {pages} {724}
  (\bibinfo {year} {2015}{\natexlab{b}})}\BibitemShut {NoStop}%
\bibitem [{\citenamefont {Xu}\ \emph {et~al.}(2015{\natexlab{c}})\citenamefont
  {Xu}, \citenamefont {Weng}, \citenamefont {Lv}, \citenamefont {Matt},
  \citenamefont {Park}, \citenamefont {Bisti}, \citenamefont {Strocov},
  \citenamefont {Pomjakushina}, \citenamefont {Conder}, \citenamefont {Plumb}
  \emph {et~al.}}]{Xu2016observation}%
  \BibitemOpen
  \bibfield  {author} {\bibinfo {author} {\bibfnamefont {N.}~\bibnamefont
  {Xu}}, \bibinfo {author} {\bibfnamefont {H.}~\bibnamefont {Weng}}, \bibinfo
  {author} {\bibfnamefont {B.}~\bibnamefont {Lv}}, \bibinfo {author}
  {\bibfnamefont {C.}~\bibnamefont {Matt}}, \bibinfo {author} {\bibfnamefont
  {J.}~\bibnamefont {Park}}, \bibinfo {author} {\bibfnamefont {F.}~\bibnamefont
  {Bisti}}, \bibinfo {author} {\bibfnamefont {V.}~\bibnamefont {Strocov}},
  \bibinfo {author} {\bibfnamefont {E.}~\bibnamefont {Pomjakushina}}, \bibinfo
  {author} {\bibfnamefont {K.}~\bibnamefont {Conder}}, \bibinfo {author}
  {\bibfnamefont {N.}~\bibnamefont {Plumb}},  \emph {et~al.},\ }\href {\doibase
  10.1038/ncomms11006} {\bibfield  {journal} {\bibinfo  {journal} {Nature
  Comm.}\ }\textbf {\bibinfo {volume} {7}} (\bibinfo {year}
  {2015}{\natexlab{c}}),\ 10.1038/ncomms11006}\BibitemShut {NoStop}%
\bibitem [{\citenamefont {Shekhar}\ \emph {et~al.}(2015)\citenamefont
  {Shekhar}, , \citenamefont {Nayak}, \citenamefont {Sun}, \citenamefont
  {Schmidt}, \citenamefont {Nicklas}, \citenamefont {Leermakers}, \citenamefont
  {Zeitler}, \citenamefont {Skourski}, \citenamefont {Wosnitza}, \citenamefont
  {Liu}, \citenamefont {Chen}, \citenamefont {Schnelle}, \citenamefont
  {Borrmann}, \citenamefont {Grin}, \citenamefont {Felser},\ and\ \citenamefont
  {Yan}}]{Shekhar2015}%
  \BibitemOpen
  \bibfield  {author} {\bibinfo {author} {\bibfnamefont {C.}~\bibnamefont
  {Shekhar}}, , \bibinfo {author} {\bibfnamefont {A.~K.}\ \bibnamefont
  {Nayak}}, \bibinfo {author} {\bibfnamefont {Y.}~\bibnamefont {Sun}}, \bibinfo
  {author} {\bibfnamefont {M.}~\bibnamefont {Schmidt}}, \bibinfo {author}
  {\bibfnamefont {M.}~\bibnamefont {Nicklas}}, \bibinfo {author} {\bibfnamefont
  {I.}~\bibnamefont {Leermakers}}, \bibinfo {author} {\bibfnamefont
  {U.}~\bibnamefont {Zeitler}}, \bibinfo {author} {\bibfnamefont
  {Y.}~\bibnamefont {Skourski}}, \bibinfo {author} {\bibfnamefont
  {J.}~\bibnamefont {Wosnitza}}, \bibinfo {author} {\bibfnamefont
  {Z.}~\bibnamefont {Liu}}, \bibinfo {author} {\bibfnamefont {Y.}~\bibnamefont
  {Chen}}, \bibinfo {author} {\bibfnamefont {W.}~\bibnamefont {Schnelle}},
  \bibinfo {author} {\bibfnamefont {H.}~\bibnamefont {Borrmann}}, \bibinfo
  {author} {\bibfnamefont {Y.}~\bibnamefont {Grin}}, \bibinfo {author}
  {\bibfnamefont {C.}~\bibnamefont {Felser}}, \ and\ \bibinfo {author}
  {\bibfnamefont {B.}~\bibnamefont {Yan}},\ }\href {\doibase 10.1038/nphys3372}
  {\bibfield  {journal} {\bibinfo  {journal} {Nature Physics}\ }\textbf
  {\bibinfo {volume} {11}} (\bibinfo {year} {2015}),\
  10.1038/nphys3372}\BibitemShut {NoStop}%
\bibitem [{\citenamefont {Parameswaran}\ \emph {et~al.}(2014)\citenamefont
  {Parameswaran}, \citenamefont {Grover}, \citenamefont {Abanin}, \citenamefont
  {Pesin},\ and\ \citenamefont {Vishwanath}}]{Parameswaran2014}%
  \BibitemOpen
  \bibfield  {author} {\bibinfo {author} {\bibfnamefont {S.~A.}\ \bibnamefont
  {Parameswaran}}, \bibinfo {author} {\bibfnamefont {T.}~\bibnamefont
  {Grover}}, \bibinfo {author} {\bibfnamefont {D.~A.}\ \bibnamefont {Abanin}},
  \bibinfo {author} {\bibfnamefont {D.~A.}\ \bibnamefont {Pesin}}, \ and\
  \bibinfo {author} {\bibfnamefont {A.}~\bibnamefont {Vishwanath}},\ }\href
  {\doibase 10.1103/PhysRevX.4.031035} {\bibfield  {journal} {\bibinfo
  {journal} {Phys. Rev. X}\ }\textbf {\bibinfo {volume} {4}},\ \bibinfo {pages}
  {031035} (\bibinfo {year} {2014})}\BibitemShut {NoStop}%
\bibitem [{\citenamefont {Potter}\ \emph {et~al.}(2014)\citenamefont {Potter},
  \citenamefont {Kimchi},\ and\ \citenamefont {Vishwanath}}]{Potter2014}%
  \BibitemOpen
  \bibfield  {author} {\bibinfo {author} {\bibfnamefont {A.~C.}\ \bibnamefont
  {Potter}}, \bibinfo {author} {\bibfnamefont {I.}~\bibnamefont {Kimchi}}, \
  and\ \bibinfo {author} {\bibfnamefont {A.}~\bibnamefont {Vishwanath}},\
  }\href {\doibase 10.1038/ncomms6161} {\bibfield  {journal} {\bibinfo
  {journal} {Nature Communications}\ }\textbf {\bibinfo {volume} {5}} (\bibinfo
  {year} {2014}),\ 10.1038/ncomms6161}\BibitemShut {NoStop}%
\bibitem [{\citenamefont {Baum}\ \emph {et~al.}(2015)\citenamefont {Baum},
  \citenamefont {Berg}, \citenamefont {Parameswaran},\ and\ \citenamefont
  {Stern}}]{Baum2015}%
  \BibitemOpen
  \bibfield  {author} {\bibinfo {author} {\bibfnamefont {Y.}~\bibnamefont
  {Baum}}, \bibinfo {author} {\bibfnamefont {E.}~\bibnamefont {Berg}}, \bibinfo
  {author} {\bibfnamefont {S.~A.}\ \bibnamefont {Parameswaran}}, \ and\
  \bibinfo {author} {\bibfnamefont {A.}~\bibnamefont {Stern}},\ }\href
  {\doibase 10.1103/PhysRevX.5.041046} {\bibfield  {journal} {\bibinfo
  {journal} {Phys. Rev. X}\ }\textbf {\bibinfo {volume} {5}},\ \bibinfo {pages}
  {041046} (\bibinfo {year} {2015})}\BibitemShut {NoStop}%
\bibitem [{\citenamefont {Huang}\ \emph {et~al.}(2015)\citenamefont {Huang},
  \citenamefont {Zhao}, \citenamefont {Long}, \citenamefont {Wang},
  \citenamefont {Chen}, \citenamefont {Yang}, \citenamefont {Liang},
  \citenamefont {Xue}, \citenamefont {Weng}, \citenamefont {Fang},
  \citenamefont {Dai},\ and\ \citenamefont {Chen}}]{Huang2015}%
  \BibitemOpen
  \bibfield  {author} {\bibinfo {author} {\bibfnamefont {X.}~\bibnamefont
  {Huang}}, \bibinfo {author} {\bibfnamefont {L.}~\bibnamefont {Zhao}},
  \bibinfo {author} {\bibfnamefont {Y.}~\bibnamefont {Long}}, \bibinfo {author}
  {\bibfnamefont {P.}~\bibnamefont {Wang}}, \bibinfo {author} {\bibfnamefont
  {D.}~\bibnamefont {Chen}}, \bibinfo {author} {\bibfnamefont {Z.}~\bibnamefont
  {Yang}}, \bibinfo {author} {\bibfnamefont {H.}~\bibnamefont {Liang}},
  \bibinfo {author} {\bibfnamefont {M.}~\bibnamefont {Xue}}, \bibinfo {author}
  {\bibfnamefont {H.}~\bibnamefont {Weng}}, \bibinfo {author} {\bibfnamefont
  {Z.}~\bibnamefont {Fang}}, \bibinfo {author} {\bibfnamefont {X.}~\bibnamefont
  {Dai}}, \ and\ \bibinfo {author} {\bibfnamefont {G.}~\bibnamefont {Chen}},\
  }\href {\doibase 10.1103/PhysRevX.5.031023} {\bibfield  {journal} {\bibinfo
  {journal} {Phys. Rev. X}\ }\textbf {\bibinfo {volume} {5}},\ \bibinfo {pages}
  {031023} (\bibinfo {year} {2015})}\BibitemShut {NoStop}%
\bibitem [{\citenamefont {Arnold}\ \emph {et~al.}(2016)\citenamefont {Arnold},
  \citenamefont {Shekhar}, \citenamefont {Wu}, \citenamefont {Sun},
  \citenamefont {dos Reis}, \citenamefont {Kumar}, \citenamefont {Naumann},
  \citenamefont {Ajeesh}, \citenamefont {Schmidt}, \citenamefont {Grushin},
  \citenamefont {Bardarson}, \citenamefont {Baenitz}, \citenamefont {Sokolov},
  \citenamefont {Borrmann}, \citenamefont {Nicklas}, \citenamefont {Felser},
  \citenamefont {Hassinger},\ and\ \citenamefont {Yan}}]{Arnold2016}%
  \BibitemOpen
  \bibfield  {author} {\bibinfo {author} {\bibfnamefont {F.}~\bibnamefont
  {Arnold}}, \bibinfo {author} {\bibfnamefont {C.}~\bibnamefont {Shekhar}},
  \bibinfo {author} {\bibfnamefont {S.-C.}\ \bibnamefont {Wu}}, \bibinfo
  {author} {\bibfnamefont {Y.}~\bibnamefont {Sun}}, \bibinfo {author}
  {\bibfnamefont {R.~D.}\ \bibnamefont {dos Reis}}, \bibinfo {author}
  {\bibfnamefont {N.}~\bibnamefont {Kumar}}, \bibinfo {author} {\bibfnamefont
  {M.}~\bibnamefont {Naumann}}, \bibinfo {author} {\bibfnamefont {M.~O.}\
  \bibnamefont {Ajeesh}}, \bibinfo {author} {\bibfnamefont {M.}~\bibnamefont
  {Schmidt}}, \bibinfo {author} {\bibfnamefont {A.~G.}\ \bibnamefont
  {Grushin}}, \bibinfo {author} {\bibfnamefont {J.~H.}\ \bibnamefont
  {Bardarson}}, \bibinfo {author} {\bibfnamefont {M.}~\bibnamefont {Baenitz}},
  \bibinfo {author} {\bibfnamefont {D.}~\bibnamefont {Sokolov}}, \bibinfo
  {author} {\bibfnamefont {H.}~\bibnamefont {Borrmann}}, \bibinfo {author}
  {\bibfnamefont {M.}~\bibnamefont {Nicklas}}, \bibinfo {author} {\bibfnamefont
  {C.}~\bibnamefont {Felser}}, \bibinfo {author} {\bibfnamefont
  {E.}~\bibnamefont {Hassinger}}, \ and\ \bibinfo {author} {\bibfnamefont
  {B.}~\bibnamefont {Yan}},\ }\href {\doibase 10.1038/ncomms11615} {\bibfield
  {journal} {\bibinfo  {journal} {Nature Communications}\ }\textbf {\bibinfo
  {volume} {7}} (\bibinfo {year} {2016}),\ 10.1038/ncomms11615}\BibitemShut
  {NoStop}%
\bibitem [{\citenamefont {Zhang}\ \emph {et~al.}(2016)\citenamefont {Zhang},
  \citenamefont {Xu}, \citenamefont {Belopolski}, \citenamefont {Yuan},
  \citenamefont {Lin}, \citenamefont {Tong}, \citenamefont {Bian},
  \citenamefont {Alidoust}, \citenamefont {Lee}, \citenamefont {Huang},
  \citenamefont {Chang}, \citenamefont {Chang}, \citenamefont {Hsu},
  \citenamefont {Jeng}, \citenamefont {Neupane}, \citenamefont {Sanchez},
  \citenamefont {Zheng}, \citenamefont {Wang}, \citenamefont {Lin},
  \citenamefont {Zhang}, \citenamefont {Lu}, \citenamefont {Shen},
  \citenamefont {Neupert}, \citenamefont {Zahid~Hasan},\ and\ \citenamefont
  {Jia}}]{Zhang2016}%
  \BibitemOpen
  \bibfield  {author} {\bibinfo {author} {\bibfnamefont {C.-L.}\ \bibnamefont
  {Zhang}}, \bibinfo {author} {\bibfnamefont {S.-Y.}\ \bibnamefont {Xu}},
  \bibinfo {author} {\bibfnamefont {I.}~\bibnamefont {Belopolski}}, \bibinfo
  {author} {\bibfnamefont {Z.}~\bibnamefont {Yuan}}, \bibinfo {author}
  {\bibfnamefont {Z.}~\bibnamefont {Lin}}, \bibinfo {author} {\bibfnamefont
  {B.}~\bibnamefont {Tong}}, \bibinfo {author} {\bibfnamefont {G.}~\bibnamefont
  {Bian}}, \bibinfo {author} {\bibfnamefont {N.}~\bibnamefont {Alidoust}},
  \bibinfo {author} {\bibfnamefont {C.-C.}\ \bibnamefont {Lee}}, \bibinfo
  {author} {\bibfnamefont {S.-M.}\ \bibnamefont {Huang}}, \bibinfo {author}
  {\bibfnamefont {T.-R.}\ \bibnamefont {Chang}}, \bibinfo {author}
  {\bibfnamefont {G.}~\bibnamefont {Chang}}, \bibinfo {author} {\bibfnamefont
  {C.-H.}\ \bibnamefont {Hsu}}, \bibinfo {author} {\bibfnamefont {H.-T.}\
  \bibnamefont {Jeng}}, \bibinfo {author} {\bibfnamefont {M.}~\bibnamefont
  {Neupane}}, \bibinfo {author} {\bibfnamefont {D.~S.}\ \bibnamefont
  {Sanchez}}, \bibinfo {author} {\bibfnamefont {H.}~\bibnamefont {Zheng}},
  \bibinfo {author} {\bibfnamefont {J.}~\bibnamefont {Wang}}, \bibinfo {author}
  {\bibfnamefont {H.}~\bibnamefont {Lin}}, \bibinfo {author} {\bibfnamefont
  {C.}~\bibnamefont {Zhang}}, \bibinfo {author} {\bibfnamefont {H.-Z.}\
  \bibnamefont {Lu}}, \bibinfo {author} {\bibfnamefont {S.-Q.}\ \bibnamefont
  {Shen}}, \bibinfo {author} {\bibfnamefont {T.}~\bibnamefont {Neupert}},
  \bibinfo {author} {\bibfnamefont {M.}~\bibnamefont {Zahid~Hasan}}, \ and\
  \bibinfo {author} {\bibfnamefont {S.}~\bibnamefont {Jia}},\ }\href {\doibase
  10.1038/ncomms10735} {\bibfield  {journal} {\bibinfo  {journal} {Nature
  Communications}\ }\textbf {\bibinfo {volume} {7}} (\bibinfo {year} {2016}),\
  10.1038/ncomms10735}\BibitemShut {NoStop}%
\bibitem [{\citenamefont {Fradkin}(1986)}]{Fradkin1986}%
  \BibitemOpen
  \bibfield  {author} {\bibinfo {author} {\bibfnamefont {E.}~\bibnamefont
  {Fradkin}},\ }\href {\doibase 10.1103/PhysRevB.33.3263} {\bibfield  {journal}
  {\bibinfo  {journal} {Phys. Rev. B}\ }\textbf {\bibinfo {volume} {33}},\
  \bibinfo {pages} {3263} (\bibinfo {year} {1986})}\BibitemShut {NoStop}%
\bibitem [{\citenamefont {Altland}\ and\ \citenamefont
  {Bagrets}(2016)}]{Altland2016}%
  \BibitemOpen
  \bibfield  {author} {\bibinfo {author} {\bibfnamefont {A.}~\bibnamefont
  {Altland}}\ and\ \bibinfo {author} {\bibfnamefont {D.}~\bibnamefont
  {Bagrets}},\ }\href {\doibase 10.1103/PhysRevB.93.075113} {\bibfield
  {journal} {\bibinfo  {journal} {Phys. Rev. B}\ }\textbf {\bibinfo {volume}
  {93}},\ \bibinfo {pages} {075113} (\bibinfo {year} {2016})}\BibitemShut
  {NoStop}%
\bibitem [{\citenamefont {Sbierski}\ \emph {et~al.}(2017)\citenamefont
  {Sbierski}, \citenamefont {Madsen}, \citenamefont {Brouwer},\ and\
  \citenamefont {Karrasch}}]{Sbierski2017}%
  \BibitemOpen
  \bibfield  {author} {\bibinfo {author} {\bibfnamefont {B.}~\bibnamefont
  {Sbierski}}, \bibinfo {author} {\bibfnamefont {K.~A.}\ \bibnamefont
  {Madsen}}, \bibinfo {author} {\bibfnamefont {P.~W.}\ \bibnamefont {Brouwer}},
  \ and\ \bibinfo {author} {\bibfnamefont {C.}~\bibnamefont {Karrasch}},\
  }\href {\doibase 10.1103/PhysRevB.96.064203} {\bibfield  {journal} {\bibinfo
  {journal} {Phys. Rev. B}\ }\textbf {\bibinfo {volume} {96}},\ \bibinfo
  {pages} {064203} (\bibinfo {year} {2017})}\BibitemShut {NoStop}%
\bibitem [{\citenamefont {Pixley}\ \emph {et~al.}(2016)\citenamefont {Pixley},
  \citenamefont {Huse},\ and\ \citenamefont {Das~Sarma}}]{Pixley2016}%
  \BibitemOpen
  \bibfield  {author} {\bibinfo {author} {\bibfnamefont {J.~H.}\ \bibnamefont
  {Pixley}}, \bibinfo {author} {\bibfnamefont {D.~A.}\ \bibnamefont {Huse}}, \
  and\ \bibinfo {author} {\bibfnamefont {S.}~\bibnamefont {Das~Sarma}},\ }\href
  {\doibase 10.1103/PhysRevX.6.021042} {\bibfield  {journal} {\bibinfo
  {journal} {Phys. Rev. X}\ }\textbf {\bibinfo {volume} {6}},\ \bibinfo {pages}
  {021042} (\bibinfo {year} {2016})}\BibitemShut {NoStop}%
\bibitem [{\citenamefont {Buchhold}\ \emph {et~al.}(2018)\citenamefont
  {Buchhold}, \citenamefont {Diehl},\ and\ \citenamefont
  {Altland}}]{Buchhold2018}%
  \BibitemOpen
  \bibfield  {author} {\bibinfo {author} {\bibfnamefont {M.}~\bibnamefont
  {Buchhold}}, \bibinfo {author} {\bibfnamefont {S.}~\bibnamefont {Diehl}}, \
  and\ \bibinfo {author} {\bibfnamefont {A.}~\bibnamefont {Altland}},\ }\href
  {\doibase 10.1103/PhysRevLett.121.215301} {\bibfield  {journal} {\bibinfo
  {journal} {Phys. Rev. Lett.}\ }\textbf {\bibinfo {volume} {121}},\ \bibinfo
  {pages} {215301} (\bibinfo {year} {2018})}\BibitemShut {NoStop}%
\bibitem [{\citenamefont {Kotov}\ \emph {et~al.}(2012)\citenamefont {Kotov},
  \citenamefont {Uchoa}, \citenamefont {Pereira}, \citenamefont {Guinea},\ and\
  \citenamefont {Castro~Neto}}]{Kotov2012}%
  \BibitemOpen
  \bibfield  {author} {\bibinfo {author} {\bibfnamefont {V.~N.}\ \bibnamefont
  {Kotov}}, \bibinfo {author} {\bibfnamefont {B.}~\bibnamefont {Uchoa}},
  \bibinfo {author} {\bibfnamefont {V.~M.}\ \bibnamefont {Pereira}}, \bibinfo
  {author} {\bibfnamefont {F.}~\bibnamefont {Guinea}}, \ and\ \bibinfo {author}
  {\bibfnamefont {A.~H.}\ \bibnamefont {Castro~Neto}},\ }\href {\doibase
  10.1103/RevModPhys.84.1067} {\bibfield  {journal} {\bibinfo  {journal} {Rev.
  Mod. Phys.}\ }\textbf {\bibinfo {volume} {84}},\ \bibinfo {pages} {1067}
  (\bibinfo {year} {2012})}\BibitemShut {NoStop}%
\bibitem [{\citenamefont {Ward}(1950)}]{Ward1950}%
  \BibitemOpen
  \bibfield  {author} {\bibinfo {author} {\bibfnamefont {J.~C.}\ \bibnamefont
  {Ward}},\ }\href {\doibase 10.1103/PhysRev.78.182} {\bibfield  {journal}
  {\bibinfo  {journal} {Phys. Rev.}\ }\textbf {\bibinfo {volume} {78}},\
  \bibinfo {pages} {182} (\bibinfo {year} {1950})}\BibitemShut {NoStop}%
\bibitem [{\citenamefont {Peskin}\ and\ \citenamefont
  {Schroeder}(1995)}]{Peskin1995}%
  \BibitemOpen
  \bibfield  {author} {\bibinfo {author} {\bibfnamefont {M.~E.}\ \bibnamefont
  {Peskin}}\ and\ \bibinfo {author} {\bibfnamefont {D.~V.}\ \bibnamefont
  {Schroeder}},\ }\href@noop {} {\emph {\bibinfo {title} {{An introduction to
  quantum field theory}}}}\ (\bibinfo  {publisher} {Westview Press},\ \bibinfo
  {year} {1995})\BibitemShut {NoStop}%
\bibitem [{\citenamefont {González}\ \emph {et~al.}(1994)\citenamefont
  {González}, \citenamefont {Guinea},\ and\ \citenamefont
  {Vozmediano}}]{Gonzalez1994}%
  \BibitemOpen
  \bibfield  {author} {\bibinfo {author} {\bibfnamefont {J.}~\bibnamefont
  {González}}, \bibinfo {author} {\bibfnamefont {F.}~\bibnamefont {Guinea}}, \
  and\ \bibinfo {author} {\bibfnamefont {M.}~\bibnamefont {Vozmediano}},\
  }\href {\doibase https://doi.org/10.1016/0550-3213(94)90410-3} {\bibfield
  {journal} {\bibinfo  {journal} {Nuclear Physics B}\ }\textbf {\bibinfo
  {volume} {424}},\ \bibinfo {pages} {595 } (\bibinfo {year}
  {1994})}\BibitemShut {NoStop}%
\bibitem [{\citenamefont {Trescher}\ \emph {et~al.}(2015)\citenamefont
  {Trescher}, \citenamefont {Sbierski}, \citenamefont {Brouwer},\ and\
  \citenamefont {Bergholtz}}]{Trescher2015}%
  \BibitemOpen
  \bibfield  {author} {\bibinfo {author} {\bibfnamefont {M.}~\bibnamefont
  {Trescher}}, \bibinfo {author} {\bibfnamefont {B.}~\bibnamefont {Sbierski}},
  \bibinfo {author} {\bibfnamefont {P.~W.}\ \bibnamefont {Brouwer}}, \ and\
  \bibinfo {author} {\bibfnamefont {E.~J.}\ \bibnamefont {Bergholtz}},\ }\href
  {\doibase 10.1103/PhysRevB.91.115135} {\bibfield  {journal} {\bibinfo
  {journal} {Phys. Rev. B}\ }\textbf {\bibinfo {volume} {91}},\ \bibinfo
  {pages} {115135} (\bibinfo {year} {2015})}\BibitemShut {NoStop}%
\bibitem [{\citenamefont {van~der Wurff}\ and\ \citenamefont
  {Stoof}(2017)}]{Wurff2017}%
  \BibitemOpen
  \bibfield  {author} {\bibinfo {author} {\bibfnamefont {E.~C.~I.}\
  \bibnamefont {van~der Wurff}}\ and\ \bibinfo {author} {\bibfnamefont
  {H.~T.~C.}\ \bibnamefont {Stoof}},\ }\href {\doibase
  10.1103/PhysRevB.96.121116} {\bibfield  {journal} {\bibinfo  {journal} {Phys.
  Rev. B}\ }\textbf {\bibinfo {volume} {96}},\ \bibinfo {pages} {121116}
  (\bibinfo {year} {2017})}\BibitemShut {NoStop}%
\bibitem [{\citenamefont {Das}\ and\ \citenamefont {Agarwal}(2019)}]{Das2019}%
  \BibitemOpen
  \bibfield  {author} {\bibinfo {author} {\bibfnamefont {K.}~\bibnamefont
  {Das}}\ and\ \bibinfo {author} {\bibfnamefont {A.}~\bibnamefont {Agarwal}},\
  }\href {\doibase 10.1103/PhysRevB.99.085405} {\bibfield  {journal} {\bibinfo
  {journal} {Phys. Rev. B}\ }\textbf {\bibinfo {volume} {99}},\ \bibinfo
  {pages} {085405} (\bibinfo {year} {2019})}\BibitemShut {NoStop}%
\bibitem [{\citenamefont {Li}\ \emph {et~al.}(2017)\citenamefont {Li},
  \citenamefont {Wen}, \citenamefont {He}, \citenamefont {Zhang}, ,
  \citenamefont {Xia}, \citenamefont {Yu}, \citenamefont {Yang}, \citenamefont
  {Zhu}, \citenamefont {Alshareef},\ and\ \citenamefont {Zhang}}]{Li2017}%
  \BibitemOpen
  \bibfield  {author} {\bibinfo {author} {\bibfnamefont {P.}~\bibnamefont
  {Li}}, \bibinfo {author} {\bibfnamefont {Y.}~\bibnamefont {Wen}}, \bibinfo
  {author} {\bibfnamefont {X.}~\bibnamefont {He}}, \bibinfo {author}
  {\bibfnamefont {Q.}~\bibnamefont {Zhang}}, , \bibinfo {author} {\bibfnamefont
  {C.}~\bibnamefont {Xia}}, \bibinfo {author} {\bibfnamefont {Z.-M.}\
  \bibnamefont {Yu}}, \bibinfo {author} {\bibfnamefont {S.~A.}\ \bibnamefont
  {Yang}}, \bibinfo {author} {\bibfnamefont {Z.}~\bibnamefont {Zhu}}, \bibinfo
  {author} {\bibfnamefont {H.~N.}\ \bibnamefont {Alshareef}}, \ and\ \bibinfo
  {author} {\bibfnamefont {X.-X.}\ \bibnamefont {Zhang}},\ }\href {\doibase
  10.1038/s41467-017-02237-1} {\bibfield  {journal} {\bibinfo  {journal}
  {Nature Communications}\ }\textbf {\bibinfo {volume} {8}} (\bibinfo {year}
  {2017}),\ 10.1038/s41467-017-02237-1}\BibitemShut {NoStop}%
\bibitem [{\citenamefont {Soluyanov}\ \emph {et~al.}(2015)\citenamefont
  {Soluyanov}, \citenamefont {Gresch}, \citenamefont {Wang}, \citenamefont
  {Wu}, \citenamefont {Troyer}, \citenamefont {Dai},\ and\ \citenamefont
  {Bernevig}}]{Soluyanov2015typeii}%
  \BibitemOpen
  \bibfield  {author} {\bibinfo {author} {\bibfnamefont {A.~A.}\ \bibnamefont
  {Soluyanov}}, \bibinfo {author} {\bibfnamefont {D.}~\bibnamefont {Gresch}},
  \bibinfo {author} {\bibfnamefont {Z.}~\bibnamefont {Wang}}, \bibinfo {author}
  {\bibfnamefont {Q.}~\bibnamefont {Wu}}, \bibinfo {author} {\bibfnamefont
  {M.}~\bibnamefont {Troyer}}, \bibinfo {author} {\bibfnamefont
  {X.}~\bibnamefont {Dai}}, \ and\ \bibinfo {author} {\bibfnamefont {B.~A.}\
  \bibnamefont {Bernevig}},\ }\href {\doibase 10.1038/nature15768} {\bibfield
  {journal} {\bibinfo  {journal} {Nature}\ }\textbf {\bibinfo {volume} {527}},\
  \bibinfo {pages} {495} (\bibinfo {year} {2015})}\BibitemShut {NoStop}%
\bibitem [{\citenamefont {Goswami}\ and\ \citenamefont
  {Chakravarty}(2011)}]{Goswami2011}%
  \BibitemOpen
  \bibfield  {author} {\bibinfo {author} {\bibfnamefont {P.}~\bibnamefont
  {Goswami}}\ and\ \bibinfo {author} {\bibfnamefont {S.}~\bibnamefont
  {Chakravarty}},\ }\href {\doibase 10.1103/PhysRevLett.107.196803} {\bibfield
  {journal} {\bibinfo  {journal} {Phys. Rev. Lett.}\ }\textbf {\bibinfo
  {volume} {107}},\ \bibinfo {pages} {196803} (\bibinfo {year}
  {2011})}\BibitemShut {NoStop}%
\bibitem [{\citenamefont {Sikkenk}\ and\ \citenamefont
  {Fritz}(2017)}]{Sikkenk2017}%
  \BibitemOpen
  \bibfield  {author} {\bibinfo {author} {\bibfnamefont {T.~S.}\ \bibnamefont
  {Sikkenk}}\ and\ \bibinfo {author} {\bibfnamefont {L.}~\bibnamefont
  {Fritz}},\ }\href {\doibase 10.1103/PhysRevB.96.155121} {\bibfield  {journal}
  {\bibinfo  {journal} {Phys. Rev. B}\ }\textbf {\bibinfo {volume} {96}},\
  \bibinfo {pages} {155121} (\bibinfo {year} {2017})}\BibitemShut {NoStop}%
\bibitem [{\citenamefont {Detassis}\ \emph {et~al.}(2017)\citenamefont
  {Detassis}, \citenamefont {Fritz},\ and\ \citenamefont
  {Grubinskas}}]{Detassis2017}%
  \BibitemOpen
  \bibfield  {author} {\bibinfo {author} {\bibfnamefont {F.}~\bibnamefont
  {Detassis}}, \bibinfo {author} {\bibfnamefont {L.}~\bibnamefont {Fritz}}, \
  and\ \bibinfo {author} {\bibfnamefont {S.}~\bibnamefont {Grubinskas}},\
  }\href {\doibase 10.1103/PhysRevB.96.195157} {\bibfield  {journal} {\bibinfo
  {journal} {Phys. Rev. B}\ }\textbf {\bibinfo {volume} {96}},\ \bibinfo
  {pages} {195157} (\bibinfo {year} {2017})}\BibitemShut {NoStop}%
\bibitem [{\citenamefont {Edwards}\ and\ \citenamefont
  {Anderson}(1975)}]{Edwards1975}%
  \BibitemOpen
  \bibfield  {author} {\bibinfo {author} {\bibfnamefont {S.~F.}\ \bibnamefont
  {Edwards}}\ and\ \bibinfo {author} {\bibfnamefont {P.~W.}\ \bibnamefont
  {Anderson}},\ }\href {\doibase 10.1088/0305-4608/5/5/017} {\bibfield
  {journal} {\bibinfo  {journal} {Journal of Physics F: Metal Physics}\
  }\textbf {\bibinfo {volume} {5}},\ \bibinfo {pages} {965} (\bibinfo {year}
  {1975})}\BibitemShut {NoStop}%
\bibitem [{\citenamefont {Altland}\ and\ \citenamefont
  {Simons}(2010)}]{Altland2010}%
  \BibitemOpen
  \bibfield  {author} {\bibinfo {author} {\bibfnamefont {A.}~\bibnamefont
  {Altland}}\ and\ \bibinfo {author} {\bibfnamefont {B.~D.}\ \bibnamefont
  {Simons}},\ }\href {\doibase 10.1017/CBO9780511789984} {\emph {\bibinfo
  {title} {Condensed Matter Field Theory}}},\ \bibinfo {edition} {2nd}\ ed.\
  (\bibinfo  {publisher} {Cambridge University Press},\ \bibinfo {year}
  {2010})\BibitemShut {NoStop}%
\bibitem [{\citenamefont {Xiong}\ \emph {et~al.}(2015)\citenamefont {Xiong},
  \citenamefont {Kushwaha}, \citenamefont {Liang}, \citenamefont {Krizan},
  \citenamefont {Wang}, \citenamefont {Cava},\ and\ \citenamefont
  {Ong}}]{Xiong2015}%
  \BibitemOpen
  \bibfield  {author} {\bibinfo {author} {\bibfnamefont {J.}~\bibnamefont
  {Xiong}}, \bibinfo {author} {\bibfnamefont {S.~K.}\ \bibnamefont {Kushwaha}},
  \bibinfo {author} {\bibfnamefont {T.}~\bibnamefont {Liang}}, \bibinfo
  {author} {\bibfnamefont {J.~W.}\ \bibnamefont {Krizan}}, \bibinfo {author}
  {\bibfnamefont {W.}~\bibnamefont {Wang}}, \bibinfo {author} {\bibfnamefont
  {R.~J.}\ \bibnamefont {Cava}}, \ and\ \bibinfo {author} {\bibfnamefont
  {N.~P.}\ \bibnamefont {Ong}},\ }\href {https://arxiv.org/abs/1503.08179}
  {\bibfield  {journal} {\bibinfo  {journal} {Preprint, arXiv:1503.08179}\ }
  (\bibinfo {year} {2015})}\BibitemShut {NoStop}%
\bibitem [{\citenamefont {Zhao}\ and\ \citenamefont {Wang}(2018)}]{Zhao2018}%
  \BibitemOpen
  \bibfield  {author} {\bibinfo {author} {\bibfnamefont {P.-L.}\ \bibnamefont
  {Zhao}}\ and\ \bibinfo {author} {\bibfnamefont {A.-M.}\ \bibnamefont
  {Wang}},\ }\href {https://arxiv.org/abs/1811.11437} {\bibfield  {journal}
  {\bibinfo  {journal} {Preprint, arXiv:1811.11437v1}\ } (\bibinfo {year}
  {2018})}\BibitemShut {NoStop}%
\bibitem [{\citenamefont {Yang}\ \emph {et~al.}(2018)\citenamefont {Yang},
  \citenamefont {Wang},\ and\ \citenamefont {Liu}}]{Yang2018}%
  \BibitemOpen
  \bibfield  {author} {\bibinfo {author} {\bibfnamefont {Z.-K.}\ \bibnamefont
  {Yang}}, \bibinfo {author} {\bibfnamefont {J.-R.}\ \bibnamefont {Wang}}, \
  and\ \bibinfo {author} {\bibfnamefont {G.-Z.}\ \bibnamefont {Liu}},\ }\href
  {\doibase 10.1103/PhysRevB.98.195123} {\bibfield  {journal} {\bibinfo
  {journal} {Phys. Rev. B}\ }\textbf {\bibinfo {volume} {98}},\ \bibinfo
  {pages} {195123} (\bibinfo {year} {2018})}\BibitemShut {NoStop}%
\bibitem [{\citenamefont {Kobayashi}\ \emph {et~al.}(2014)\citenamefont
  {Kobayashi}, \citenamefont {Ohtsuki}, \citenamefont {Imura},\ and\
  \citenamefont {Herbut}}]{Kobayashi2014}%
  \BibitemOpen
  \bibfield  {author} {\bibinfo {author} {\bibfnamefont {K.}~\bibnamefont
  {Kobayashi}}, \bibinfo {author} {\bibfnamefont {T.}~\bibnamefont {Ohtsuki}},
  \bibinfo {author} {\bibfnamefont {K.-I.}\ \bibnamefont {Imura}}, \ and\
  \bibinfo {author} {\bibfnamefont {I.~F.}\ \bibnamefont {Herbut}},\ }\href
  {\doibase 10.1103/PhysRevLett.112.016402} {\bibfield  {journal} {\bibinfo
  {journal} {Phys. Rev. Lett.}\ }\textbf {\bibinfo {volume} {112}},\ \bibinfo
  {pages} {016402} (\bibinfo {year} {2014})}\BibitemShut {NoStop}%
\bibitem [{\citenamefont {Roy}\ and\ \citenamefont
  {Das~Sarma}(2014)}]{Roy2014}%
  \BibitemOpen
  \bibfield  {author} {\bibinfo {author} {\bibfnamefont {B.}~\bibnamefont
  {Roy}}\ and\ \bibinfo {author} {\bibfnamefont {S.}~\bibnamefont
  {Das~Sarma}},\ }\href {\doibase 10.1103/PhysRevB.90.241112} {\bibfield
  {journal} {\bibinfo  {journal} {Phys. Rev. B}\ }\textbf {\bibinfo {volume}
  {90}},\ \bibinfo {pages} {241112} (\bibinfo {year} {2014})}\BibitemShut
  {NoStop}%
\bibitem [{\citenamefont {Abrikosov}\ and\ \citenamefont
  {Beneslavskii}(1971)}]{Abrikosov1971}%
  \BibitemOpen
  \bibfield  {author} {\bibinfo {author} {\bibfnamefont {A.~A.}\ \bibnamefont
  {Abrikosov}}\ and\ \bibinfo {author} {\bibfnamefont {S.}~\bibnamefont
  {Beneslavskii}},\ }\href
  {http://www.jetp.ac.ru/cgi-bin/e/index/e/32/4/p699?a=list} {\bibfield
  {journal} {\bibinfo  {journal} {Soviet Physics JETP}\ }\textbf {\bibinfo
  {volume} {32}},\ \bibinfo {pages} {699} (\bibinfo {year} {1971})}\BibitemShut
  {NoStop}%
\bibitem [{\citenamefont {Gonz\'alez}(2017)}]{Gonzalez2017}%
  \BibitemOpen
  \bibfield  {author} {\bibinfo {author} {\bibfnamefont {J.}~\bibnamefont
  {Gonz\'alez}},\ }\href {\doibase 10.1103/PhysRevB.96.081104} {\bibfield
  {journal} {\bibinfo  {journal} {Phys. Rev. B}\ }\textbf {\bibinfo {volume}
  {96}},\ \bibinfo {pages} {081104} (\bibinfo {year} {2017})}\BibitemShut
  {NoStop}%
\bibitem [{\citenamefont {Vozmediano}\ \emph {et~al.}(2010)\citenamefont
  {Vozmediano}, \citenamefont {Katsnelson},\ and\ \citenamefont
  {Guinea}}]{Vozmediano2010}%
  \BibitemOpen
  \bibfield  {author} {\bibinfo {author} {\bibfnamefont {M.}~\bibnamefont
  {Vozmediano}}, \bibinfo {author} {\bibfnamefont {M.}~\bibnamefont
  {Katsnelson}}, \ and\ \bibinfo {author} {\bibfnamefont {F.}~\bibnamefont
  {Guinea}},\ }\href {\doibase https://doi.org/10.1016/j.physrep.2010.07.003}
  {\bibfield  {journal} {\bibinfo  {journal} {Physics Reports}\ }\textbf
  {\bibinfo {volume} {496}},\ \bibinfo {pages} {109 } (\bibinfo {year}
  {2010})}\BibitemShut {NoStop}%
\end{thebibliography}

%

\end{document}